\documentclass[aps,pra,superscriptaddress,twocolumn,longbibliography]{revtex4-1}

\usepackage{amssymb}
\usepackage{graphicx}
\usepackage{amsmath}
\usepackage{amsthm}
\usepackage{amsfonts}
\usepackage{color}
\usepackage{bm}
\usepackage{bbm}
\usepackage{url}
\usepackage{amssymb}
\usepackage[driverfallback=dvipdfm]{hyperref}
\hypersetup{pdfpagemode=FullScreen,colorlinks=true,breaklinks,urlcolor=blue,linkcolor=blue,citecolor=blue}

\usepackage{txfontsb}
\usepackage[OT1]{fontenc}

\usepackage{hyperref}
\usepackage{physics}
\usepackage{mathtools}
\usepackage{verbatim}

\begin{document}

\makeatletter
\newcommand{\Rmnum}[1]{\expandafter\@slowromancap\romannumeral #1@}
\makeatother

\global\long\def\id{\mathbbm{1}}
\global\long\def\ui{\mathbbm{i}}
\global\long\def\ud{\mathrm{d}}

\title{Ultracold atomic lattice systems for simulating topological phases: A review}

\author{Bei-Bei Wang}
\affiliation{School of Physics and Institute for Quantum Science and Engineering, Huazhong University of Science and Technology, Wuhan 430074, China}
\author{Xiao-Dong Lin}
\affiliation{Hefei National Research Center for Physical Sciences at the Microscale and School of Physical Sciences, University of Science and Technology of China, Hefei 230026, China}
\affiliation{School of Physics and Institute for Quantum Science and Engineering, Huazhong University of Science and Technology, Wuhan 430074, China}
\author{Jinyi Zhang}
\email{jinyi@ustc.edu.cn}
\affiliation{Hefei National Research Center for Physical Sciences at the Microscale and School of Physical Sciences, University of Science and Technology of China, Hefei 230026, China}
\affiliation{Shanghai Research Center for Quantum Sciences and CAS Center for Excellence in Quantum Information and Quantum Physics, University of Science and Technology of China, Shanghai 201315, China}
\affiliation{Hefei National Laboratory, Hefei 230088, China}
\author{Long Zhang}
\email{lzhangphys@hust.edu.cn}
\affiliation{School of Physics and Institute for Quantum Science and Engineering, Huazhong University of Science and Technology, Wuhan 430074, China}
\affiliation{Hefei National Laboratory, Hefei 230088, China}

\begin{abstract}

Owing to rapid recent progress, ultracold atomic lattice systems for simulating topological phases are now at a pivotal stage, 
evolving from established paradigms into increasingly versatile and programmable quantum simulators. In this review, we survey recent experimental advances across four major classes of platforms: optical lattices, including optical lattices with laser-assisted tunneling and optical Raman lattices; synthetic lattices in momentum or internal-state space; Floquet-engineered lattices; and optical tweezer arrays, all of which offer distinct capabilities for realizing and probing topological matter. For each class, we highlight representative experimental breakthroughs, the topological models that have been realized, and the advanced detection and characterization techniques employed, emphasizing how these complementary approaches collectively expand the frontier of quantum simulation. We also discuss emerging directions in strongly correlated and nonequilibrium topological phases, and conclude with an outlook on future prospects.

\end{abstract}

\maketitle

\section{Introduction}

\begin{figure*}
	\includegraphics[width=0.99\textwidth]{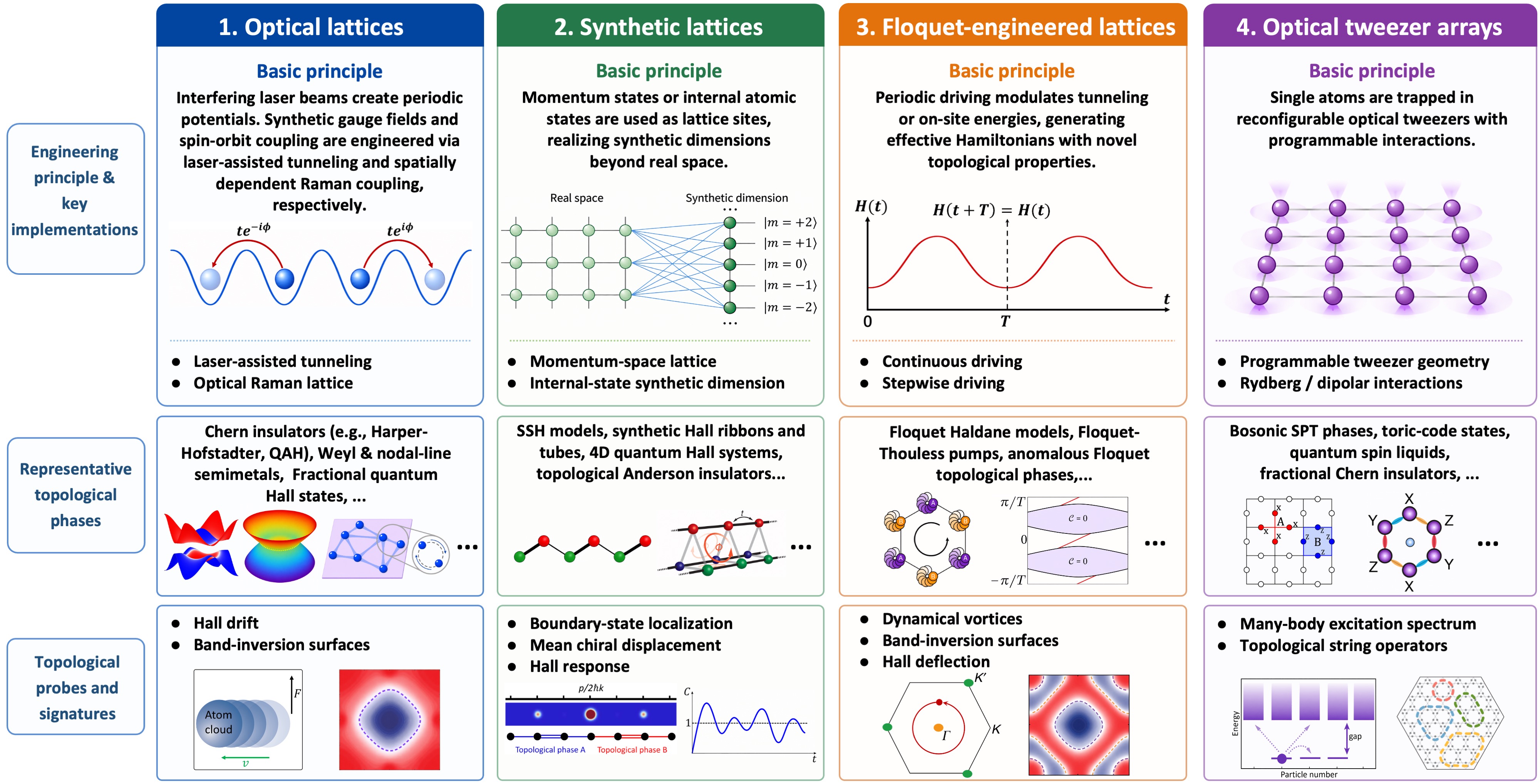}
	\caption{Roadmap of the four major ultracold-atom platforms for realizing and probing topological quantum matter. For each platform, we summarize the underlying physical principle, key implementation schemes, representative topological phases, and characteristic topological probes and signatures. The figure highlights the complementary capabilities of optical lattices, synthetic lattices, Floquet-engineered lattices, and optical tweezer arrays for exploring topological phenomena in ultracold atomic systems.
	}\label{Fig1}
\end{figure*}

Topological quantum matter represents a broad class of phases characterized by global, rather than local, properties of the many-body wavefunction. 
Owing to their topological nature, these systems exhibit remarkable robustness against local perturbations, disorder, and imperfections, 
leading to phenomena such as topologically protected and dissipationless edge or surface transport~\cite{Hasan2010_Review,Qi2011_Review}. 
Since the discovery of topological insulators and related phases~\cite{Schnyder2008,Kitaev2009,Ludwig2015_Review,Chiu2016_Review}, extensive theoretical and experimental efforts have revealed a rich landscape of topological states, including symmetry-protected topological (SPT) phases~\cite{Chen2012,Chen2013,Senthil2015_review}, topological semimetals~\cite{Yan2017_Review,Armitage2018_Review,Lv2021_Review}, and intrinsic topological orders~\cite{Wen1990,Wen2017}. 
These exotic properties not only deepen our understanding of quantum phases of matter but also open promising avenues for technological applications. 
In particular, topological protection provides a natural platform for fault-tolerant quantum computation~\cite{Nayak2008}, as exemplified by ongoing efforts toward topological qubits based on Majorana zero modes~\cite{Sarma2015}. Furthermore, the dissipationless transport properties and robustness against environmental noise hold great potential for developing low-power electronic and spintronic devices~\cite{Igor2004_review,Pesin2012_review,Lin2019_review}. 
Consequently, the search for novel topological quantum matter and the exploration of their physical properties have become central themes in contemporary condensed matter physics.

Despite significant progress in solid-state realizations~\cite{Hasan2010_Review,Qi2011_Review}, natural materials often suffer from limited tunability, uncontrolled disorder, and complex interactions, which pose challenges for the systematic exploration and precise characterization of topological phases. Ultracold atomic gases in optical lattices provide an alternative and highly versatile platform that overcomes many of these limitations. In these systems, key parameters---including lattice geometry, dimensionality, interaction strength, and {synthetic} gauge fields---can be engineered with high precision~\cite{Bloch2005_review,Bloch2008_Review,Windpassinger2013_Review,Dalibard2011_Review,Goldman2014_Review,Gross2017_Review,Schafer2020_Review,Zohar2016,Aidelsburger2021_review,Halimeh2025}, enabling the realization of a wide variety of model Hamiltonians that are difficult or impossible to access in conventional materials. Moreover, advanced experimental techniques, such as quantum gas microscopy~\cite{Kuhr2016_review,Gross2021_review} and interferometric measurements~\cite{Vale2021_review}, allow for the direct detection of topological invariants and dynamical observables. 
For instance, recent experiments have successfully realized both dynamical~\cite{Lohse2018} and genuine~\cite{Bouhiron2024} four-dimensional (4D) quantum Hall systems in ultracold atomic platforms, enabling direct measurement of the second Chern number---a key high-dimensional topological invariant---in both systems, as well as the observation of non-planar cyclotron orbits and anisotropic hyperedge modes~\cite{Bouhiron2024}, phenomena that are inherently inaccessible in conventional materials. 
These capabilities establish ultracold atomic systems as a powerful quantum simulation platform for exploring topological phenomena.

In recent years, the field has witnessed rapid progress on multiple fronts, particularly driven by experimental breakthroughs. 
First, there has been a clear shift from the study of single-particle topological band structures toward strongly correlated topological phases. 
Notable experimental advances include the realization of interacting SPT phases~\cite{De2019,Sompet2022,Yue2026} and the exploration of fractional quantum Hall (FQH) states~\cite{Leonard2023FQH}, chiral spin liquids~\cite{Semeghini2021,Evered2025}, and other topologically ordered phases in engineered quantum systems. 
In parallel, theoretical frameworks have also advanced significantly, with a rich variety of exotic topological states having been proposed~\cite{Potirniche2017,Zhou2017,Li2021,Verresen2021,Tarabunga2022,Verresen2022,Weber2022,Kalinowski2023,Sun2023,Kamal2024,Chen2024,Mogerle2025}. 
These developments mark an important transition from band topology to genuinely many-body topological phenomena, where entanglement and interactions play a central role.

Second, significant conceptual and technical advances have broadened the scope of experimental platforms and methodologies. 
On the one hand, there is a growing interest in nonequilibrium topological physics, including Floquet engineering~\cite{Eckardt2017_review,Rudner2020_review,Weitenberg2021_review,Harper2020_Review}, quantum quenches~\cite{McGinley2018,McGinley2019a,Gong2018a,Wang2017,ZhangLin2018,ZhangLin2022}, 
and non-Hermitian physics~\cite{Bergholtz2021_Review,Ding2022_Review,Okuma2023_Review}, 
which enable the realization of dynamical topological phases beyond equilibrium constraints~\cite{Wintersperger2020,Zhang2023tuning,ZhangH2024,Ren2022,Zhao2025}. 
On the other hand, experimental platforms are evolving from conventional optical lattices to more flexible architectures, 
with optical tweezer arrays emerging as a particularly powerful example~\cite{Browaeys2020_review,Morgado2021_review,Kaufman2021,Wu2021,Cheng2024}. 
These new platforms offer enhanced control at the {single-site} level and facilitate the engineering of tailored interactions and geometries, 
thereby opening new directions for quantum simulation.

Given these rapid developments, the field is currently at a pivotal stage, transitioning from established paradigms toward more versatile and programmable quantum simulators. 
Many of these emerging directions remain in their infancy, with substantial opportunities for both theoretical exploration and experimental realization. 
Therefore, although several earlier reviews on quantum simulation of topological phases with ultracold atoms exist~\cite{Goldman2016,ZhangDW2018,Cooper2019},
most of them date back approximately seven to eight years and, while still valuable, do not capture the significant experimental advances of the past few years. Furthermore, existing reviews have largely focused on individual platform classes or specific phenomena; a concise, structured overview that systematically compares the four complementary experimental platforms---optical lattices, synthetic lattices, Floquet-engineered lattices, and optical tweezer arrays---under a unified framework is currently lacking. Therefore, a timely synthesis of recent progress---with a particular focus on experimental achievements---along with a forward-looking perspective on future research directions, is essential for guiding further advances in ultracold-atom-based topological quantum matter.

\section{Recent experimental progress in diverse lattice systems}

In this section, we survey recent experimental advances in simulating topological quantum matter using different ultracold atomic lattice systems. 
Specifically, we cover four major classes of platforms: 
(i) optical lattices, including optical lattices with laser-assisted tunneling and optical Raman lattices, 
which provide periodic potentials and synthetic gauge fields for realizing topological band structures; 
(ii) synthetic lattices, implemented in momentum space or via internal atomic states, 
which enable site-resolved control or the engineering of higher-dimensional and flexible lattice geometries that are difficult to achieve in real space; 
(iii) Floquet-engineered lattices, where explicit periodic driving is used to create stroboscopic effective Hamiltonians with tunable topological properties, 
including conventional topological models and anomalous Floquet insulators; 
and (iv) optical tweezer arrays, a rapidly developing reconfigurable platform that offers single-site addressability and flexible geometry, 
facilitating the preparation and manipulation of strongly correlated topological states. 
Figure \ref{Fig1} provides a concise overview of the four experimental platforms and serves as a roadmap for the organization of the following subsections.

\subsection{Optical lattices}

In this subsection, we focus on topological phases realized in pure optical lattice platforms---all-optical systems 
where interfering lasers create periodic potentials that serve as clean, controllable analogues of crystalline solids~\cite{Schafer2020_Review}. 
These platforms enable precise engineering of lattice geometries, tunneling amplitudes, and effective gauge fields, 
making them a primary setting for exploring topological band structures. 
The discussion is organized into two categories: optical lattices with laser-assisted tunneling and optical Raman lattices.
Laser-assisted tunneling is included here rather than under Floquet-engineered systems, 
as its time-dependent couplings can be described by an effective static Hamiltonian via the rotating-wave approximation (RWA)~\cite{Cooper2019}, 
and it does not require additional periodic modulation beyond the all-optical setup. 
We reserve the term ``Floquet-engineered lattices'' for cases where explicit periodic driving (e.g., lattice shaking) plays an essential role (see Sec.~\ref{Sec: Floquet-engineered lattices}).

\begin{figure*}
	\includegraphics[width=0.98\textwidth]{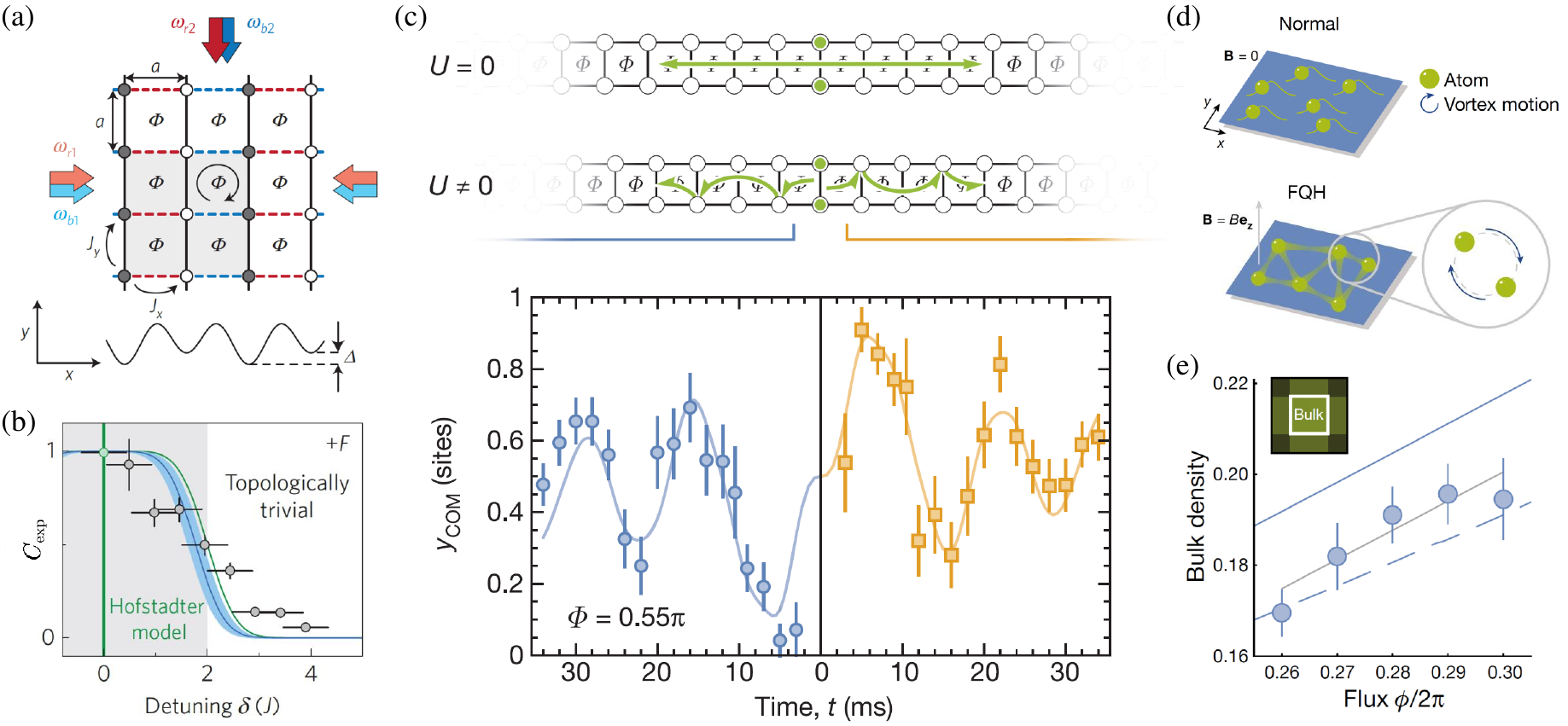}
	\caption{(a) Experimental setup for generating a uniform artificial magnetic field via laser-assisted tunneling. A 2D optical lattice with inhibited tunneling along $x$ is combined with two pairs of laser beams (red and blue) that restore tunneling independently on selected red and blue links. This produces a rectified flux $\Phi=\pi/2$ per plaquette and a magnetic unit cell four times larger than the original lattice cell. 
	(b) Measured Chern number $C_{\text{exp}}$ versus staggered detuning $\delta$. The topological phase transition near $\delta=2J$  is broadened by experimental uncertainties. Panels (a) and (b) are reprinted from Ref.~\cite{Aidelsburger2015}.
	(c) Interaction‑induced chiral dynamics for two particles in a Harper‑Hofstadter ladder. Top: When interactions are present ($U\neq0$), chirality emerges for particles initially placed on opposite sides of the central rung. Bottom: This chirality is manifested in the center-of-mass coordinate $y_{\rm COM}$ (blue and orange data for left and right halves of the system, respectively). Solid lines represent results from exact diagonalization. Panel (c) is reprinted from Ref.~\cite{Tai2017}.
	(d) In the presence of a strong effective magnetic field, strong correlations emerge in Laughlin-type states, characterized by correlated vortex motion between atom pairs, in contrast to the case without a magnetic field, where the system remains in a superfluid state with only weak correlations.
	(e) Fractional Hall conductance extracted from bulk density versus magnetic flux. A linear fit gives $\sigma_H / \sigma_0=0.6(2)$, consistent with the expected  $1/2$ in the thermodynamic limit, confirming the topological response of the FQH state.
	Panels (d) and (e) are reprinted from Ref.~\cite{Leonard2023FQH}.
	}
	\label{Fig2}
\end{figure*}

\subsubsection{Laser-assisted tunneling}

Laser-assisted tunneling provides a powerful and versatile technique for engineering synthetic magnetic flux in optical lattices. 
The basic idea is to restore otherwise suppressed intersite hopping using auxiliary laser fields, 
which simultaneously imprint a controllable phase onto the tunneling matrix elements~\cite{Jaksch2003,Gerbier2010}. 
In a typical implementation, a static tilt or energy offset is introduced to inhibit bare tunneling.
Raman or other laser-assisted couplings supply the required energy and momentum, 
enabling atoms to hop resonantly while acquiring a Peierls phase. 
As a result, the atoms experience an effective gauge field analogous to a magnetic field acting on charged particles~\cite{Jaksch2003}. The accumulated Peierls phases around a plaquette mimic magnetic fluxes threading the lattice, naturally realizing Harper-Hofstadter-type models. Depending on the lattice geometry and phase engineering, 
this approach can generate either staggered or uniform synthetic flux patterns~\cite{Aidelsburger2011,Aidelsburger2013,Miyake2013,Aidelsburger2015}.

We illustrate the realization of the Harper-Hofstadter model using laser-assisted tunneling, following the experiment reported in Ref.~\cite{Aidelsburger2015}.
As shown in Fig.~\ref{Fig2}(a), an ultracold gas of $^{87}$Rb atoms was loaded into a 2D optical lattice formed by two orthogonal standing waves with the same wavelength $\lambda_s$.
Confinement along the perpendicular direction was provided by a weak harmonic potential. 
To suppress bare tunneling along the $x$‑direction, an additional standing wave of wavelength $\lambda_L = 2\lambda_s$ is applied, 
creating a staggered potential that introduced an energy offset $\Delta$ much larger than the bare nearest-neighbor tunneling $J_x$. 
Resonant tunneling was subsequently restored using two pairs of laser beams, 
whose frequency differences were tuned close to the resonance condition $\hbar\omega=\Delta+\delta$,
where $\delta$ denotes the residual detuning from exact Raman resonance.
This near-resonant laser-assisted process imparts a controlled tunneling phase via momentum transfer from the Raman beams, yielding direction-dependent Peierls phases and complex hopping amplitudes~\cite{Aidelsburger2013,Miyake2013}.
Each beam pair consisted of a running wave along $y$ and a retro‑reflected beam along $x$, 
enabling control of tunneling on successive links [Fig.~\ref{Fig2}(a)].
In the regime $\hbar\omega\gg J_x,J_y$, where the RWA is valid, and at exact resonance ($\delta=0$), this configuration realizes the Harper-Hofstadter Hamiltonian
\begin{align}
H_{\rm HH}=-J\sum_{mn}\left(e^{\ui n\Phi} a^\dagger_{m+1,n}a_{m,n}+a^\dagger_{m,n+1}a_{m,n}+{\rm H.c.}\right),
\end{align}
with a uniform synthetic magnetic flux $\Phi = \pi / 2$ per plaquette.
Here, $a_{m,n}$ ($a^\dagger_{m,n}$) annihilates (creates) a particle on site $(m,n)$, 
and the effective coupling strengths along the $x$ and $y$ directions are taken to be equal, with value $J$.
The lowest Hofstadter band of this model carries a Chern number $C=+1$, a topological invariant defined by the integral of the Berry curvature over the first Brillouin zone.
In the experiment, the Chern number was extracted from the transverse drift of the atomic cloud under an applied force induced by an optical gradient~\cite{Aidelsburger2015}.
As shown in Fig.~\ref{Fig2}(b), the measured value, $C_{\rm exp}=0.99(5)$, is in excellent agreement with the theoretical prediction for small $\delta$.

Using laser‑assisted tunneling, experiments have successfully implemented and probed the Hofstadter-Harper model~\cite{Aidelsburger2013,Miyake2013,Aidelsburger2015,Tai2017}.
A defining feature of this model is the emergence of cyclotron orbits associated with the Aharonov-Bohm phase, whose chirality and dynamics are directly controlled by the imposed flux. 
In an early experiment~\cite{Aidelsburger2013}, the time evolution of the center-of-mass position of spin-resolved $^{87}$Rb atoms revealed clear signatures of such quantum cyclotron motion, providing an intuitive and direct confirmation of the underlying synthetic gauge field.

Building on this foundation, subsequent work moved beyond single-particle physics to investigate the interplay between interactions and magnetic flux, 
which is essential for accessing strongly correlated topological phases. 
In particular, an interacting Harper-Hofstadter ladder was realized in the few-body regime~\cite{Tai2017}, where two bosons were initialized on adjacent sites of a single rung.
While in the noninteracting limit the population of chiral bands remains symmetric, resulting in no net chiral motion, the introduction of finite repulsive interactions breaks this balance and induces pronounced chiral dynamics, characterized by the transverse center-of-mass displacement $y_{\rm COM}$ [see Fig.~\ref{Fig2}(c)].
The resulting chirality was shown to depend sensitively on both the magnetic flux $\Phi$ and the interaction strength $U$~\cite{Tai2017}, 
thereby providing a minimal and highly controlled setting for understanding interaction-induced chirality in topological systems.

A major milestone along this direction was the experimental realization of a lattice analogue of a bosonic $\nu=1/2$ Laughlin state in a $4\times4$ lattice~\cite{Leonard2023FQH}.  
In this experiment, a strong synthetic magnetic field generated via laser-assisted tunneling drove the system into a FQH regime despite the small particle number. The hallmark correlations of the Laughlin state manifest as collective vortex motion shared among particle pairs, accompanied by an effective screening of on-site interactions, as illustrated in Fig.~\ref{Fig2}(d). Experimentally, these features were revealed through the observation of characteristic density-density correlations and a strong suppression of doublon probability~\cite{Leonard2023FQH}. Furthermore, a fractional Hall response was extracted from the bulk density response to an increasing flux, with the measured Hall conductance 
$\sigma_H\approx 0.6\sigma_0$, close to the expected value of $1/2$ [Fig.~\ref{Fig2}(e)]. 
More recently, this line of research has been extended toward non-Abelian FQH physics, including Pfaffian-type states of ultracold bosons in engineered synthetic gauge fields~\cite{Kwan2026}.
Together, these results constitute compelling evidence for the emergence of strongly correlated FQH physics in laser-engineered optical lattices, highlighting laser-assisted tunneling as a powerful tool for accessing exotic topological states.

\subsubsection{Optical Raman lattices}

Another distinct class of optical lattice schemes that has been successfully implemented in recent experiments is that of optical Raman lattices~\cite{Liu2013,Liu2014,ZhangBook}. 
The core concept of an optical Raman lattice involves the simultaneous generation of a conventional optical lattice and a periodic Raman coupling using the same laser fields, typically in the form of standing waves. 
In this scheme, the lattice potential governs normal (spin-conserving) tunneling of atoms, whereas the Raman coupling induces spin-flip tunneling. 
{These two components are intrinsically integrated, with their relative spatial configuration naturally enforcing fixed spatial symmetries without external fine-tuning.
This approach was first proposed by Liu {\it et al.}~\cite{Liu2013,Liu2014} and has since become a versatile platform for exploring topological quantum matter~\cite{WangBZ2018,Lu2020,Liu2016,ZhangDW2016,Zheng2019,WangJT2024,LinXD2026}.} A key advantage of this architecture is that its built-in spatial symmetries greatly facilitate the realization of multi-dimensional spin-orbit (SO) coupling, enabling the emergence of a wide range of topological phases---from one-dimensional (1D) and 2D topological insulators~\cite{Song2018,Wu2016,Sun2018a,Liang2023} to three-dimensional (3D) nodal-line~\cite{Song2019} and Weyl semimetals~\cite{Wang2021}.

\begin{figure*}
	\includegraphics[width=0.95\textwidth]{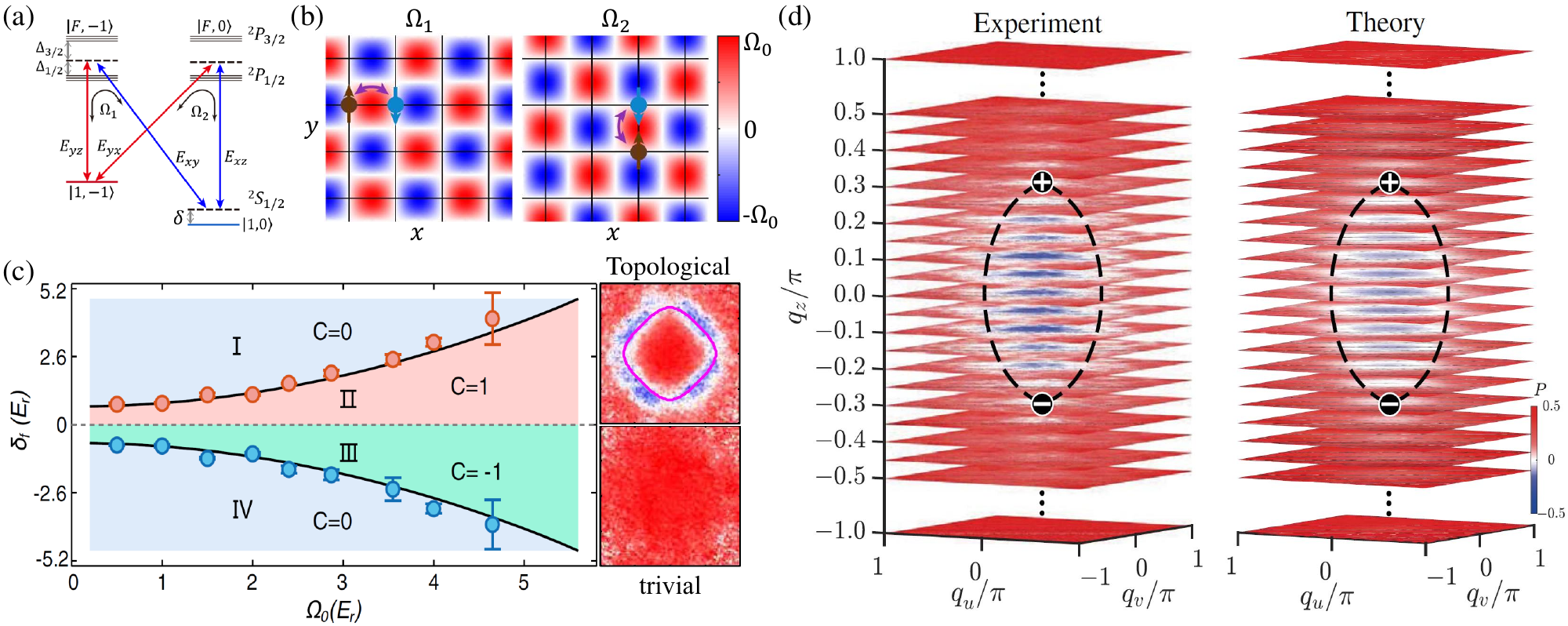}
	\caption{(a) Level structure and Raman coupling scheme for realizing 2D QAH model. 
	(b) Real‑space antisymmetric structure of the two Raman couplings $\Omega_1$ and $\Omega_2$. The grid represent the square optical lattice $V_{\text{latt}}(x, y)$.
	Panels (a) and (b) are reprinted from Ref.~\cite{Sun2018a}. 
	(c) Left: Topological phase diagram of the lowest band for the QAH model measured by quench dynamics. Here $\delta_f$ denotes the post-quench value of the two-photon detuning, while $\Omega_0$ denotes the Raman coupling strength, with $\Omega_{01}=\Omega_{02}=\Omega_0$. Experimental data (blue and red dots) agree with theoretical calculations (solid lines), identifying trivial areas ($C=0$) and nontrivial areas ($C=1$ and $C=-1$). Right: Time-evolved spin texture in the first Brillouin zone at fixed evolution time for two detuning values, where the emergence of a band-inversion ring marks a topologically nontrivial regime. Panel (c) is reprinted from Ref.~\cite{Sun2018b}.
	(d) 3D spin texture reconstruction directly revealing two Weyl points in the realization of the ideal Weyl semimetal. Experimental data (left) match theoretical calculations (right), with Weyl points marked by $\oplus$ and $\ominus$ and the {band-inversion rings} indicated by dashed lines. Panel (d) is reprinted from Ref.~\cite{Wang2021}.}
	\label{Fig3}
\end{figure*}

The construction of an optical Raman lattice can be illustrated through its application to realizing 2D topological insulators, such as the quantum anomalous Hall (QAH) model in a bosonic $^{87}$Rb system~\cite{Sun2018a}. In this implementation, a 2D spin-independent square lattice is formed by two linearly polarized laser beams, each comprising two orthogonally polarized components ($E_{xz}, E_{xy}$ and $E_{yz}, E_{yx}$, respectively), with a shared wavelength $\lambda$.
The lattice potential is given by $V_{\text{latt}}(x, y) = V_{0x} \cos^2(k_0 x) + V_{0y} \cos^2(k_0 y)$, where $k_0 = 2\pi/\lambda$, and the depths $V_{0x}\propto|E_{xy}|^2-|E_{xz}|^2$ and $V_{0y}\propto|E_{yx}|^2-|E_{yz}|^2$ are tunable. 
As depicted in Fig.~\ref{Fig3}(a), the Raman couplings between the two spin states, 
which can be written as $(\Omega_1+\Omega_2 e^{i\delta\varphi})|\!\uparrow\rangle\langle\downarrow\!|+{\rm H.c.}$,
are generated by the orthogonal polarization pairs ($E_{xz}, E_{yx}$) and ($E_{xy}, E_{yz}$) in a double-$\Lambda$ configuration, resulting in spatially periodic coupling strengths $\Omega_1 = \Omega_{01} \sin(k_0 x) \cos(k_0 y)$ and $\Omega_2 = \Omega_{02} \cos(k_0 x) \sin(k_0 y)$, as shown in Fig.~\ref{Fig3}(b). 
Here, $\Omega_{01} \propto |E_{yz}| |E_{xy}|$ and $\Omega_{02} \propto |E_{yx}| |E_{xz}|$. 
A critical feature is the antisymmetric structure of these Raman couplings along one spatial direction, which ensures that spin-flip tunneling occurs only between nearest neighbors. The relative phase $\delta\varphi$ between the two couplings is controlled by an electro-optic phase modulator.
When $\delta\varphi=\pm\pi/2$, the two spatially dependent Raman potentials $\Omega_1$ and $\Omega_2$ couple to two orthogonal spin components,
thereby realizing a genuine 2D SO coupling~\cite{Sun2018a}.
The total Hamiltonian is given by
\begin{equation}
H_{\rm 2D}= \frac{{\bf p}^2}{2m} + V_{\text{latt}}(x, y) + \Omega_1(x, y)\sigma_x+ \Omega_2(x, y)\sigma_y+ \frac{\hbar\delta}{2} \sigma_z,
\end{equation}
where $m$ is the atomic mass and $\delta$ is the two-photon detuning.
This configuration realizes a highly controllable 2D SO coupling and, 
together with the Raman detuning term acting as an effective Zeeman field, forms an effective QAH model, with its topological phase diagram shown in Fig.~\ref{Fig3}(c).
The topological phase boundaries are governed by the relative strength of the Raman-induced SO coupling and the effective Zeeman field set by the detuning.

Extending this approach to three dimensions enables the simulation of gapless topological semimetals. The realization of an ideal Weyl semimetal band in a $^{87}\text{Rb}$ system~\cite{Wang2021} serves as a prime example, in which three laser beams of wavelength $\lambda$ are retroreflected along the $x$, $y$, and $z$ directions to construct a 3D optical Raman lattice. In the $xy$ plane, the beams form a checkerboard lattice. Transforming to rotated coordinates $u=(x+y)/\sqrt{2}$ and $v=(x-y)/\sqrt{2}$, the 3D lattice potential becomes $V_{\text{latt}}(u,\nu,z)=V_{\rm 2D}(\cos^2 k_1 u + \cos^2 k_1 v)+V_{z}\cos^2 k_0 z$, where $k_1=k_0/\sqrt{2}$. The simultaneously generated Raman couplings exhibit 3D structures, given by $\Omega_u(u, v, z) \sigma_x + \Omega_v(u, v, z) \sigma_y$, with $\Omega_u(u, v, z) = \Omega_0 \cos k_0 z \sin k_1 u \cos k_1 v$ and $\Omega_v = \Omega_u (u \leftrightarrow v, z)$. The antisymmetry of $\Omega_u$ ($\Omega_v$) along $u$ ($v$) direction facilitates spin-flip hopping in the $u$ ($v$) direction. The resulting tight-binding Hamiltonian in Bloch momentum space is equivalent, for each fixed  $q_z$, to a 2D QAH model in the $uv$ plane, with its topology evolving as $q_z$ varies, leading to the emergence of Weyl points. This scheme enabled the experimental realization of an ideal Weyl semimetal {hosting only two Weyl nodes, which were clearly resolved through} virtual slicing imaging and quench dynamics~\cite{Wang2021}. As shown in Fig.~\ref{Fig3}(d), the 3D spin distribution can be reconstructed from 2D spin textures by scanning the two-photon detuning, a procedure referred to as virtual slicing imaging~\cite{Lu2020,Song2019}. The results show that, as $q_z$ is tuned away from zero, the band-inversion rings in the 2D slices shrink and eventually disappear at the Weyl nodes, providing a direct way to locate the Weyl points.

The unique characteristics of optical Raman lattices---particularly the intrinsic spatial symmetry and the precise control over Raman couplings---have made them an exceptional platform for exploring a wide array of topological phases.
Early work realized a 1D topological phase in fermionic $^{173}\text{Yb}$ atoms, demonstrating a SPT phase and associated quench dynamics~\cite{Song2018}. Subsequently, 2D topological phases were achieved. The first realization of the QAH model in a bosonic $^{87}\text{Rb}$ system~\cite{Wu2016} was followed by a more advanced scheme that addressed prior limitations, such as limited tunability and symmetry constraints, resulting in a highly stable and controllable 2D SO-coupled Bose-Einstein condensate (BEC) with an extended topological region~\cite{Sun2018a}. More recently, the QAH model was realized in a fermionic $^{87}\text{Sr}$ system~\cite{Liang2023}, where a pump-probe quench measurement protocol was developed to characterize the topology. In parallel, optical Raman lattices have been instrumental in realizing gapless topological semimetals. Beyond the ideal Weyl semimetal~\cite{Wang2021}, this platform has enabled the realization of nodal-line semimetals~\cite{Song2019}. In such systems, researchers have successfully characterized 3D topological band structures using spin texture measurements and topological quench dynamics, allowing for the observation of bulk nodal lines and the full reconstruction of 3D topological bands~\cite{Song2019}.

Notably, nonequilibrium approaches implemented in optical Raman lattices have proven highly effective for topological characterization. 
{Following a quench from a trivial to a topological regime, characteristic ring structures emerge in the spin dynamics, directly revealing the band-inversion surfaces (BISs) of the post-quench Hamiltonian~\cite{ZhangLin2018}. Since BISs constitute the fundamental manifolds that encode the bulk topology, the emergence and configuration of the ring structures provide direct dynamical signatures of topological phases. A dynamical measurement of the phase diagram of the QAH model is shown in Fig.~\ref{Fig3}(c), where the disappearance of the ring pattern at the $\Gamma$ ($M$) point marks the upper (lower) boundary of the topological region, enabling an experimental determination of the phase diagram and the associated Chern numbers~\cite{Sun2018b}.}
Furthermore, this platform has facilitated fundamental studies of topological quench dynamics, including verification of the Kibble-Zurek mechanism~\cite{Yuan2025} and demonstration of dynamical bulk-surface correspondence~\cite{Yi2019}, {which shed light on the universal nonequilibrium dynamics of topological systems}.
The combination of versatile SO-coupling engineering with powerful nonequilibrium probes continues to {place optical Raman lattices at the forefront of research on topological quantum matter}.

\subsection{Synthetic lattices}

Synthetic lattices refer to artificial lattice structures constructed by utilizing momentum states~\cite{Gadway2015,Meier2016,Yuan2023,Dong2025,Xie2019,Paladugu2024,Xiao2021,Meier2018,Zeng2024,Li2022}, atomic internal states~\cite{Boada2012,Celi2014,Stuhl2015,Mancini2015,Chalopin2020,Kanungo2022,Tsuno2025,Roell2023,Zhou2023,Zhou2025,Han2019,Li2022bose,Fabre2022laughlin,Bouhiron2024}, time~\cite{Lohse2018,Nakajima2021,Minguzzi2022,Walter2023,Viebahn2024,Zhu2024}, or other degrees of freedom as effective ``spatial dimensions.'' By transcending the constraints inherent to real-space lattices---such as fixed geometries, limited dimensionality, and restricted tunability---these synthetic platforms enable the creation of entirely new lattice geometries with arbitrary dimensionality and offer unprecedented control over Hamiltonian parameters. This flexibility opens the door to realizing topological phases that are otherwise inaccessible in conventional settings~\cite{Ozawa2019_review,Arguello-Luengo2024,Fabre2024}. Within this broad framework, two particularly prominent and complementary approaches have emerged: momentum-space synthetic lattices, which exploit discretized atomic momentum states as lattice sites, and internal-state synthetic dimensions, which leverage atomic spin or hyperfine states to construct synthetic lattices. The principles, implementations, and key topological applications of each platform are discussed in the following.

\subsubsection{Momentum-space synthetic lattices}

Momentum lattices are artificial structures constructed in momentum space, where discrete momentum states serve as synthetic lattice sites. 
Tunneling between these sites is {typically} induced by laser-driven two-photon Bragg transitions, enabling the emulation of tight-binding models in a highly controllable fashion. 
The concept of using a discrete set of free-particle momentum states to construct a single-particle quantum transport model was first introduced in Ref.~\cite{Gadway2015}, 
laying a theoretical foundation for topological simulations in momentum space. 
Unlike traditional real-space optical lattices, which are constrained by fixed geometries and site-resolved manipulation challenges, momentum lattices offer exceptional flexibility. 
They allow for independent control of individual sites via precise adjustments of laser phase, intensity, and detuning. 
Furthermore, experimental detection is greatly simplified: instead of requiring complex {\it in situ} imaging, time-of-flight (TOF) imaging provides direct access to the momentum distribution, enabling {site-resolved readout of the population in the synthetic lattice}. This combination of high controllability and convenient measurement {has established} momentum lattices as a powerful platform for investigating topology, disorder, and their interplay, {leading} to a series of experimental advances.

\begin{figure}
	\includegraphics[width=0.49\textwidth]{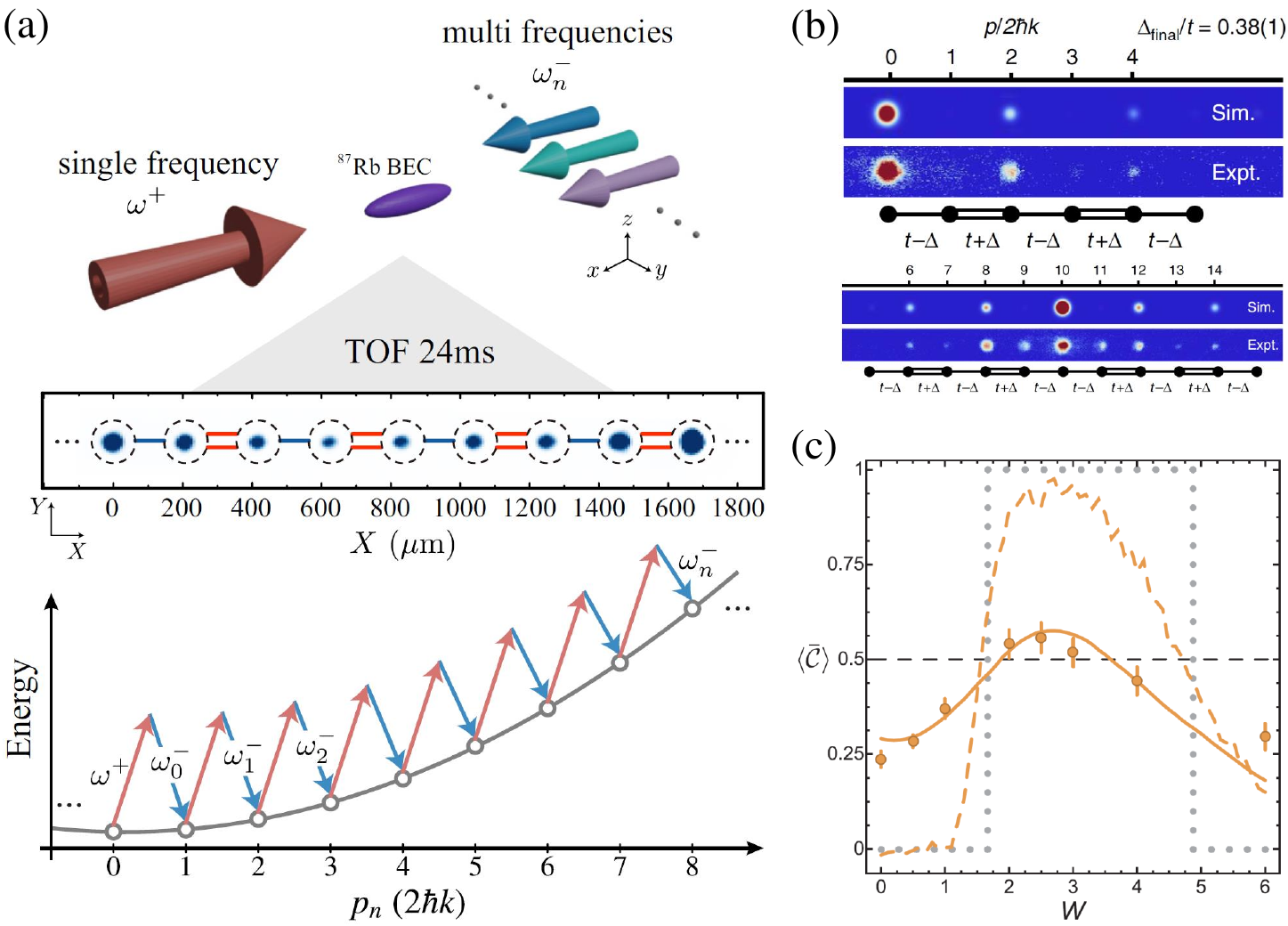}
	\caption{(a) Experimental setup of the momentum lattice. A BEC is coupled by Bragg beams with multiple frequency components, creating a synthetic lattice that implements the SSH model. The middle panel shows a typical experimental absorption image of the atomic density distribution in momentum states after TOF expansion, while the bottom panel depicts the dispersion relation and Bragg transition pathways for nearest-neighbor couplings. Panel (a) is reprinted from Ref.~\cite{Yuan2023}.
	(b) Direct imaging of topological soliton states prepared adiabatically. Experimental absorption images (top: edge defect; bottom: central defect) show good agreement with theoretical simulations. Panel (b) is reprinted from Ref.~\cite{Meier2016}.
	(c) Observation of the topological Anderson insulator phase in a momentum lattice with controllable disorder. The time- and disorder-averaged mean chiral displacement is measured as a function of disorder strength $W$. The data, supported by numerical simulations (solid lines), show that the topological index reaches 0.5 at the critical points, consistent with the expected topological phase transition. Panel (c) is reprinted from Ref.~\cite{Meier2018}.}
	\label{Fig4}
\end{figure}

The construction and operational principle of momentum lattices can be illustrated through the experimental realization of edge-to-edge topological transport of atomic momentum states in a BEC~\cite{Yuan2023}. As shown in Fig.~\ref{Fig4}(a), a weakly trapped $^{87}$Rb BEC serves as the experimental system, where a synthetic momentum lattice is created using a pair of counterpropagating Bragg lasers. One beam is a single-frequency laser, while the other is modulated by an acousto-optic modulator to generate multiple discrete frequency components. These two beams drive two-photon Bragg transitions, which alter the atomic momentum in steps of $2\hbar k$. 
The resulting discrete momentum states $p_n=2n\hbar k$ are encoded as lattice sites, forming a quantum chain that realizes the 1D Su-Schrieffer-Heeger (SSH) model~\cite{Su1980}, with the Hamiltonian given by
\begin{align}
H_{\rm SSH}=-\sum_{n=0}^{L-2} [J-(-1)^n\delta J]\left(a^\dagger_{n+1} a_n+{\rm H.c.}\right),
\end{align}
where $a^\dagger_{n}$ and $a_n$ are the {creation and annihilation operators} acting on the lattice site of $n$, respectively. Here, $L$ denotes the
length of the SSH chain.
By precisely controlling the intensity and phase of the Bragg laser fields associated with different momentum-state transitions, 
alternating nearest-neighbor couplings $J\pm\delta J$ are engineered, placing the system in the topologically nontrivial regime ($\delta J>0$).
{In this regime, the system hosts zero-energy edge states protected by chiral symmetry through the bulk-edge correspondence.}
In the experiment, atoms {were} initially prepared at the left-edge lattice site, and the edge state {was} adiabatically transferred to the right edge via optimized modulation of the coupling strengths~\cite{Yuan2023}. The density distribution at each momentum lattice site can be measured by TOF imaging, as shown in middle panel of Fig.~\ref{Fig4}(a).

The advantages of momentum lattices in experimental implementation, including {independent control of sites}, straightforward detection via TOF imaging, and well-defined edges achieved through truncation of the momentum-state chain, have established them as a powerful platform for studying topological {boundary} modes and disorder-driven topological transitions. A series of key experimental advances highlights their capabilities.

{\it Topological boundary states.} 1D and 2D SSH models have been successfully implemented in ultracold-atom momentum lattices, and their {topological boundary states} have been experimentally demonstrated~\cite{Meier2016,Yuan2023,Dong2025}. 
In early work, topological soliton states were directly imaged and quantified via adiabatic preparation~\cite{Meier2016}. 
As shown in Fig.~\ref{Fig4}(b), the atomic population localizes at an edge defect site ($n=0$) with exponential decay into the bulk, 
occupying only every other site of the synthetic SSH chain (i.e., even values of the lattice index $n$) and reproducing the characteristic wavefunction of a topological soliton. 
Similarly, when the defect is placed at the center, the population exhibits symmetric decay on both sides while maintaining the same sublattice-selective occupation. 
These observations confirm { the existence of topological soliton states bound to interfaces between distinct topological phases}, consistent with the predictions of the SSH model.
More recently, disorder-robust edge-to-edge transport in the 1D SSH model~\cite{Yuan2023} and the localized dynamics of corner and edge states in the 2D SSH model~\cite{Dong2025} have been experimentally demonstrated. The 1D extended SSH model with four sublattices has also been realized in momentum lattices~\cite{Xie2019}, 
where the existence of topological edge states was verified through quench dynamics. 
Moreover, injection spectroscopy provides a powerful tool for probing energy spectra in synthetic momentum lattices~\cite{Paladugu2024}. 
{Numerical simulations} indicate that this technique {could} resolve the Hofstadter butterfly fractal spectrum and detect topological edge states of the Aubry-Andr\'e-Harper-Hofstadter model~\cite{Paladugu2024}. 


{\it Interplay between topology and disorder.} The highly controllable nature of momentum lattices makes them particularly well suited for exploring the interplay between topology and disorder. 
Representative studies include the realization of the generalized Aubry-Andr\'e model with both diagonal and off-diagonal quasiperiodic disorder in momentum space~\cite{Xiao2021}, where dynamical observables were used to characterize localization and topology, revealing a critically localized topological phase that remains stable under weak interactions.
In another experiment~\cite{Meier2018}, the topological Anderson insulator phase was observed in disordered atomic wires, demonstrating a disorder-induced transition from a trivial to a topologically nontrivial phase. Figure~\ref{Fig4}(c) shows the dependence of the time- and disorder-averaged mean chiral displacement, defined as the expectation value of position weighted by the chiral symmetry operator~\cite{Cardano2017}, on the strength of disorder. The observed non-monotonic behavior, featuring an initial increase followed by a decrease, indicates successive transitions from trivial wires to a topological Anderson insulator phase and then to a strongly localized Anderson phase.
Additionally, the crossover between flat-band localization and Anderson localization in a 1D Tasaki lattice was investigated in the momentum dimension~\cite{Zeng2024}.

%

\begin{figure*}
	\includegraphics[width=0.95\textwidth]{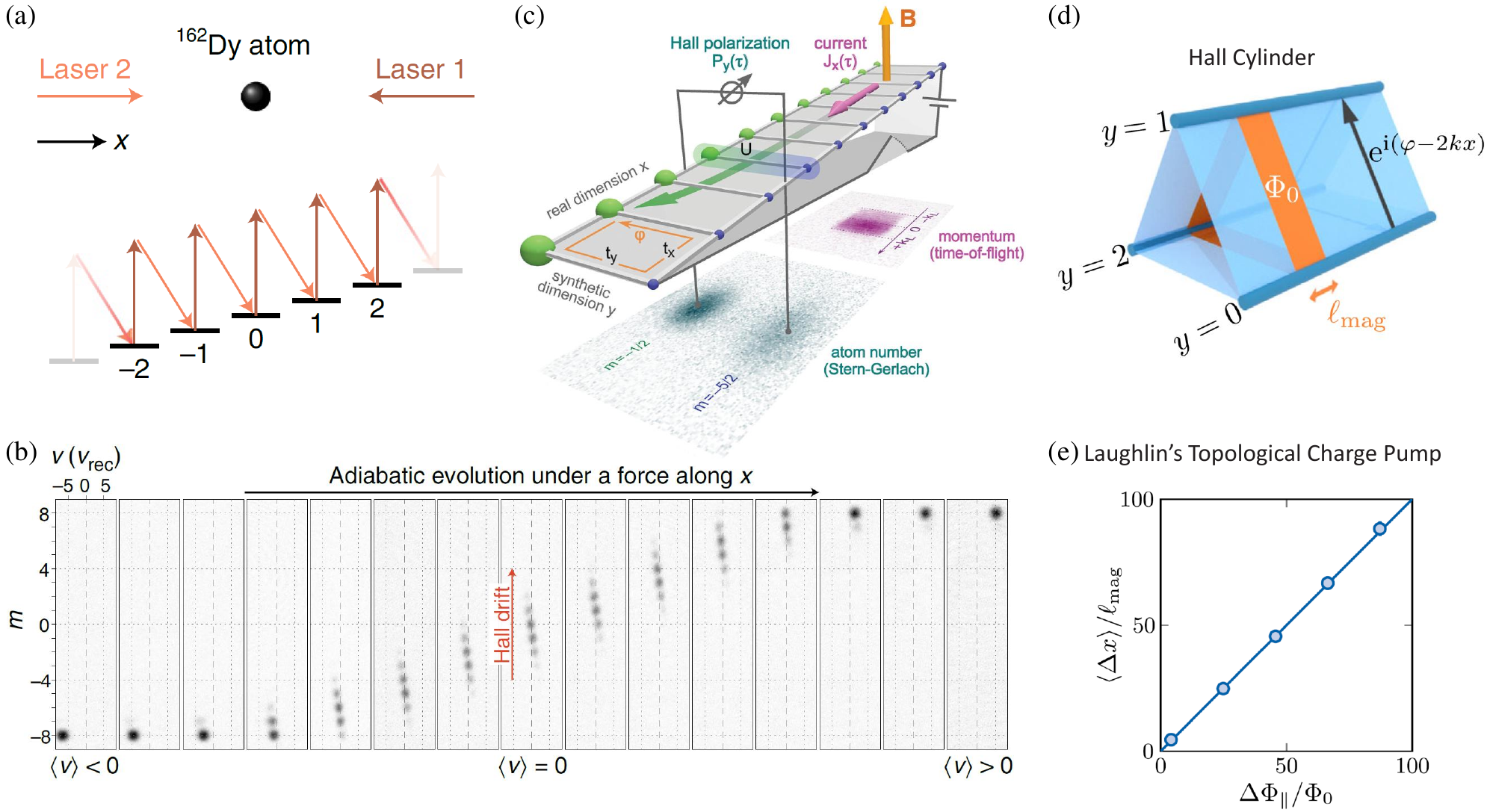}
	\caption{(a) Laser configuration coupling magnetic sublevels of $^{162}$Dy atoms to realize a synthetic dimension with $-J\le m\le J$. 
	(b) Chiral edge states observed via directional motion: negative velocity on the bottom edge ($m=-J$), positive velocity on the top edge ($m=J$), and zero average velocity in the bulk. Single-shot images show the three distinct behaviors. Panels (a) and (b) are reprinted from Ref.~\cite{Chalopin2020}.
	(c) Synthetic ladder implemented with fermionic $^{173}$Yb atoms in a 1D optical lattice. Raman coupling between nuclear spins with a position‑dependent phase simulates a magnetic field. An applied gradient drives an atomic current $J_x$; the resulting Hall polarization $P_y$ (leg population imbalance) is measured via TOF imaging and optical Stern‑Gerlach detection. Panel (c) is reprinted from Ref.~\cite{Zhou2023}.
	(d) Quantum Hall cylinder using three spin states to realize Laughlin's pump protocol. Complex hopping phases encode a radial magnetic field $B_\perp$ and an axial field $B_\parallel$. The orange region (with length $\ell_{\rm mag}$) is threaded by one flux quantum $\Phi_0$. 
	(e) Topological charge pumps: displacement of the center of mass $\langle \Delta x \rangle$ versus axial flux variation $\Delta\Phi_\parallel$, averaged over the magnetic Brillouin zone. The linear fit (blue line) confirms the quantized topological transport. Panels (d) and (e) are reprinted from Ref.~\cite{Fabre2022laughlin}.
	}\label{Fig5}
\end{figure*}

\subsubsection{Internal-state synthetic dimensions}

While momentum lattices leverage discrete momentum states as synthetic sites, an alternative and highly complementary approach is to use atomic internal states---such as Zeeman sublevels {of} a hyperfine manifold---to construct synthetic dimensions. 
This concept, first proposed in Refs.~\cite{Boada2012, Celi2014}, enables the simulation of higher-dimensional quantum systems using lower-dimensional experimental geometries. In contrast to momentum lattices, where tunneling is mediated by Bragg transitions that change atomic momentum, internal-state synthetic dimensions rely on laser-induced Raman or microwave transitions that couple different internal states {and can simultaneously} imparting momentum kicks.
These two platforms offer distinct yet complementary advantages. Momentum lattices excel in providing flexible, site-resolved control in 1D or 2D synthetic spaces with precise engineering of tunneling amplitudes and on-site potentials. Internal-state synthetic dimensions, on the other hand, naturally incorporate spin degrees of freedom, allow for the creation of closed loops (e.g., synthetic Hall tubes) with periodic boundary conditions (PBCs), and enable the simulation of {effectively} higher-dimensional physics (e.g., 4D quantum Hall systems) within compact experimental setups. Together, they form a powerful toolkit for quantum simulation of topological matter.

A representative implementation of this approach is the experimental realization of a synthetic Hall system in ultracold bosonic $^{162}$Dy atoms~\cite{Chalopin2020}. 
As shown in Fig.~\ref{Fig5}(a), the 17 magnetic sublevels of the electronic spin $J=8$ serve as a 1D synthetic dimension. 
When combined with a 1D real-space geometry, this configuration forms a 2D synthetic Hall system. 
Two-photon optical transitions are driven by counterpropagating laser beams along the $x$-direction, coupling atomic motion with spin flips. 
The momentum kick accompanying each spin transition introduces an artificial magnetic field, 
rendering the system Hamiltonian analogous to the Landau Hamiltonian for electrons in a perpendicular magnetic field. 
The extreme spin projections $m=\pm J$ naturally act as edges, providing an intrinsic distinction between bulk and boundary.
Experimentally, atoms {were} initially spin-polarized in $m=-J$ and adiabatically loaded into {the lowest energy band at a finite canonical momentum} $p$. 
A weak force $F_x$ along $x$ drives adiabatic evolution {within the lowest band}, satisfying $\dot{p}=F_x$.
The atomic velocity and spin distributions are measured via expansion imaging under a magnetic field gradient. 
As shown in Fig.~\ref{Fig5}(b), the observations reveal the main features of Landau level physics: 
while bulk atoms exhibit nearly zero average velocity accompanied by a transverse Hall drift along the synthetic dimension, 
atoms at the upper ($m=J$) and lower ($m=-J$) edges undergo chiral ballistic motion with opposite velocities.
Remarkably, the local Chern marker~\cite{Bianco2011}---previously experimentally inaccessible---{was} reconstructed by integrating the Hall mobility and spin projection probability in this experiment~\cite{Chalopin2020}. For the 11 central states ($-5 \le m \le 5$), the marker reaches $98(5)\%$ of the quantized theoretical value, accurately characterizing the robust bulk topology.

Like momentum lattices, synthetic lattices constructed from atomic internal states provide high controllability, precise parameter tunability, and direct observability of atomic distributions { across lattice sites}. 
{Accordingly}, they have also been employed to implement the SSH model, complementing the momentum-lattice realizations discussed previously.
For instance, three-site $\Lambda$-type and five-site SSH-type systems were realized via microwave coupling of hyperfine levels in a BEC, 
achieving preparation in dark states and topological edge states, thereby effectively realizing a topological atom laser~\cite{Tsuno2025}. 
Similarly, the SSH model was implemented in Rydberg-atom synthetic dimensions, where topological edge states and band structures were experimentally characterized, 
and zero-energy SPT edge states were directly observed~\cite{Kanungo2022}. 
Compared to momentum-lattice implementations of the SSH model~\cite{Meier2016,Yuan2023,Dong2025,Xie2019}, which require careful engineering of Bragg transitions, 
internal-state approaches leverage couplings that often provide more straightforward control over {tunneling amplitudes, phases, and effective on-site potentials}.

In addition to the SSH model, a series of representative works has explored quantum Hall physics 
in hybrid real-synthetic lattices threaded by artificial magnetic fields, often referred to as synthetic Hall ribbons.
In these systems, {tunneling} along the real spatial direction is provided by the optical lattice, while Raman-induced couplings between neighboring internal states generate tunneling along the synthetic dimension. 
The tunneling amplitude in the synthetic dimension is set by the Raman coupling strength, whereas the momentum transfer of the Raman beams imprints a spatially dependent phase on the coupling matrix elements. 
Together, these laser-controlled amplitudes and phases define bond-dependent Peierls phases {for tunneling along} the synthetic dimension, 
whose accumulated phase around a plaquette yields an effective artificial magnetic flux.
Early experiments used a $^{87}$Rb BEC in a 2D hybrid lattice formed by a 1D optical lattice and three internal states, 
observing chiral currents and skipping orbits along the system’s edges, as well as the dynamical Hall effect for bulk excitations~\cite{Stuhl2015}. 
A complementary study observed similar physics, including chiral edge currents and edge-cyclotron orbits, 
with fermionic $^{173}\text{Yb}$ atoms~\cite{Mancini2015}, where both two- and three-leg ladder configurations were employed.
In addition to visualizing edge states, the $^{162}\text{Dy}$ system described above~\cite{Chalopin2020} 
{extended these studies to bulk topology, demonstrating quantized topological characterization through the local Chern marker}. 
Similar studies on edge currents and nontrivial bulk topology were also performed in a $^{168}\text{Er}$ BEC, 
{using $J=6$ Zeeman sublevels as a synthetic dimension~\cite{Roell2023}}.

Beyond {the noninteracting regime}, a synthetic two-leg ribbon was constructed using fermionic $^{173}\text{Yb}$ atoms in a 1D optical lattice to investigate the Hall effect in the presence of strong interactions~\cite{Zhou2023}. Addressing the breakdown of the conventional Hall relation in strongly correlated systems, this study tested a theoretical prediction that beyond a certain interaction threshold the Hall response becomes universal---independent of interaction strength. 
To this end, {a quench of a linear potential generated} an effective electric field to drive a longitudinal current $J_x$, 
with the induced transverse Hall polarization $P_y$ measured via optical Stern-Gerlach techniques, as sketched in Fig.~\ref{Fig5}(c). 
The Hall imbalance $\Delta_{\rm H}=P_y/J_x$ serves as a proxy for the Hall coefficient. 
The results show that for fixed strong interactions, the Hall imbalance $\Delta_{\rm H}$ saturates with increasing synthetic tunneling $t_y$; 
more importantly, for fixed tunneling, it converges above an interaction threshold to a constant, interaction‑independent value that matches theoretical predictions~\cite{Zhou2023}. 
Subsequent quench protocols enabled the first direct measurement of the Hall resistance in strongly interacting ultracold fermions, 
confirming the universal $1/n$ dependence~\cite{Zhou2025}. 
These works demonstrate the power of such platforms to access strongly correlated regimes that {remain challenging for theoretical and numerical simulations}.

A particularly powerful geometry enabled by internal-state dimensions---and one that is {generally more challenging} to realize in momentum lattices----is the synthetic Hall tube, formed by combining real space with a closed synthetic dimension (e.g., a set of cyclically coupled internal states), which naturally {implements} PBCs [Fig. \ref{Fig5}(d)]. This setup is well suited for studying quantum Hall physics, Thouless pumping, interaction effects, and topology in curved spaces. Using ultracold fermionic atoms in a 1D optical lattice with three hyperfine spin states as synthetic legs, researchers observed a band-gap closing at a critical magnetic flux, signaling a topological phase transition~\cite{Han2019}. A BEC on a synthetic cylindrical surface subject to a radial synthetic magnetic flux revealed an SPT band structure that emerges on the cylinder but is absent in the planar counterpart~\cite{Li2022bose}. In a synthetic quantum Hall cylinder formed with $^{162}\text{Dy}$ atoms, Laughlin's quantized pumping picture~\cite{Laughlin1981} was experimentally tested~\cite{Fabre2022laughlin}. By adiabatically varying the axial magnetic flux $\Phi_{||}$ via the laser phase difference, the atomic center-of-mass displacement $\langle\Delta x\rangle$ was measured. As shown in Fig. \ref{Fig5}(e), $\langle\Delta x\rangle$ exhibits a {linear dependence on $\Phi_{||}$ with a quantized} coefficient $C=1.00(4)$, directly demonstrating quantized center-of-mass transport---a hallmark of Laughlin's topological charge pump.

Finally, internal-state synthetic dimensions enable the simulation of {higher-dimensional systems beyond three spatial dimensions}---a capability that also distinguishes them from momentum lattices, which are typically limited to 1D or 2D synthetic spaces. By combining two real spatial dimensions with two synthetic dimensions derived from the $J=8$ electronic spin manifold, researchers constructed a 4D quantum Hall system~\cite{Bouhiron2024}. This platform allowed the measurement of a second Chern number {with a value} close to unity, the observation of anisotropic hyperedge modes and nonplanar cyclotron orbits, and the direct verification of nonlinear topological responses.



Beyond the two primary approaches discussed above, hybrid synthetic lattices combining momentum and internal-state degrees of freedom~\cite{Lauria2022,Cooper2012,He2021}, as well as synthetic dimensions implemented using highly excited Rydberg-state manifolds~\cite{Kanungo2022,Lu2024,ChenT2024,Trautmann2024} or rotational-state manifolds of polar molecules~\cite{Raghuram2026,Sundar2018}, have further expanded the scope of experimentally accessible synthetic lattice implementations.
Despite these advantages, a key limitation of synthetic lattices is the finite number of accessible internal or momentum states, which restricts the achievable system size and may hinder the exploration of {thermodynamic-limit behavior}.

\subsection{Floquet-engineered lattices}\label{Sec: Floquet-engineered lattices}

As noted in the previous subsection, we use ``Floquet-engineered lattices'' to refer specifically to systems where explicit periodic driving (e.g., lattice shaking) plays an essential role.
Beyond static optical or synthetic lattices, Floquet engineering---the use of time-periodic modulations to control quantum systems---provides a general and powerful strategy for engineering topological band structures in ultracold atoms~\cite{Eckardt2017_review,Rudner2020_review,Weitenberg2021_review}. Using Floquet theory, a time-periodic Hamiltonian $H(t)=H(t+T)$ with period $T=2\pi/\omega$ can be mapped to an effective time-independent Hamiltonian $H_F$ that governs the stroboscopic dynamics via~\cite{Eckardt2017_review}
\begin{align}{
U(T)={\cal T}\exp\left[-\frac{\ui}{\hbar}\int_{0}^T H(t)dt\right]=e^{-\ui H_F T/\hbar},}
\end{align}
where ${\cal T}$ denotes time-ordering operator. 
{ The eigenvalues of $H_F$ define the quasienergies, which are periodic modulo $\hbar\omega$.}
In the high-frequency regime, where {$\hbar\omega$} is large compared to the characteristic energy scales of the system, $H_F$ can be systematically derived via expansions in powers of $1/\omega$, capturing the leading effects of the time-periodic drive~\cite{Goldman2014,Bukov2015}. Floquet engineering thus provides a versatile toolbox for constructing effective Hamiltonians that are difficult or impossible to realize in static systems. Moreover, it can enrich phase diagrams and give rise to anomalous topological phases that have no equilibrium counterparts~\cite{Kitagawa2010,Rudner2013}. The exceptional controllability of ultracold atomic lattice systems---allowing precise tuning of lattice geometries, driving amplitudes, frequencies, and phases---makes them ideal platforms for implementing Floquet-engineered topological phases.

\begin{figure*}
	\includegraphics[width=0.95\textwidth]{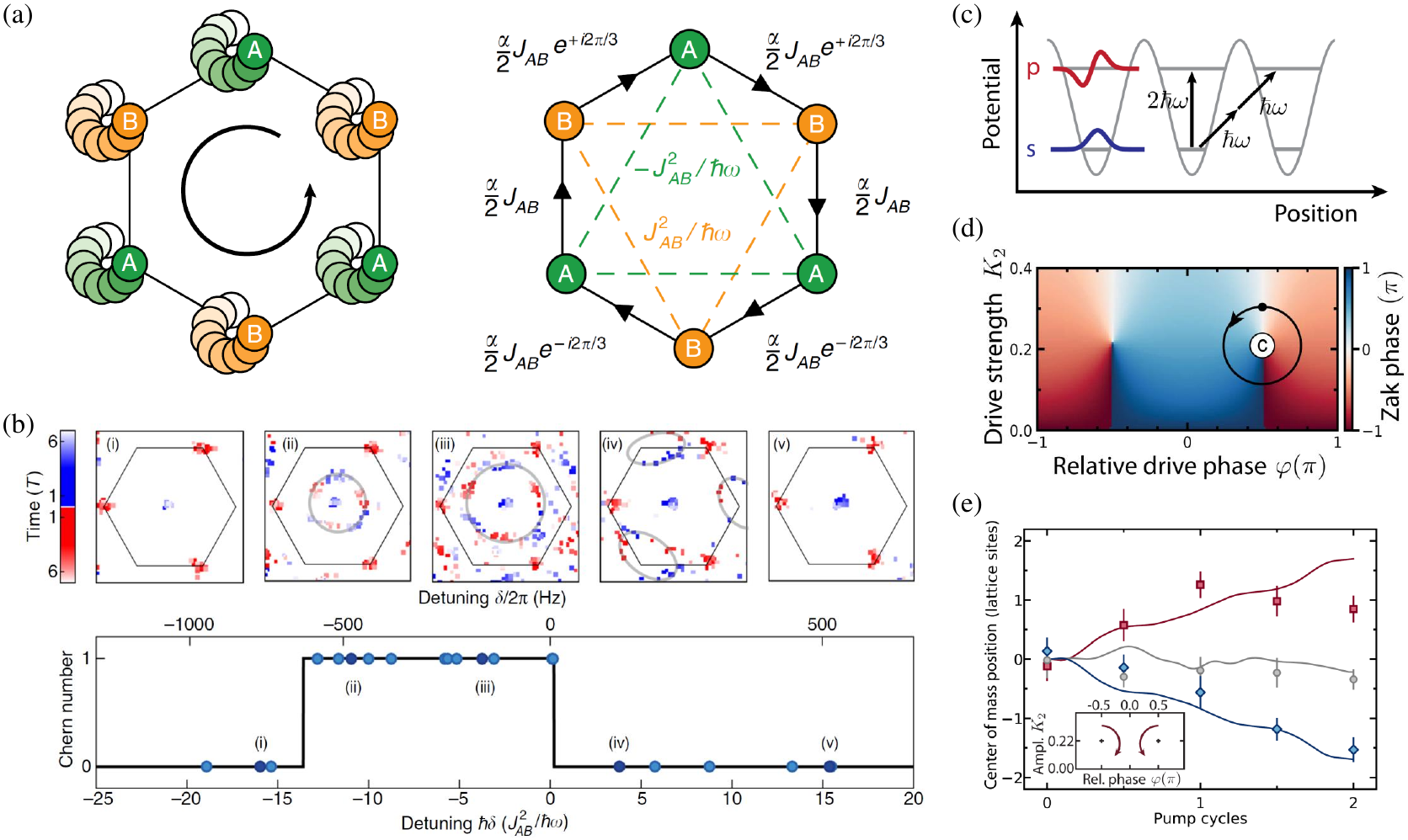}
	\caption{(a) Scheme for realizing a Haldane-like model in a driven hexagonal lattice. Circular shaking of the lattice renormalizes tunneling, yielding an effective Hamiltonian in which nearest-neighbor tunneling carries Peierls phases (right). 
	(b) Topological phase diagram mapped via dynamical linking number. Vortex trajectories in the Brillouin zone (red/blue dots) form closed contours whose linking number yields the Chern number. The extracted Chern number (bottom) agrees with numerical calculations, identifying the nontrivial regime.
	Panels (a) and (b) are reprinted from Ref.~\cite{Tarnowski2019}.
	(c) Generalized Creutz ladder (Shockley model) realized in a 1D optical lattice with two‑tone driving (frequencies $\omega$ and $2\omega$). $s$ and $p$ orbitals are coupled on‑site via one‑photon ($2\omega$) resonances and between neighboring sites via two‑photon ($\omega+\omega$) resonances.
	(d) Zak phase of the lower Floquet‑Bloch band in parameter space. A counterclockwise orbit around the critical point ``$c$'' winds the Zak phase by $+2\pi$.
	(e) Averaged center-of-mass shifts for paired orbits as a function of pump cycles. Parameter cycles induce Zak-phase changes of $-2\pi$ (blue), 0 (gray), and $+2\pi$ (red). The inset shows the paired clockwise and counterclockwise red orbits, which accumulate a Zak-phase winding of $+2\pi$.
	Panels (c)-(e) are reprinted from Ref.~\cite{Minguzzi2022}. 
	}
	\label{Fig6}
\end{figure*}

\subsubsection{Conventional topological models}

As time-periodic modulation can renormalize tunneling and generate effective complex hoppings, periodic driving provides an experimentally accessible route to {realizing topological Hamiltonians that are challenging to implement} in equilibrium~\cite{Eckardt2017_review}. For example, realistic schemes have been proposed to realize Floquet topological states by periodically shaking optical lattices~\cite{Zheng2014}. Specifically, in the 1D case, shaking under a two-photon resonance condition can create an effective Hamiltonian analogous to the topological SSH model. In the 2D honeycomb lattice, circular shaking
{acts as an effective circularly polarized field}, leading to an effective Haldane model~\cite{Haldane1988} that exhibits the QAH effect with chiral edge states; this idea was later experimentally confirmed in ultracold-atom realizations of the Haldane model~\cite{Jotzu2014} and Haldane-like models~\cite{Flaschner2016,Flaschner2018,Tarnowski2019,Asteria2019}.

As a paradigmatic example, we consider the Floquet-engineered realization of Haldane-type models, 
as demonstrated in an experiment with ultracold fermionic $^{40}$K atoms in a periodically driven hexagonal lattice~\cite{Tarnowski2019}. 
This implementation clearly illustrates the underlying working principle.
In the experiment, a hexagonal optical lattice is first created by three interfering laser beams, 
giving rise to two sublattices, $A$ and $B$, with an initial offset $\Delta_{AB}$. 
The lattice is then subjected to a circular shaking motion via phase modulation of the lasers, as illustrated in the left panel of Fig.~\ref{Fig6}(a), 
which introduces a time-periodic inertial force of frequency $\omega=\Delta_{AB}/\hbar-\delta$.
In the comoving frame of the lattice, this inertial force acts as an effective time-periodic vector potential, leading to time-dependent Peierls phases in the tunneling matrix elements. {After} one driving period, this results in effective direction-dependent Peierls phases, leading to a renormalized nearest-neighbor tunneling in the Floquet effective Hamiltonian $J_{AB}^{\rm eff}\approx \frac{\alpha}{2}J_{AB}e^{i\phi_{l'l}}$,
where $\alpha$ denotes the dimensionless driving strength and $\phi_{l'l}$ is the bond-dependent Peierls phase acquired during tunneling from site $l$ to $l'$.
At the same time, {next-nearest-neighbor} tunneling terms are generated through second-order Floquet processes, yielding
$J_{AA}^{\rm eff}=-J_{BB}^{\rm eff}\approx J_{AB}^2/\hbar\omega$.
The effective sublattice offset $\Delta_{AB}^{\rm eff}$ is tuned by the shaking detuning $\delta$.
As summarized in the right panel of Fig.~\ref{Fig6}(a), this Floquet-engineering scheme breaks time-reversal symmetry and realizes an effective Haldane-like model.

Early work reported the first experimental {realization of a Floquet-engineered Haldane model and the observation of its topological phase transition}~\cite{Jotzu2014}.  
Subsequent advances included the first fully momentum-resolved reconstruction of Berry curvature in Floquet Bloch bands of a Haldane-like model~\cite{Flaschner2016} and the observation of dynamical vortices { via time- and momentum-resolved full-state tomography~\cite{Flaschner2018}. In the latter work, the generation, motion, and annihilation of momentum-space vortices---phase singularities in the reconstructed Bloch-state texture---following a sudden quench were tracked in real time, revealing a close connection between non-equilibrium dynamics and the underlying band topology~\cite{Flaschner2018}.}
Building on these results, a direct quantitative mapping was established between a dynamical linking number---{reflecting whether the trajectories of dynamical vortices} enclose static vortices---and the ground-state Chern number~\cite{Tarnowski2019}. As shown in Fig.~\ref{Fig6}(b), when the closed {dynamical-vortex trajectory surrounds} a static vortex, as in the subfigures (ii) and (iii), the linking number equals 1, corresponding to a topological phase with Chern number $C=1$, whereas otherwise it is 0 for the trivial phase. 
This provides a direct dynamical route for extracting the Chern number from vortex linking, thereby determining the topological phase diagram shown in the bottom panel of Fig.~\ref{Fig6}(b).
Furthermore, quantized circular dichroism, as predicted in Ref.~\cite{Tran2017}, was also demonstrated in this system~\cite{Asteria2019}, 
establishing depletion-rate spectroscopy as a novel tool for extracting the Chern number and probing the geometry of Bloch bands.

Beyond honeycomb lattices, Floquet engineering has enabled the realization of a variety of other lattice models. 
One notable example is the Creutz ladder, implemented in a resonantly driven one-dimensional optical lattice~\cite{Kang2020a}, 
where two-photon resonant coupling between the $s$ and $p$ orbital states generates the {cross inter-leg} links characteristic of the ladder geometry. 
The pseudospin winding structure of the resulting energy bands was experimentally revealed using momentum-resolved Ramsey interferometry~\cite{Kang2020a}. 
Building on this platform, a two-tone driving scheme was subsequently proposed to further control the inter-leg couplings and realize topological charge pumping~\cite{Kang2020a,Kang2020b}.

Motivated by this proposal, topological pumping in Floquet-Bloch bands was later demonstrated experimentally in a 1D resonantly shaken optical lattice~\cite{Minguzzi2022},
where the individual band gaps were independently controlled using a two-tone drive with frequencies $\omega$ and $2\omega$.
The resulting system can be accurately described by a cross-linked two-leg ladder model, also referred to as a Shockley model or a generalized Creutz ladder.
In this description, the two legs correspond to the $s$ and $p$ orbitals and are coupled via both direct and diagonal inter-leg links, induced by one-photon ($2\omega$) and two-photon ($\omega+\omega$) resonances, respectively, as illustrated in Fig.~\ref{Fig6}(c).
The relative phase $\varphi$ between the $\omega$ and $2\omega$ drives determines whether time-reversal symmetry is preserved ($\varphi=0,\pi$) or broken ($\varphi\neq0,\pi$). When time-reversal symmetry is broken, the Floquet-Bloch bands become asymmetric, and the individual band gaps can close at specific critical points in parameter space~\cite{Sandholzer2022}. Such a gap closure signals a topological transition characterized by a $2\pi$ winding of the Zak phase---
the Berry phase acquired by a Floquet-Bloch band over the 1D Brillouin zone~\cite{Zak1989}---as shown in Fig.~\ref{Fig6}(d).
Topological pumping then emerges by cyclically varying the driving parameters adiabatically around such a critical point.
In the experiment~\cite{Minguzzi2022}, the center-of-mass shift was averaged over paired orbits---clockwise and counterclockwise paths around $\varphi=+\pi/2$ and $-\pi/2$, respectively---so as to cancel group velocity effects while accumulating the same Zak phase winding, as illustrated in the inset of Fig.~\ref{Fig6}(e). 
The measured results are consistent with topological pumping [Fig.~\ref{Fig6}(e)].
This experiment demonstrates that topological pumping can be achieved without complex superlattice potentials, relying instead on a simple sinusoidal lattice combined with Floquet engineering.

\subsubsection{Anomalous Floquet topological phases}

A distinctive feature of Floquet systems is their ability to host { genuinely nonequilibrium} topological phases that have no static counterparts. 
Among them, anomalous Floquet topological phases are characterized by {trivial bulk band invariants, such as vanishing Chern numbers in 2D systems,
yet supporting topologically protected boundary modes}~\cite{Rudner2013}. 
These phases lie beyond the conventional bulk-boundary correspondence and cannot be fully characterized by quasienergy-band invariants alone. 
Instead, a complete topological characterization of two-band Floquet systems requires two gap-dependent winding numbers, { $W_0$ and $W_\pi$, 
which characterize the topology of the quasienergy gaps centered at $0$ and $\hbar\omega/2$, respectively (referred to as the 0 gap and $\pi$ gap), 
and determine the net number of chiral edge modes in 2D systems~\cite{Rudner2013}.}
Unlike the Chern number, which is determined solely by the Floquet bands, 
these winding numbers are defined from the full time-evolution operator over one driving period and characterize the topology of the micromotion.
A variety of theoretical proposals have put forward feasible driving protocols to engineer such phases in ultracold atoms~\cite{Groh2016,Quelle2017,Budich2017,Zhou2018a,LiuH2019,Zhang2022,Koyama2023}.
Building on these proposals, experimental realizations in ultracold atomic platforms have been reported in recent years~\cite{Wintersperger2020,Xie2020,Lu2022,Zhang2023tuning,ZhangH2024}.

\begin{figure*}
	\includegraphics[width=0.95\textwidth]{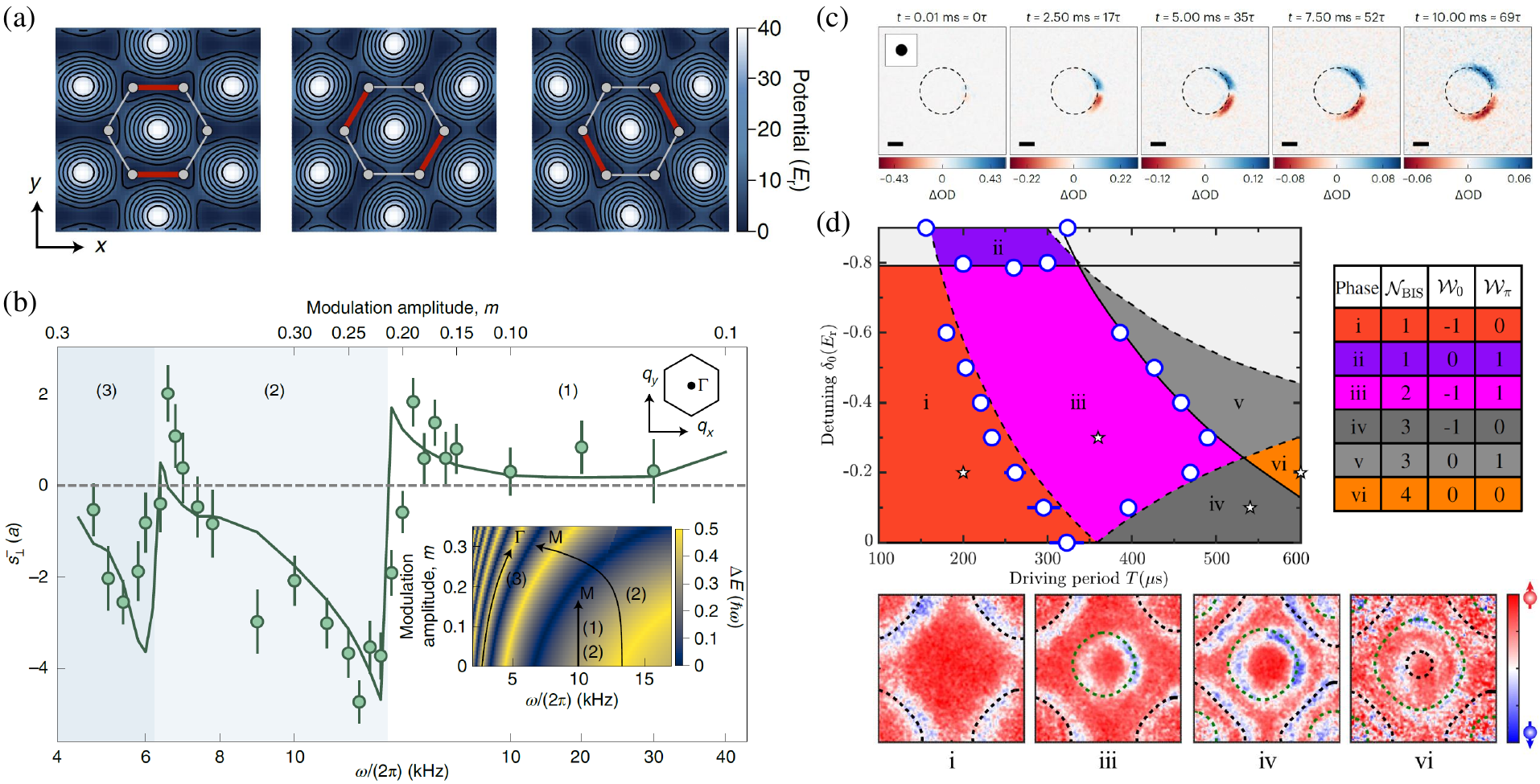}
	\caption{(a) Stepwise modulation protocol of tunnel couplings over one driving period to realize an anomalous Floquet topological system. Red lines indicate regions of enhanced tunnel coupling.
	(b) Local Hall deflections used to probe the Berry curvature distribution. Measured transverse deflections $s_{\perp}^{-}$ along specific quasimomentum paths for different modulation regimes. The inset shows the ramp‑up parameters (frequency $\omega$ and modulation amplitude $m$) for the three regimes (denoted by the numbers).
	Panels (a) and (b) are reprinted from Ref.~\cite{Wintersperger2020}.
	(c) Observation of anomalous Floquet topological edge modes. Difference images reveal edge‑state propagation at increasing evolution times. Each image is averaged over a 100--300 realizations. Dashed lines indicate the programmed pattern from the digital micromirror device. Panel (c) is reprinted from Ref.~\cite{Braun2024}.
	(d) Anomalous Floquet phases in a periodically driven optical Raman lattice. Top: Phase diagram in the $T-\delta_0$ plane. Experimental data (blue circles) agree with calculated phase boundaries (solid: 0 gap, dashed: $\pi$ gap). Six topological phases (i-vi) are identified, with topological invariants listed. Bottom: Spin textures $\langle \sigma_z\rangle_z$ for representative phases; the configuration of band-inversion rings (dashed curves) fully distinguishes the topology. Panel (d) is reprinted from Ref.~\cite{Zhang2023tuning}.
	}
	\label{Fig7}
\end{figure*}

{\it Stepwise driving.} A stepwise drive consists of a driving cycle divided into multiple equal-duration steps, with each step governed by a distinct system configuration. 
A representative implementation uses stepwise periodic modulation of the tunneling couplings in a honeycomb lattice~\cite{Kitagawa2010,Wintersperger2020}. 
As shown in Fig.~\ref{Fig7}(a), the three laser beams forming the hexagonal lattice are each subjected to periodic intensity modulation. 
During one driving period $T$, the stronger tunneling links (marked in red) are sequentially shifted across the lattice, 
enabling spatiotemporal control of nearest-neighbor hopping. 
This periodic driving breaks time-reversal symmetry and induces anomalous Floquet topological phases with vanishing Chern numbers but robust chiral edge states.

Experimental breakthroughs have been achieved using $^{39}$K bosonic atoms in a periodically driven honeycomb lattice~\cite{Wintersperger2020}. 
By measuring the transverse deflection $s_{\perp}^{-}$ of the atomic cloud for the lowest {Floquet} band, the local Berry curvature distribution was mapped. 
When the system evolved from a Haldane-like regime [regime (1)] to an anomalous Floquet topological regime [regime (2)], 
the transverse deflection $s_{\perp}^{-}$ exhibited a sign change [Fig.~\ref{Fig7}(b)], consistent with a topological phase transition {driven by a $\pi$-gap closing}. 
Analysis of topological invariants revealed that the winding numbers changed from $(W_0,W_{\pi})=(1,0)$ to $(1,1)$, 
indicating an anomalous Floquet phase with zero Chern number but protected chiral edge states in both quasienergy gaps.

Building on this, researchers employed a digital micromirror device to create a periodically modulated honeycomb lattice with a step-like edge potential, enabling direct detection and manipulation of Floquet topological chiral edge modes~\cite{Braun2024}. As shown in the differential optical density images in Fig.~\ref{Fig7}(c), atoms initially localized at the release point ($t=0.01$ ms) exhibit clear unidirectional chiral propagation along the edge of a repulsive potential at later times ($t=7.50$ms, $10.00$ms), with negligible bulk scattering. This direct observation of topologically protected chiral edge modes confirmed the successful preparation of atoms into anomalous Floquet topological edge states. Subsequent studies investigated the emergence of these edge modes across three distinct topological regimes and characterized their group velocities~\cite{Braun2024}. Further experiments explored disorder-driven topological phase transitions, revealing that disorder can stabilize anomalous Floquet topological regimes, paving the way toward exotic nonequilibrium phases such as the anomalous Floquet-Anderson insulator~\cite{Hesse2025}.

Other notable advances include the realization of discrete-time topological quantum walks via stroboscopic driving in momentum-space lattices, where topological invariants were directly measured using time-averaged mean chiral displacements, and interaction-induced localization was observed~\cite{Xie2020}. By switching between two configurations of a 1D spin-dependent bipartite lattice, near-ideal topological Dirac bands characterized by a nonzero {quasienergy-band} winding number were realized~\cite{Lu2022}. 

{\it Continuous driving.} Besides stepwise driving, continuous driving protocols also provide a route to anomalous Floquet topological phases. 
In cold-atom experiments, a common strategy is to apply periodic continuous modulation to an already gapped band structure in an optical Raman lattice.
Such driving can induce new band crossings in the quasienergy spectrum and thereby generate novel Floquet topological phases~\cite{Zhang2020,Zhang2022}.

In a periodically driven optical Raman lattice of ultracold $^{87}$Rb atoms, BISs were systematically 
engineered to realize and detect anomalous Floquet topological states with distinct BIS configurations~\cite{Zhang2023tuning}.
As established in earlier work~\cite{Sun2018b}, the appearance or disappearance of a BIS signals a topological transition, 
and each BIS can be {associated with a winding number and corresponds to the existence of a chiral edge mode}~\cite{Zhang2022}. 
Consequently, {in this periodically driven system, the emergence of a new BIS associated with the 0 or $\pi$ gap signals a transition to a new Floquet topological phase.
The full topological phase diagram} was therefore experimentally determined, as shown in Fig.~\ref{Fig7}(d).

{
Since the emergence and evolution of BISs encode additional information about the underlying local band-inversion structure beyond the global topological invariants,
a complete characterization of the experimentally observed phase diagram requires three quantities}: the gap invariants $W_0$ and $W_\pi$,
together with the number of BISs, $N_{\rm BIS}$. Experimentally, $N_{\rm BIS}$ was extracted from quench dynamics, 
while $W_0$ and $W_\pi$ were obtained from the BIS winding numbers in the corresponding gaps~\cite{Zhang2023tuning,Zhang2020}. 
Notably, this platform enabled the observation of an anomalous Floquet valley-Hall state with $W_0=0$ and $W_\pi=0$, 
yet featuring helical-like edge modes protected by valley degrees of freedom {associated with distinct BISs ($N_{\rm BIS}\neq0$) [Fig.~\ref{Fig7}(d), region vi], demonstrating that nontrivial BIS topology can exist even when the global gap invariants remain trivial.}

Recently, a 2D anomalous Floquet topological Fermi gas was realized, in which the effective temporal modulation arises from the interference between two Raman coupling processes beyond the RWA~\cite{ZhangH2024}. The beyond-RWA mechanism was confirmed through the observation of a deterministic phase relation between the engineered SO couplings, while simultaneously enabling a significantly extended system lifetime. 
Pump-probe quench measurements of BISs further revealed anomalous Floquet topological phases {characterized by high-Chern-number Floquet bands}~\cite{ZhangH2024}.
These experiments~\cite{Zhang2023tuning,ZhangH2024} highlight distinct continuous-driving mechanisms for realizing anomalous Floquet topology, ranging from BIS-based band engineering to beyond-RWA interference schemes, while also underscoring the importance of mitigating heating and extending coherence times for future developments.

 \begin{figure*}
	\includegraphics[width=0.95\textwidth]{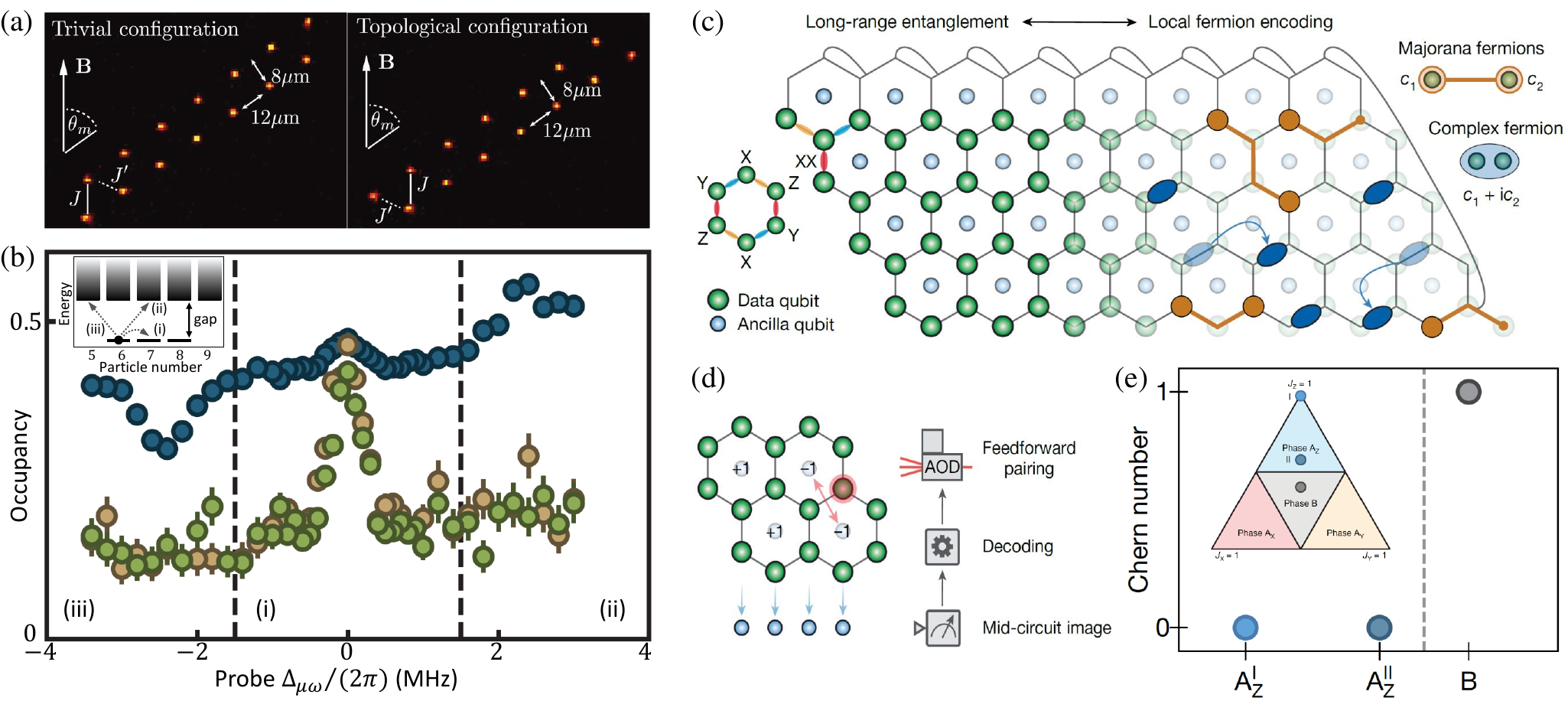}
	\caption{(a) SSH model for hard‑core bosons implemented with Rydberg atoms in a chain of alternating strong and weak couplings. Two configurations are realized: trivial (weak links at the edges) and topological (strong links at the edges). Single‑shot fluorescence images show the atomic arrangement; a tilt by the magic angle $\theta_m$ cancels unwanted couplings between sites in the same sublattice.
	(b) Measured occupancies of bulk (blue circles) and edge sites (green and brown circles), showing the three expected transitions indicated in the energy spectrum (inset). Inset:  Low-energy many-body excitation spectrum in the topological configuration. Starting from the six-particle ground state (solid circle), three types of excitations are resolved: (i) edge excitations connect different states within the degenerate ground-state manifold at (near) zero energy, while (ii) particle addition and (iii) particle removal in the bulk probe a finite excitation gap.
	Panels (a) and (b) are reprinted from Ref.~\cite{De2019}.
	(c) Optical tweezer‑reconfigured honeycomb lattice with 104 atomic qubits ($^{87}$Rb) arranged on a cylinder. Rydberg interactions enable entangling gates. Fermion statistics are encoded through a long‑range entangled state defined by hexagonal plaquette operators $W_p$. Majorana fermions reside at vertices, and conventional fermions are formed by pairing Majoranas along chosen links.
	(d) Topological state preparation using mid‑circuit measurement of ancilla qubits. Plaquettes are deterministically flipped to $+1$ via conditional single‑qubit gates (red circles), relying on acousto‑optic deflectors for qubit control.
	(e) Chern number measurement of topological phases. Using learned parent Hamiltonians from string distributions, the Chern number is found to be $C=0$ in phase A and 
$C=1$ in phase B. Phase B is identified as non‑Abelian, characterized by both topological order and an odd Chern number~\cite{Kitaev2006}. Inset: topological phase diagram of the Kitaev model (anisotropic interactions $J_X+J_Y+J_Z=1$) showing three Abelian phases (A) and one non‑Abelian phase B.
	Panels (c)-(e) are reprinted from Ref.~\cite{Evered2025}.
	}
	\label{Fig8}
\end{figure*}

\subsection{Optical tweezer arrays}\label{Sec: Optical tweezer arrays}

Optical tweezer arrays have emerged as a powerful and highly programmable platform for quantum simulation~\cite{Browaeys2020_review,Morgado2021_review,Kaufman2021,Wu2021,Cheng2024}. 
In this approach, {individual neutral atoms, often prepared in Rydberg states,} are trapped in reconfigurable arrays of optical tweezers, 
enabling precise control over atomic positions, internal states, and interactions. 
By combining the flexibility of single-site addressing with tools such as the Rydberg blockade, dipolar exchange interactions, 
and quantum control techniques, optical tweezer arrays provide a versatile playground for realizing a wide range of topological phases 
and quantum many-body states that are difficult to access with conventional optical lattices.

Theoretically, a rich variety of {strongly correlated} topological states has been proposed for Rydberg atom systems trapped in optical tweezer arrays, with particular emphasis on fractional Chern insulators and various forms of topological order. For fractional Chern insulators, a theoretically accessible scheme was proposed in Ref.~\cite{Weber2022}, where the topological properties of a fractional Chern insulator were studied, experimentally accessible parameters were identified, and the system was shown to exhibit topological features analogous to 
$\nu=1/2$ Laughlin states. The work also outlined an adiabatic preparation scheme and a fractional charge detection method~\cite{Weber2022}. In the domain of topological order, diverse theoretical proposals have been advanced, including the realization of toric code topological order~\cite{Verresen2021}, non-Abelian Floquet spin liquids~\cite{Kalinowski2023}, and Kitaev quantum spin liquids~\cite{Chen2024} in Rydberg tweezer arrays. For toric code order, researchers demonstrated that $Z_2$ topological order can be realized via Rydberg blockade, identified a topological quantum liquid phase, verified its stability against long-range interactions, and proposed a detection scheme using string operators~\cite{Verresen2021}. For non-Abelian Floquet spin liquids, periodic driving was employed to simulate the Kitaev model, {providing routes for preparing, controlling, and detecting} such phases, with the potential to braid and fuse Majorana zero modes for topological quantum computation~\cite{Kalinowski2023}. For the Kitaev quantum spin liquid, theoretical studies explored its realization via laser-assisted dipole-dipole and van der Waals interactions, achieving a high-precision Kitaev model with a gapped non-Abelian phase, and proposed detection schemes for chiral edge modes~\cite{Chen2024}.
{These theoretical proposals establish optical tweezer arrays as a promising platform for exploring strongly correlated topological phases beyond conventional band descriptions.}

Experimental progress in optical tweezer arrays has also accelerated rapidly, with key advances encompassing 
the realization of bosonic SPT phases~\cite{De2019,Yue2026} and topological order~\cite{Semeghini2021,Evered2025}, as detailed below.

{\it Bosonic SPT phases.} 
The initial realization of a bosonic SPT phase used $^{87}$Rb atoms excited to Rydberg levels in a 1D optical tweezer array, 
simulating the SSH model for hard-core bosons~\cite{De2019}.
Here, we take it as a representative example to illustrate the working principle of achieving SPT phases with this platform.
As shown in Fig.~\ref{Fig8}(a), the researchers individually assembled 14 atoms in an optical tweezer array and configured two distinct geometries: 
a topological configuration with weak links at the chain ends and a trivial configuration with strong links at the ends. 
{The atoms were initialized in a Rydberg $s$-state $| 60 S_{1/2}, m_J=1/2\rangle$, defining the vacuum state, while microwave coupling to a $p$-state $|60 P_{1/2}, m_J=-1/2 \rangle$ created mobile hard-core bosonic excitations.}
 The hard-core constraint {arises naturally from the restriction to a single Rydberg excitation per site.}
The key ``hopping'' mechanism originates from resonant dipolar exchange: two atoms in $s$ and $p$ states respectively exchange excitation via the dipole-dipole interaction, 
transferring a ``particle'' from site $i$ to site $j$.
By tuning interatomic distances $R_{ij}$, the nearest-neighbor couplings are engineered into alternating strong $J$ and weak $J'$ values, as shown in Fig.~\ref{Fig8}(a).
The crucial sublattice (chiral) symmetry is enforced using the angular dependence of dipolar exchange, $J_{ij}=d^{2}(3\cos^2\theta_{ij}-1)/{R_{ij}^3}$, with $d$ the transition
dipole moment. 
By arranging the two subchains along the magic angle $\theta_m \approx 54.7^\circ$, same-sublattice couplings are suppressed, yielding an effective AB bipartite SSH geometry.

In this experiment~\cite{De2019}, microwave spectroscopy on a single particle in the chain revealed that zero-energy edge modes emerged only in the topological configuration. 
More importantly, after adiabatic preparation of the half-filled state, the many-body excitation spectrum shown in Fig.~\ref{Fig8}(b) directly reveals the characteristic structure of the interacting bosonic SPT phase: a bulk excitation gap coexisting with a degenerate ground-state manifold associated with edge modes. In particular, edge excitations connect states within this manifold at essentially zero energy cost, whereas bulk particle addition or removal requires a finite gap, clearly separating edge and bulk responses within the many-body spectrum. This provides a clear many-body signature beyond the single-particle SSH description, reflecting the interacting hard-core bosonic nature of the system. Beyond spectroscopy, the experiment further confirms the bosonic SPT phase through the observation of a finite string order parameter and the robustness of the ground-state degeneracy against symmetry-preserving perturbations~\cite{De2019}.

More recently, a disorder-induced bosonic average SPT phase was realized in a programmable 1D Rydberg atom array~\cite{Yue2026}.
In this setup, structural disorder was introduced by applying random offsets to the positions of the optical tweezers from their regular lattice sites, 
which in turn created fluctuating long-range dipolar interactions between the atoms. 
Similar to the previous study~\cite{De2019}, this system~\cite{Yue2026} was also probed at both single-particle and many-body levels. 
At the single-particle level, structural disorder beyond a critical strength induced edge modes, 
signifying a topological phase transition characterized by a quantized average topological invariant. 
At many-body half-filling, measurements of ground-state degeneracy and correlation functions 
confirmed the emergence of a topological phase in disordered lattices, in contrast to the trivial phase in regular lattices. 
Furthermore, quench dynamics from a highly excited N\'eel state demonstrated the robustness of edge modes.
Collectively, these results established that disorder can induce an average SPT phase, 
whose topological stability is guaranteed by average inversion symmetry.

{\it Topological order.} 
The reconfigurable nature of optical tweezer arrays offers unique advantages in lattice adaptability and {controlled} state preparation, 
which have been pivotal in the experimental realization of intrinsic topological order. 
In a pioneering work~\cite{Semeghini2021}, a programmable quantum simulator with 219 $^{87}$Rb atoms 
placed on the links of a kagome lattice was used to realize a dimer model via Rydberg blockade. 
Using this platform, topological string operators were measured, and the onset of a toric-code type quantum spin liquid phase 
with long-range coherence and topological order was detected. 
Subsequently, the Kitaev honeycomb model was simulated on a quantum computer with reconfigurable neutral atom arrays~\cite{Evered2025},
demonstrating the first experimental {realization} of the non-Abelian chiral spin-liquid phase and paving the way for digital quantum simulations of exotic topological matter including topological order.

The realization of the Kitaev honeycomb model in Ref.~\cite{Evered2025} serves as a representative example that highlights the advantages of optical tweezer arrays for studying topological order. 
 As shown in Fig.~\ref{Fig8}(c), the programmability of optical tweezers enables the precise arrangement of $^{87}$Rb atoms into the honeycomb lattice required by the Kitaev model, with vertices serving as qubits and PBCs applied along the shorter direction to form a cylindrical geometry. 
 This configuration perfectly matches the topological structure requirements of $XX$, $YY$ and $ZZ$ anisotropic interactions, a level of control {that is difficult to achieve} with traditional fixed-lattice platforms. Building on this, as illustrated in Fig.~\ref{Fig8}(d), the experiment employed optical tweezer-assisted mid-circuit measurement and real-time feedforward technology. 
The hexagonal plaquette operator $W_p=X_{1}Z_{2}Y_{3}X_{4}Z_{5}Y_{6}$ was measured by separating ancilla qubits from data qubits, followed by deterministic correction of plaquette operator defects to $+1$ via conditional single-qubit gates. This enabled efficient preparation of the long-range entangled topological state required by the Kitaev model.

To characterize the phases, the Chern number was measured by reconstructing {the effective Majorana Hamiltonian in momentum space} from correlation data, 
yielding $C=0$ for the Abelian phases and {an odd Chern number ($C=1$)} for the non-Abelian phase, as shown in Fig.~\ref{Fig8}(e). 
Additionally, Pauli string correlations (two-point correlations of Majorana fermions) were measured across different phases~\cite{Evered2025}, 
revealing a significant enhancement of long-string correlations in the non-Abelian phase in excellent agreement with theoretical predictions. 
This work successfully observed the key physical characteristics of the Kitaev model, thereby confirming its experimental realization.

{These experiments demonstrate that optical tweezer arrays provide a unique route to studying strongly correlated topological matter, combining programmable geometry, tunable interactions, and advanced measurement capabilities.
Beyond equilibrium topological phases}, 
a recent experiment reported the demonstration of interaction-assisted topological pumping of self-bound states in Rydberg synthetic dimensions, 
bridging observations from nonlinear photonics to the few- and many-body quantum limit~\cite{Huang2025}. 
At present, practical implementations are constrained by finite coherence times and system sizes relative to large-scale optical lattice experiments.
As the platform continues to scale in qubit number and operational fidelity, 
it holds great promise for exploring more complex topological states, including non-Abelian anyons and their braiding dynamics, 
with potential applications in topological quantum computation~\cite{Nayak2008}.

\begin{table*}[t]
\centering
\caption{Comparison of the four ultracold-atom platforms discussed in this review.}\label{tab:comparison}
\renewcommand{\arraystretch}{1.1}
\begin{tabular}{p{0.23\linewidth}p{0.22\linewidth}p{0.23\linewidth}p{0.23\linewidth}}
\hline\hline
\textbf{Platform} &
\textbf{Key strength} &
\textbf{Main limitation} &
\textbf{Particularly suitable for} \\
\hline

\textbf{Optical lattices} &
Band engineering &
Limited programmability &
Band topology \\

\textbf{Synthetic lattices} &
Flexible dimensions &
Finite size &
Higher-dimensional topology \\

\textbf{Floquet-engineered lattices} &
Dynamical control &
Heating  &
Nonequilibrium topology \\

\textbf{Optical tweezer arrays} &
Programmability &
Coherence limits &
Strongly correlated topology \\

\hline\hline
\end{tabular}
\end{table*}

\section{Outlook}

The ultracold atomic lattice platforms reviewed in this work provide complementary approaches for realizing and probing topological phases. 
Table~\ref{tab:comparison} provides a concise comparison of the four ultracold-atom platforms, highlighting their key strengths, main limitations, and areas where they are particularly well suited. Together, these platforms provide a versatile toolbox for exploring topological quantum matter in ultracold atomic systems and have enabled remarkable progress in both theory and experiment. Such achievements indicate that the study of topological phases is entering a new stage, marked by a transition from proof-of-principle realizations of band topology toward the exploration of increasingly complex quantum matter and exotic topological phenomena. From this perspective, future developments will be driven by the interplay of nonequilibrium dynamics, strong correlations, and the enhanced controllability intrinsic to cold-atom platforms.


A foremost direction is the extension from equilibrium settings to nonequilibrium quantum systems. 
Ultracold atoms, owing to their high degree of isolation and tunability, provide an ideal platform for studying quantum dynamics far from equilibrium, 
including relaxation, thermalization, and prethermalization phenomena~\cite{Polkovnikov2011,Langen2015,Ueda2020}. 
In this context, periodically driven systems enable the realization of topological phases without static counterparts~\cite{Harper2020_Review,Kitagawa2010,Rudner2013,Zhang2022}, while quench dynamics allows topology to manifest in time-evolved states rather than ground-state wavefunctions~\cite{McGinley2018,McGinley2019a,ZhangLin2018,ZhangLin2022}. 
Furthermore, controlled dissipation and measurement backaction open the door to non-Hermitian~\cite{Bergholtz2021_Review,Ding2022_Review,Okuma2023_Review} and dissipative topological phases~\cite{Diehl2011,Bardyn2013}, 
where topology is encoded in biorthogonal eigenstructures or steady states. 
A key challenge for the future lies in understanding the stability of such phases in the presence of interactions and heating, 
as well as developing a unified classification of topology in dynamical and open systems.

In parallel, the field is expected to evolve from predominantly single-particle physics toward strongly correlated topological matter. 
The intrinsic tunability of interatomic interactions---ranging from short-range contact interactions to engineered long-range couplings---provides 
a powerful avenue to explore interaction-driven phases beyond conventional band theory.
This includes the realization of interacting SPT phases~\cite{De2019,Sompet2022,Yue2026}, {FQH states}~\cite{Leonard2023FQH}, quantum spin liquids~\cite{Semeghini2021,Evered2025}, and other exotic many-body states. 
Of particular interest is the emergence of topological excitations, especially anyonic quasiparticles with braiding statistics distinct from conventional bosonic or fermionic particles~\cite{Stern2008_review}, including non-Abelian cases such as Majorana zero modes in topological superfluids~\cite{Zhu2011,Jiang2011,Liu2012}.
These developments are poised to bridge the gap between topological band theory and quantum many-body physics, 
while laying the groundwork for the long-term goal of using topologically protected states for quantum information processing.

A closely related frontier is the implementation of lattice gauge theories with dynamical degrees of freedom~\cite{Zohar2016,Aidelsburger2021_review,Cheng2024,Halimeh2025}. 
This direction is not only important for quantum simulations of high-energy physics, but is also closely connected to the study of strongly correlated topological matter. 
In several strongly correlated topological phases, including quantum spin liquids and FQH states, emergent gauge fields provide an effective low-energy description of the underlying many-body physics~\cite{Wen1990,Wen2007,Zhang1989,Lee2006,Savary2017}.
Recent advances in cold-atom quantum simulators have demonstrated the feasibility of realizing gauge-invariant dynamics and scaling toward larger systems, enabling the study of phenomena such as confinement, string breaking, and ergodicity breaking in controlled settings~\cite{Halimeh2025}. 
Looking ahead, the ability to engineer dynamical gauge fields together with strong interactions may provide a powerful route toward exploring topological order, fractionalized excitations, and new forms of quantum matter arising from the interplay between topology and dynamical gauge fields.

Finally, the continued development of ultracold-atom platforms and techniques will further expand the scope of accessible topological phenomena. 
One important direction lies in the development of hybrid and programmable quantum architectures, 
particularly through the combination of optical lattices and optical tweezer arrays~\cite{Spar2022,Trisnadi2022,Young2022,Young2024,Tao2024,Gyger2024,Lim2025}. 
Optical lattices offer scalable and highly homogeneous structures, whereas tweezer arrays provide single-site addressability and flexible geometry; 
their integration thus enables the realization of designer Hamiltonians with controlled inhomogeneity, defects, and boundaries, 
facilitating the preparation and manipulation of strongly correlated topological states. 
Such hybrid platforms are expected to play a crucial role in studying low-entropy many-body phases 
and in bridging analog quantum simulation with emerging quantum information processing capabilities~\cite{Daley2022_review,Gonzalez-Cuadra2023}. 
Complementing these developments, high-resolution detection techniques such as quantum gas microscopy~\cite{Kuhr2016_review,Gross2021_review}
and interferometric probes~\cite{Vale2021_review} enable direct access to Berry curvature, edge modes, and entanglement properties.
Collectively, these capabilities will be essential for fully characterizing topological order and its dynamical manifestations, 
positioning the field to address increasingly complex quantum phenomena.

In summary, the field of quantum simulation of topological phases with ultracold atoms stands at an exciting juncture. 
The platforms developed over the past decade---ranging from optical and synthetic lattices to Floquet-engineered systems and reconfigurable tweezer arrays---provide a versatile toolkit for exploring topological matter. 
The coming years will likely witness a convergence of three overarching trends: the transition from equilibrium to nonequilibrium topology, from single-particle to strongly correlated physics, and from simple implementations to highly controllable, hybrid quantum platforms. As experimental capabilities continue to advance, ultracold atomic systems will remain a uniquely powerful setting for exploring topological quantum matter beyond the reach of conventional materials and classical computation.

\begin{acknowledgments}
This work was supported by the National Natural Science Foundation of China (Grant No. 12574294, No. 12374248, and No. 12204187), 
and the Quantum Science and Technology-National Science and Technology Major Project (Grant No. 2021ZD0302000). 
J. Z. acknowledges support from the Fundamental Research Funds for the Central Universities 
(Grant No. WK9990000122).
\end{acknowledgments}

\bibliographystyle{apsrev}

\bibliography{UltracoldAtomicLatticeSystems_Review}

@article{Liu2012,
	author = {Liu, Xia-Ji and Jiang, Lei and Pu, Han and Hu, Hui},
	doi = {10.1103/PhysRevA.85.021603},
	issue = {2},
	journal = {Phys. Rev. A},
	month = {Feb},
	numpages = {4},
	pages = {021603(R)},
	publisher = {American Physical Society},
	title = {Probing Majorana fermions in spin-orbit-coupled atomic Fermi gases},
	url = {https://link.aps.org/doi/10.1103/PhysRevA.85.021603},
	volume = {85},
	year = {2012}}

@article{Jiang2011,
	author = {Jiang, Liang and Kitagawa, Takuya and Alicea, Jason and Akhmerov, A. R. and Pekker, David and Refael, Gil and Cirac, J. Ignacio and Demler, Eugene and Lukin, Mikhail D. and Zoller, Peter},
	doi = {10.1103/PhysRevLett.106.220402},
	issue = {22},
	journal = {Phys. Rev. Lett.},
	month = {Jun},
	numpages = {4},
	pages = {220402},
	publisher = {American Physical Society},
	title = {Majorana Fermions in Equilibrium and in Driven Cold-Atom Quantum Wires},
	url = {https://link.aps.org/doi/10.1103/PhysRevLett.106.220402},
	volume = {106},
	year = {2011}}

@article{Zheng2019,
	author = {Zheng, Zhen and Lin, Zhi and Zhang, Dan-Wei and Zhu, Shi-Liang and Wang, Z. D.},
	doi = {10.1103/PhysRevResearch.1.033102},
	issue = {3},
	journal = {Phys. Rev. Res.},
	month = {Nov},
	numpages = {7},
	pages = {033102},
	publisher = {American Physical Society},
	title = {Chiral magnetic effect in three-dimensional optical lattices},
	url = {https://link.aps.org/doi/10.1103/PhysRevResearch.1.033102},
	volume = {1},
	year = {2019}}

@article{Liu2016,
	author = {Liu, Xiong-Jun and Liu, Zheng-Xin and Law, K T and Liu, W Vincent and Ng, T K},
	doi = {10.1088/1367-2630/18/3/035004},
	journal = {New Journal of Physics},
	month = {mar},
	number = {3},
	pages = {035004},
	publisher = {IOP Publishing},
	title = {Chiral topological orders in an optical Raman lattice},
	url = {https://doi.org/10.1088/1367-2630/18/3/035004},
	volume = {18},
	year = {2016}}

@book{Wen2007,
	author = {Wen, Xiao-Gang},
	doi = {10.1093/acprof:oso/9780199227259.001.0001},
	month = {09},
	publisher = {Oxford University Press},
	title = {Quantum Field Theory of Many-Body Systems: From the Origin of Sound to an Origin of Light and Electrons},
	url = {https://doi.org/10.1093/acprof:oso/9780199227259.001.0001},
	year = {2007}}

@article{Flaschner2018,
	author = {Fl{\"a}schner, N. and Vogel, D. and Tarnowski, M. and Rem, B. S. and L{\"u}hmann, D. -S. and Heyl, M. and Budich, J. C. and Mathey, L. and Sengstock, K. and Weitenberg, C.},
	date = {2018/03/01},
	doi = {10.1038/s41567-017-0013-8},
	id = {Fl{\"a}schner2018},
	journal = {Nature Physics},
	number = {3},
	pages = {265--268},
	title = {Observation of dynamical vortices after quenches in a system with topology},
	url = {https://doi.org/10.1038/s41567-017-0013-8},
	volume = {14},
	year = {2018}}

@article{Tsuno2025,
	author = {Tsuno, Takuto and Taie, Shintaro and Takasu, Yosuke and Yamashita, Kazuya and Ozawa, Tomoki and Takahashi, Yoshiro},
	date = {2025/12/13},
	doi = {10.1038/s41467-025-67106-8},
	id = {Tsuno2025},
	journal = {Nature Communications},
	number = {1},
	pages = {421},
	title = {Gain engineering and atom lasing in a topological edge state in synthetic dimensions},
	url = {https://doi.org/10.1038/s41467-025-67106-8},
	volume = {17},
	year = {2025}}

@article{Kwan2026,
	author = {Joyce Kwan and Perrin Segura and Yanfei Li and Tizian Blatz and Annie Zhi and Brice Bakkali-Hassani and Annabelle Bohrdt and Martin Greiter and Fabian Grusdt and Markus Greiner},
	journal = {arXiv: 2606.12409},
	primaryclass = {cond-mat.quant-gas},
	title = {A Pfaffian quantum Hall state of ultracold bosons},
	url = {https://arxiv.org/abs/2606.12409},
	year = {2026}}

@article{Savary2017,
	author = {Savary, Lucile and Balents, Leon},
	doi = {10.1088/0034-4885/80/1/016502},
	journal = {Reports on Progress in Physics},
	month = {nov},
	number = {1},
	pages = {016502},
	publisher = {IOP Publishing},
	title = {Quantum spin liquids: a review},
	url = {https://doi.org/10.1088/0034-4885/80/1/016502},
	volume = {80},
	year = {2016}}

@article{Lee2006,
	author = {Lee, Patrick A. and Nagaosa, Naoto and Wen, Xiao-Gang},
	doi = {10.1103/RevModPhys.78.17},
	issue = {1},
	journal = {Rev. Mod. Phys.},
	month = {Jan},
	numpages = {0},
	pages = {17--85},
	publisher = {American Physical Society},
	title = {Doping a Mott insulator: Physics of high-temperature superconductivity},
	url = {https://link.aps.org/doi/10.1103/RevModPhys.78.17},
	volume = {78},
	year = {2006}}

@article{Zhang1989,
	author = {Zhang, S. C. and Hansson, T. H. and Kivelson, S.},
	doi = {10.1103/PhysRevLett.62.82},
	issue = {1},
	journal = {Phys. Rev. Lett.},
	month = {Jan},
	numpages = {0},
	pages = {82--85},
	publisher = {American Physical Society},
	title = {Effective-Field-Theory Model for the Fractional Quantum Hall Effect},
	url = {https://link.aps.org/doi/10.1103/PhysRevLett.62.82},
	volume = {62},
	year = {1989}}

@article{Lu2024,
	author = {Lu, Y. and Wang, C. and Kanungo, S. K. and Dunning, F. B. and Killian, T. C.},
	doi = {10.1103/PhysRevA.110.023318},
	issue = {2},
	journal = {Phys. Rev. A},
	month = {Aug},
	numpages = {6},
	pages = {023318},
	publisher = {American Physical Society},
	title = {Probing the topological phase transition in the Su-Schrieffer-Heeger Hamiltonian using Rydberg-atom synthetic dimensions},
	url = {https://link.aps.org/doi/10.1103/PhysRevA.110.023318},
	volume = {110},
	year = {2024}}

@article{Trautmann2024,
	author = {Trautmann, Martin and Sodemann Villadiego, Inti and Deiglmayr, Johannes},
	doi = {10.1103/PhysRevA.110.L040601},
	issue = {4},
	journal = {Phys. Rev. A},
	month = {Oct},
	numpages = {6},
	pages = {L040601},
	publisher = {American Physical Society},
	title = {Realization of topological Thouless pumping in a synthetic Rydberg dimension},
	url = {https://link.aps.org/doi/10.1103/PhysRevA.110.L040601},
	volume = {110},
	year = {2024}}

@article{ChenT2024,
	author = {Chen, Tao and Huang, Chenxi and Velkovsky, Ivan and Hazzard, Kaden R. A. and Covey, Jacob P. and Gadway, Bryce},
	date = {2024/03/27},
	doi = {10.1038/s41467-024-46823-6},
	id = {Chen2024},
	journal = {Nature Communications},
	number = {1},
	pages = {2675},
	title = {Strongly interacting Rydberg atoms in synthetic dimensions with a magnetic flux},
	url = {https://doi.org/10.1038/s41467-024-46823-6},
	volume = {15},
	year = {2024}}

@article{Sundar2018,
	author = {Sundar, Bhuvanesh and Gadway, Bryce and Hazzard, Kaden R. A.},
	date = {2018/02/21},
	doi = {10.1038/s41598-018-21699-x},
	id = {Sundar2018},
	journal = {Scientific Reports},
	number = {1},
	pages = {3422},
	title = {Synthetic dimensions in ultracold polar molecules},
	url = {https://doi.org/10.1038/s41598-018-21699-x},
	volume = {8},
	year = {2018}}

@article{Raghuram2026,
	author = {Adarsh P. Raghuram and Francesca M. Blondell and Jonathan M. Mortlock and Benjamin P. Maddox and Sohail Dasgupta and Holly A. J. Middleton-Spencer and Kaden R. A. Hazzard and Hannah M. Price and Philip D. Gregory and Simon L. Cornish},
	journal = {arXiv:2604.00745},
	primaryclass = {physics.atom-ph},
	title = {Probing topological edge states in a molecular synthetic dimension},
	url = {https://arxiv.org/abs/2604.00745},
	year = {2026}}

@article{Lim2025,
	author = {Lim, Kelvin and Mancois, Vincent and Wu, Haijun and Shen, Yijie and Wilkowski, David},
	doi = {10.1103/nj9v-gvzb},
	issue = {5},
	journal = {Phys. Rev. A},
	month = {Nov},
	numpages = {6},
	pages = {L051307},
	publisher = {American Physical Society},
	title = {Superresolution optical trapping of multiple cold atoms},
	url = {https://link.aps.org/doi/10.1103/nj9v-gvzb},
	volume = {112},
	year = {2025}}

@article{Li2022,
	author = {Li, Yuqing and Zhang, Jiahui and Wang, Yunfei and Du, Huiying and Wu, Jizhou and Liu, Wenliang and Mei, Feng and Ma, Jie and Xiao, Liantuan and Jia, Suotang},
	date = {2022/01/07},
	doi = {10.1038/s41377-021-00702-7},
	id = {Li2022},
	journal = {Light: Science \& Applications},
	number = {1},
	pages = {13},
	title = {Atom-optically synthetic gauge fields for a noninteracting Bose gas},
	url = {https://doi.org/10.1038/s41377-021-00702-7},
	volume = {11},
	year = {2022}}

@article{He2021,
	author = {He, Yanyan and Mao, Ruosong and Cai, Han and Zhang, Jun-Xiang and Li, Yongqiang and Yuan, Luqi and Zhu, Shi-Yao and Wang, Da-Wei},
	doi = {10.1103/PhysRevLett.126.103601},
	issue = {10},
	journal = {Phys. Rev. Lett.},
	month = {Mar},
	numpages = {7},
	pages = {103601},
	publisher = {American Physical Society},
	title = {Flat-Band Localization in {C}reutz Superradiance Lattices},
	url = {https://link.aps.org/doi/10.1103/PhysRevLett.126.103601},
	volume = {126},
	year = {2021}}

@article{Cooper2012,
	author = {Cooper, N. R. and Moessner, R.},
	doi = {10.1103/PhysRevLett.109.215302},
	issue = {21},
	journal = {Phys. Rev. Lett.},
	month = {Nov},
	numpages = {5},
	pages = {215302},
	publisher = {American Physical Society},
	title = {Designing Topological Bands in Reciprocal Space},
	url = {https://link.aps.org/doi/10.1103/PhysRevLett.109.215302},
	volume = {109},
	year = {2012}}

@article{Lauria2022,
	author = {Lauria, Paul and Kuo, Wei-Ting and Cooper, Nigel R. and Barreiro, Julio T.},
	doi = {10.1103/PhysRevLett.128.245301},
	issue = {24},
	journal = {Phys. Rev. Lett.},
	month = {Jun},
	numpages = {7},
	pages = {245301},
	publisher = {American Physical Society},
	title = {Experimental Realization of a Fermionic Spin-Momentum Lattice},
	url = {https://link.aps.org/doi/10.1103/PhysRevLett.128.245301},
	volume = {128},
	year = {2022}}

@article{ZhangDW2016,
	author = {Zhang, Dan-Wei and Zhao, Y. X. and Liu, Rui-Bin and Xue, Zheng-Yuan and Zhu, Shi-Liang and Wang, Z. D.},
	doi = {10.1103/PhysRevA.93.043617},
	issue = {4},
	journal = {Phys. Rev. A},
	month = {Apr},
	numpages = {10},
	pages = {043617},
	publisher = {American Physical Society},
	title = {Quantum simulation of exotic $\mathcal{PT}$-invariant topological nodal loop bands with ultracold atoms in an optical lattice},
	url = {https://link.aps.org/doi/10.1103/PhysRevA.93.043617},
	volume = {93},
	year = {2016}}

@article{Zhu2011,
	author = {Zhu, Shi-Liang and Shao, L.-B. and Wang, Z. D. and Duan, L.-M.},
	doi = {10.1103/PhysRevLett.106.100404},
	issue = {10},
	journal = {Phys. Rev. Lett.},
	month = {Mar},
	numpages = {4},
	pages = {100404},
	publisher = {American Physical Society},
	title = {Probing Non-{A}belian Statistics of {M}ajorana Fermions in Ultracold Atomic Superfluid},
	url = {https://link.aps.org/doi/10.1103/PhysRevLett.106.100404},
	volume = {106},
	year = {2011}}

@article{Cardano2017,
	author = {Cardano, Filippo and D'Errico, Alessio and Dauphin, Alexandre and Maffei, Maria and Piccirillo, Bruno and de Lisio, Corrado and De Filippis, Giulio and Cataudella, Vittorio and Santamato, Enrico and Marrucci, Lorenzo and Lewenstein, Maciej and Massignan, Pietro},
	date = {2017/06/01},
	doi = {10.1038/ncomms15516},
	id = {Cardano2017},
	journal = {Nature Communications},
	number = {1},
	pages = {15516},
	title = {Detection of {Z}ak phases and topological invariants in a chiral quantum walk of twisted photons},
	url = {https://doi.org/10.1038/ncomms15516},
	volume = {8},
	year = {2017}}

@article{Zak1989,
	author = {Zak, J.},
	doi = {10.1103/PhysRevLett.62.2747},
	issue = {23},
	journal = {Phys. Rev. Lett.},
	month = {Jun},
	numpages = {0},
	pages = {2747--2750},
	publisher = {American Physical Society},
	title = {Berry's phase for energy bands in solids},
	url = {https://link.aps.org/doi/10.1103/PhysRevLett.62.2747},
	volume = {62},
	year = {1989}}

@article{LinXD2026,
	author = {Lin, Xiao-Dong and Zhang, Jinyi and Zhang, Long},
	doi = {10.1103/2x7q-48m1},
	issue = {6},
	journal = {Phys. Rev. A},
	month = {Jun},
	numpages = {13},
	pages = {063303},
	publisher = {American Physical Society},
	title = {Proposal for realizing unpaired Weyl points in a three-dimensional periodically driven optical Raman lattice},
	url = {https://link.aps.org/doi/10.1103/2x7q-48m1},
	volume = {113},
	year = {2026}}

@article{Yue2026,
	author = {Yue, Zongpei and Mao, Yu-Feng and Liang, Xinhui and Hua, Zhen-Xing and Ge, Peiyun and Chao, Yu-Xin and Li, Kai and Jia, Chen and Tey, Meng Khoon and Xu, Yong and You, Li},
	date = {2026/04/20},
	doi = {10.1038/s41567-026-03271-x},
	id = {Yue2026},
	isbn = {1745-2481},
	journal = {Nature Physics},
	title = {Average topological phase in a disordered Rydberg atom array},
	url = {https://doi.org/10.1038/s41567-026-03271-x},
	year = {2026}}

@article{Viebahn2024,
	author = {Viebahn, Konrad and Walter, Anne-Sophie and Bertok, Eric and Zhu, Zijie and G\"achter, Marius and Aligia, Armando A. and Heidrich-Meisner, Fabian and Esslinger, Tilman},
	doi = {10.1103/PhysRevX.14.021049},
	issue = {2},
	journal = {Phys. Rev. X},
	month = {Jun},
	numpages = {14},
	pages = {021049},
	publisher = {American Physical Society},
	title = {Interactions Enable Thouless Pumping in a Nonsliding Lattice},
	url = {https://link.aps.org/doi/10.1103/PhysRevX.14.021049},
	volume = {14},
	year = {2024}}

@article{Nakajima2021,
	author = {Nakajima, Shuta and Takei, Nobuyuki and Sakuma, Keita and Kuno, Yoshihito and Marra, Pasquale and Takahashi, Yoshiro},
	date = {2021/07/01},
	doi = {10.1038/s41567-021-01229-9},
	id = {Nakajima2021},
	isbn = {1745-2481},
	journal = {Nature Physics},
	number = {7},
	pages = {844--849},
	title = {Competition and interplay between topology and quasi-periodic disorder in Thouless pumping of ultracold atoms},
	url = {https://doi.org/10.1038/s41567-021-01229-9},
	volume = {17},
	year = {2021}}

@article{Walter2023,
	author = {Walter, Anne-Sophie and Zhu, Zijie and G{\"a}chter, Marius and Minguzzi, Joaqu{\'\i}n and Roschinski, Stephan and Sandholzer, Kilian and Viebahn, Konrad and Esslinger, Tilman},
	date = {2023/10/01},
	doi = {10.1038/s41567-023-02145-w},
	id = {Walter2023},
	isbn = {1745-2481},
	journal = {Nature Physics},
	number = {10},
	pages = {1471--1475},
	title = {Quantization and its breakdown in a Hubbard--Thouless pump},
	url = {https://doi.org/10.1038/s41567-023-02145-w},
	volume = {19},
	year = {2023}}

@article{Zhu2024,
	author = {Zijie Zhu and Marius G{\"a}chter and Anne-Sophie Walter and Konrad Viebahn and Tilman Esslinger},
	doi = {10.1126/science.adg3848},
	journal = {Science},
	number = {6693},
	pages = {317-320},
	title = {Reversal of quantized Hall drifts at noninteracting and interacting topological boundaries},
	url = {https://www.science.org/doi/abs/10.1126/science.adg3848},
	volume = {384},
	year = {2024}}

@article{Laughlin1981,
	author = {Laughlin, R. B.},
	doi = {10.1103/PhysRevB.23.5632},
	issue = {10},
	journal = {Phys. Rev. B},
	month = {May},
	numpages = {0},
	pages = {5632--5633},
	publisher = {American Physical Society},
	title = {Quantized Hall conductivity in two dimensions},
	url = {https://link.aps.org/doi/10.1103/PhysRevB.23.5632},
	volume = {23},
	year = {1981}}

@article{Minguzzi2022,
	author = {Minguzzi, Joaqu\'{\i}n and Zhu, Zijie and Sandholzer, Kilian and Walter, Anne-Sophie and Viebahn, Konrad and Esslinger, Tilman},
	doi = {10.1103/PhysRevLett.129.053201},
	issue = {5},
	journal = {Phys. Rev. Lett.},
	month = {Jul},
	numpages = {6},
	pages = {053201},
	publisher = {American Physical Society},
	title = {Topological Pumping in a Floquet-Bloch Band},
	url = {https://link.aps.org/doi/10.1103/PhysRevLett.129.053201},
	volume = {129},
	year = {2022}}

@article{Sandholzer2022,
	author = {Sandholzer, Kilian and Walter, Anne-Sophie and Minguzzi, Joaqu\'{\i}n and Zhu, Zijie and Viebahn, Konrad and Esslinger, Tilman},
	doi = {10.1103/PhysRevResearch.4.013056},
	issue = {1},
	journal = {Phys. Rev. Res.},
	month = {Jan},
	numpages = {16},
	pages = {013056},
	publisher = {American Physical Society},
	title = {Floquet engineering of individual band gaps in an optical lattice using a two-tone drive},
	url = {https://link.aps.org/doi/10.1103/PhysRevResearch.4.013056},
	volume = {4},
	year = {2022}}

@article{Kang2020b,
	author = {Kang, Jin Hyoun and Shin, Yong-il},
	doi = {10.1103/PhysRevA.102.063315},
	issue = {6},
	journal = {Phys. Rev. A},
	month = {Dec},
	numpages = {10},
	pages = {063315},
	publisher = {American Physical Society},
	title = {Topological Floquet engineering of a one-dimensional optical lattice via resonant shaking with two harmonic frequencies},
	url = {https://link.aps.org/doi/10.1103/PhysRevA.102.063315},
	volume = {102},
	year = {2020}}

@article{Tran2017,
	author = {Duc Thanh Tran and Alexandre Dauphin and Adolfo G. Grushin and Peter Zoller and Nathan Goldman},
	doi = {10.1126/sciadv.1701207},
	journal = {Science Advances},
	number = {8},
	pages = {e1701207},
	title = {Probing topology by ``heating'': Quantized circular dichroism in ultracold atoms},
	url = {https://www.science.org/doi/abs/10.1126/sciadv.1701207},
	volume = {3},
	year = {2017}}

@article{Koyama2023,
	author = {Koyama, Yusuke and Fujimoto, Kazuya and Nakajima, Shuta and Kawaguchi, Yuki},
	doi = {10.1103/PhysRevResearch.5.043167},
	issue = {4},
	journal = {Phys. Rev. Res.},
	month = {Nov},
	numpages = {17},
	pages = {043167},
	publisher = {American Physical Society},
	title = {Designing nontrivial one-dimensional Floquet topological phases using a spin-1/2 double-kicked rotor},
	url = {https://link.aps.org/doi/10.1103/PhysRevResearch.5.043167},
	volume = {5},
	year = {2023}}

@article{Zhou2018a,
	author = {Zhou, Longwen and Gong, Jiangbin},
	doi = {10.1103/PhysRevA.97.063603},
	issue = {6},
	journal = {Phys. Rev. A},
	month = {Jun},
	numpages = {10},
	pages = {063603},
	publisher = {American Physical Society},
	title = {Floquet topological phases in a spin-$1/2$ double kicked rotor},
	url = {https://link.aps.org/doi/10.1103/PhysRevA.97.063603},
	volume = {97},
	year = {2018}}

@article{Groh2016,
	author = {Groh, Thorsten and Brakhane, Stefan and Alt, Wolfgang and Meschede, Dieter and Asb\'oth, Janos K. and Alberti, Andrea},
	doi = {10.1103/PhysRevA.94.013620},
	issue = {1},
	journal = {Phys. Rev. A},
	month = {Jul},
	numpages = {15},
	pages = {013620},
	publisher = {American Physical Society},
	title = {Robustness of topologically protected edge states in quantum walk experiments with neutral atoms},
	url = {https://link.aps.org/doi/10.1103/PhysRevA.94.013620},
	volume = {94},
	year = {2016}}

@article{LiuH2019,
	author = {Liu, Hui and Xiong, Tian-Shi and Zhang, Wei and An, Jun-Hong},
	doi = {10.1103/PhysRevA.100.023622},
	issue = {2},
	journal = {Phys. Rev. A},
	month = {Aug},
	numpages = {9},
	pages = {023622},
	publisher = {American Physical Society},
	title = {Floquet engineering of exotic topological phases in systems of cold atoms},
	url = {https://link.aps.org/doi/10.1103/PhysRevA.100.023622},
	volume = {100},
	year = {2019}}

@article{Li2021,
	author = {Li, Kai and Wang, Jiong-Hao and Yang, Yan-Bin and Xu, Yong},
	doi = {10.1103/PhysRevLett.127.263004},
	issue = {26},
	journal = {Phys. Rev. Lett.},
	month = {Dec},
	numpages = {6},
	pages = {263004},
	publisher = {American Physical Society},
	title = {Symmetry-Protected Topological Phases in a Rydberg Glass},
	url = {https://link.aps.org/doi/10.1103/PhysRevLett.127.263004},
	volume = {127},
	year = {2021}}

@article{Verresen2022,
	author = {Verresen, Ruben and Vishwanath, Ashvin},
	doi = {10.1103/PhysRevX.12.041029},
	issue = {4},
	journal = {Phys. Rev. X},
	month = {Dec},
	numpages = {32},
	pages = {041029},
	publisher = {American Physical Society},
	title = {Unifying Kitaev Magnets, Kagom\'e Dimer Models, and Ruby Rydberg Spin Liquids},
	url = {https://link.aps.org/doi/10.1103/PhysRevX.12.041029},
	volume = {12},
	year = {2022}}

@article{Tarabunga2022,
	author = {Tarabunga, P. S. and Surace, F. M. and Andreoni, R. and Angelone, A. and Dalmonte, M.},
	doi = {10.1103/PhysRevLett.129.195301},
	issue = {19},
	journal = {Phys. Rev. Lett.},
	month = {Nov},
	numpages = {7},
	pages = {195301},
	publisher = {American Physical Society},
	title = {Gauge-Theoretic Origin of Rydberg Quantum Spin Liquids},
	url = {https://link.aps.org/doi/10.1103/PhysRevLett.129.195301},
	volume = {129},
	year = {2022}}

@article{Mogerle2025,
	author = {M\"ogerle, J. and Brechtelsbauer, K. and Gea-Caballero, A.T. and Prior, J. and Emperauger, G. and Bornet, G. and Chen, C. and Lahaye, T. and Browaeys, A. and B\"uchler, H.P.},
	doi = {10.1103/PRXQuantum.6.020332},
	issue = {2},
	journal = {PRX Quantum},
	month = {May},
	numpages = {13},
	pages = {020332},
	publisher = {American Physical Society},
	title = {Spin-1 Haldane Phase in a Chain of Rydberg Atoms},
	url = {https://link.aps.org/doi/10.1103/PRXQuantum.6.020332},
	volume = {6},
	year = {2025}}

@article{Kamal2024,
	author = {Kamal, Helia and Kemp, Jack and He, Yin-Chen and Fuji, Yohei and Aidelsburger, Monika and Zoller, Peter and Yao, Norman Y.},
	doi = {10.1103/PhysRevLett.133.163403},
	issue = {16},
	journal = {Phys. Rev. Lett.},
	month = {Oct},
	numpages = {10},
	pages = {163403},
	publisher = {American Physical Society},
	title = {Floquet Flux Attachment in Cold Atomic Systems},
	url = {https://link.aps.org/doi/10.1103/PhysRevLett.133.163403},
	volume = {133},
	year = {2024}}

@article{Aidelsburger2011,
	author = {Aidelsburger, M. and Atala, M. and Nascimb\`ene, S. and Trotzky, S. and Chen, Y.-A. and Bloch, I.},
	doi = {10.1103/PhysRevLett.107.255301},
	issue = {25},
	journal = {Phys. Rev. Lett.},
	month = {Dec},
	numpages = {5},
	pages = {255301},
	publisher = {American Physical Society},
	title = {Experimental Realization of Strong Effective Magnetic Fields in an Optical Lattice},
	url = {https://link.aps.org/doi/10.1103/PhysRevLett.107.255301},
	volume = {107},
	year = {2011}}

@article{Gerbier2010,
	author = {Gerbier, Fabrice and Dalibard, Jean},
	doi = {10.1088/1367-2630/12/3/033007},
	journal = {New Journal of Physics},
	month = {mar},
	number = {3},
	pages = {033007},
	title = {Gauge fields for ultracold atoms in optical superlattices},
	url = {https://doi.org/10.1088/1367-2630/12/3/033007},
	volume = {12},
	year = {2010}}

@article{Kitaev2006,
	author = {Alexei Kitaev},
	doi = {https://doi.org/10.1016/j.aop.2005.10.005},
	journal = {Annals of Physics},
	number = {1},
	pages = {2-111},
	title = {Anyons in an exactly solved model and beyond},
	url = {https://www.sciencedirect.com/science/article/pii/S0003491605002381},
	volume = {321},
	year = {2006}}

@article{Tai2017,
	author = {Tai, M. Eric and Lukin, Alexander and Rispoli, Matthew and Schittko, Robert and Menke, Tim and Dan Borgnia and Preiss, Philipp M. and Grusdt, Fabian and Kaufman, Adam M. and Greiner, Markus},
	date = {2017/06/01},
	doi = {10.1038/nature22811},
	id = {Tai2017},
	isbn = {1476-4687},
	journal = {Nature},
	number = {7659},
	pages = {519--523},
	title = {Microscopy of the interacting Harper--Hofstadter model in the two-body limit},
	url = {https://doi.org/10.1038/nature22811},
	volume = {546},
	year = {2017}}

@article{Budich2017,
	author = {Budich, Jan Carl and Hu, Ying and Zoller, Peter},
	doi = {10.1103/PhysRevLett.118.105302},
	issue = {10},
	journal = {Phys. Rev. Lett.},
	month = {Mar},
	numpages = {5},
	pages = {105302},
	publisher = {American Physical Society},
	title = {Helical {F}loquet Channels in {1D} Lattices},
	url = {https://link.aps.org/doi/10.1103/PhysRevLett.118.105302},
	volume = {118},
	year = {2017}}

@article{Stern2008_review,
	author = {Ady Stern},
	doi = {https://doi.org/10.1016/j.aop.2007.10.008},
	journal = {Annals of Physics},
	number = {1},
	pages = {204-249},
	title = {Anyons and the quantum Hall effect---A pedagogical review},
	url = {https://www.sciencedirect.com/science/article/pii/S0003491607001674},
	volume = {323},
	year = {2008}}

@article{Asteria2019,
	author = {Asteria, Luca and Tran, Duc Thanh and Ozawa, Tomoki and Tarnowski, Matthias and Rem, Benno S. and Fl{\"a}schner, Nick and Sengstock, Klaus and Goldman, Nathan and Weitenberg, Christof},
	date = {2019/05/01},
	doi = {10.1038/s41567-019-0417-8},
	id = {Asteria2019},
	isbn = {1745-2481},
	journal = {Nature Physics},
	number = {5},
	pages = {449--454},
	title = {Measuring quantized circular dichroism in ultracold topological matter},
	url = {https://doi.org/10.1038/s41567-019-0417-8},
	volume = {15},
	year = {2019}}

@article{Su1980,
	author = {Su, W. P. and Schrieffer, J. R. and Heeger, A. J.},
	doi = {10.1103/PhysRevB.22.2099},
	issue = {4},
	journal = {Phys. Rev. B},
	month = {Aug},
	numpages = {0},
	pages = {2099--2111},
	publisher = {American Physical Society},
	title = {Soliton excitations in polyacetylene},
	url = {https://link.aps.org/doi/10.1103/PhysRevB.22.2099},
	volume = {22},
	year = {1980}}

@article{Bianco2011,
	author = {Bianco, Raffaello and Resta, Raffaele},
	doi = {10.1103/PhysRevB.84.241106},
	issue = {24},
	journal = {Phys. Rev. B},
	month = {Dec},
	numpages = {4},
	pages = {241106},
	publisher = {American Physical Society},
	title = {Mapping topological order in coordinate space},
	url = {https://link.aps.org/doi/10.1103/PhysRevB.84.241106},
	volume = {84},
	year = {2011}}

@article{Fabre2024,
	author = {Fabre, A. and Nascimbene, S.},
	doi = {10.1209/0295-5075/ad2ff6},
	journal = {Europhysics Letters},
	month = {mar},
	number = {6},
	pages = {65001},
	publisher = {EDP Sciences, IOP Publishing and Societ{\`a} Italiana di Fisica},
	title = {Atomic topological quantum matter using synthetic dimensions},
	url = {https://doi.org/10.1209/0295-5075/ad2ff6},
	volume = {145},
	year = {2024}}

@article{Ozawa2019_review,
	author = {Ozawa, Tomoki and Price, Hannah M.},
	date = {2019/05/01},
	doi = {10.1038/s42254-019-0045-3},
	id = {Ozawa2019},
	isbn = {2522-5820},
	journal = {Nature Reviews Physics},
	number = {5},
	pages = {349--357},
	title = {Topological quantum matter in synthetic dimensions},
	url = {https://doi.org/10.1038/s42254-019-0045-3},
	volume = {1},
	year = {2019}}

@article{Arguello-Luengo2024,
	author = {Arg{\"u}ello-Luengo, Javier and Bhattacharya, Utso and Celi, Alessio and Chhajlany, Ravindra W. and Grass, Tobias and P{\l}odzie{\'n}, Marcin and Rakshit, Debraj and Salamon, Tymoteusz and Stornati, Paolo and Tarruell, Leticia and Lewenstein, Maciej},
	date = {2024/05/04},
	doi = {10.1038/s42005-024-01636-3},
	id = {Arg{\"u}ello-Luengo2024},
	isbn = {2399-3650},
	journal = {Communications Physics},
	number = {1},
	pages = {143},
	title = {Synthetic dimensions for topological and quantum phases},
	url = {https://doi.org/10.1038/s42005-024-01636-3},
	volume = {7},
	year = {2024}}

@article{WangJT2024,
	author = {Wang, Jian-Te and Liu, Jing-Xin and Ding, Hai-Tao and He, Peng},
	doi = {10.1103/PhysRevA.109.053314},
	issue = {5},
	journal = {Phys. Rev. A},
	month = {May},
	numpages = {11},
	pages = {053314},
	publisher = {American Physical Society},
	title = {Proposal for implementing Stiefel-Whitney insulators in an optical Raman lattice},
	url = {https://link.aps.org/doi/10.1103/PhysRevA.109.053314},
	volume = {109},
	year = {2024}}

@article{Lu2020,
	author = {Yue-Hui Lu and Bao-Zong Wang and Xiong-Jun Liu},
	doi = {https://doi.org/10.1016/j.scib.2020.09.036},
	journal = {Science Bulletin},
	number = {24},
	pages = {2080-2085},
	title = {Ideal Weyl semimetal with 3D spin-orbit coupled ultracold quantum gas},
	url = {https://www.sciencedirect.com/science/article/pii/S2095927320306526},
	volume = {65},
	year = {2020}}

@article{WangBZ2018,
	author = {Wang, Bao-Zong and Lu, Yue-Hui and Sun, Wei and Chen, Shuai and Deng, Youjin and Liu, Xiong-Jun},
	doi = {10.1103/PhysRevA.97.011605},
	issue = {1},
	journal = {Phys. Rev. A},
	month = {Jan},
	numpages = {5},
	pages = {011605},
	publisher = {American Physical Society},
	title = {Dirac-, Rashba-, and Weyl-type spin-orbit couplings: Toward experimental realization in ultracold atoms},
	url = {https://link.aps.org/doi/10.1103/PhysRevA.97.011605},
	volume = {97},
	year = {2018}}

@inbook{ZhangBook,
	author = {Long Zhang and Xiong-Jun Liu},
	booktitle = {Synthetic Spin-Orbit Coupling in Cold Atoms},
	chapter = {1},
	doi = {10.1142/9789813272538_0001},
	pages = {1-87},
	publisher = {World Scientific},
	title = {Spin-orbit Coupling and Topological Phases for Ultracold Atoms},
	url = {https://www.worldscientific.com/doi/abs/10.1142/9789813272538_0001},
	year = {2018}}

@article{Zhou2017,
	author = {Zhou, Xiaofan and Pan, Jian-Song and Liu, Zheng-Xin and Zhang, Wei and Yi, Wei and Chen, Gang and Jia, Suotang},
	doi = {10.1103/PhysRevLett.119.185701},
	issue = {18},
	journal = {Phys. Rev. Lett.},
	month = {Nov},
	numpages = {6},
	pages = {185701},
	publisher = {American Physical Society},
	title = {Symmetry-Protected Topological States for Interacting Fermions in Alkaline-Earth-Like Atoms},
	url = {https://link.aps.org/doi/10.1103/PhysRevLett.119.185701},
	volume = {119},
	year = {2017}}

@article{Potirniche2017,
	author = {Potirniche, I.-D. and Potter, A. C. and Schleier-Smith, M. and Vishwanath, A. and Yao, N. Y.},
	doi = {10.1103/PhysRevLett.119.123601},
	issue = {12},
	journal = {Phys. Rev. Lett.},
	month = {Sep},
	numpages = {6},
	pages = {123601},
	publisher = {American Physical Society},
	title = {Floquet Symmetry-Protected Topological Phases in Cold-Atom Systems},
	url = {https://link.aps.org/doi/10.1103/PhysRevLett.119.123601},
	volume = {119},
	year = {2017}}

@article{Sompet2022,
	author = {Sompet, Pimonpan and Hirthe, Sarah and Bourgund, Dominik and Chalopin, Thomas and Bibo, Julian and Koepsell, Joannis and Bojovi{\'c}, Petar and Verresen, Ruben and Pollmann, Frank and Salomon, Guillaume and Gross, Christian and Hilker, Timon A. and Bloch, Immanuel},
	date = {2022/06/01},
	doi = {10.1038/s41586-022-04688-z},
	id = {Sompet2022},
	isbn = {1476-4687},
	journal = {Nature},
	number = {7914},
	pages = {484--488},
	title = {Realizing the symmetry-protected Haldane phase in Fermi--Hubbard ladders},
	url = {https://doi.org/10.1038/s41586-022-04688-z},
	volume = {606},
	year = {2022}}

@article{Sun2023,
	author = {Sun, Bo-Ye and Goldman, Nathan and Aidelsburger, Monika and Bukov, Marin},
	doi = {10.1103/PRXQuantum.4.020329},
	issue = {2},
	journal = {PRX Quantum},
	month = {May},
	numpages = {24},
	pages = {020329},
	publisher = {American Physical Society},
	title = {Engineering and Probing Non-Abelian Chiral Spin Liquids Using Periodically Driven Ultracold Atoms},
	url = {https://link.aps.org/doi/10.1103/PRXQuantum.4.020329},
	volume = {4},
	year = {2023}}

@article{Daley2022_review,
	author = {Daley, Andrew J. and Bloch, Immanuel and Kokail, Christian and Flannigan, Stuart and Pearson, Natalie and Troyer, Matthias and Zoller, Peter},
	date = {2022/07/01},
	doi = {10.1038/s41586-022-04940-6},
	id = {Daley2022},
	isbn = {1476-4687},
	journal = {Nature},
	number = {7920},
	pages = {667--676},
	title = {Practical quantum advantage in quantum simulation},
	url = {https://doi.org/10.1038/s41586-022-04940-6},
	volume = {607},
	year = {2022}}

@article{Gonzalez-Cuadra2023,
	author = {D. Gonz{\'a}lez-Cuadra and D. Bluvstein and M. Kalinowski and R. Kaubruegger and N. Maskara and P. Naldesi and T. V. Zache and A. M. Kaufman and M. D. Lukin and H. Pichler and B. Vermersch and Jun Ye and P. Zoller},
	doi = {10.1073/pnas.2304294120},
	journal = {Proceedings of the National Academy of Sciences},
	number = {35},
	pages = {e2304294120},
	title = {Fermionic quantum processing with programmable neutral atom arrays},
	url = {https://www.pnas.org/doi/abs/10.1073/pnas.2304294120},
	volume = {120},
	year = {2023}}

@article{Zohar2016,
	author = {Zohar, Erez and Cirac, J Ignacio and Reznik, Benni},
	doi = {10.1088/0034-4885/79/1/014401},
	journal = {Reports on Progress in Physics},
	month = {dec},
	number = {1},
	pages = {014401},
	publisher = {IOP Publishing},
	title = {Quantum simulations of lattice gauge theories using ultracold atoms in optical lattices},
	url = {https://doi.org/10.1088/0034-4885/79/1/014401},
	volume = {79},
	year = {2015}}

@article{Aidelsburger2021_review,
	author = {Aidelsburger, Monika and Barbiero, Luca and Bermudez, Alejandro and Chanda, Titas and Dauphin, Alexandre and Gonz{\'a}lez-Cuadra, Daniel and Grzybowski, Przemys{\l}aw R. and Hands, Simon and Jendrzejewski, Fred and J{\"u}nemann, Johannes and Juzeli{\=u}nas, Gediminas and Kasper, Valentin and Piga, Angelo and Ran, Shi-Ju and Rizzi, Matteo and Sierra, Germ{\'a}n and Tagliacozzo, Luca and Tirrito, Emanuele and Zache, Torsten V. and Zakrzewski, Jakub and Zohar, Erez and Lewenstein, Maciej},
	doi = {10.1098/rsta.2021.0064},
	journal = {Philosophical Transactions of the Royal Society A: Mathematical, Physical and Engineering Sciences},
	month = {12},
	number = {2216},
	pages = {20210064},
	title = {Cold atoms meet lattice gauge theory},
	url = {https://doi.org/10.1098/rsta.2021.0064},
	volume = {380},
	year = {2021}}

@article{Bardyn2013,
	author = {Bardyn, C-E and Baranov, M A and Kraus, C V and Rico, E and {\.I}mamo{\u g}lu, A and Zoller, P and Diehl, S},
	doi = {10.1088/1367-2630/15/8/085001},
	journal = {New Journal of Physics},
	month = {aug},
	number = {8},
	pages = {085001},
	publisher = {IOP Publishing},
	title = {Topology by dissipation},
	url = {https://doi.org/10.1088/1367-2630/15/8/085001},
	volume = {15},
	year = {2013}}

@article{Diehl2011,
	author = {Diehl, Sebastian and Rico, Enrique and Baranov, Mikhail A. and Zoller, Peter},
	date = {2011/12/01},
	doi = {10.1038/nphys2106},
	id = {Diehl2011},
	isbn = {1745-2481},
	journal = {Nature Physics},
	number = {12},
	pages = {971--977},
	title = {Topology by dissipation in atomic quantum wires},
	url = {https://doi.org/10.1038/nphys2106},
	volume = {7},
	year = {2011}}

@article{Harper2020_Review,
	author = {Harper, Fenner and Roy, Rahul and Rudner, Mark S. and Sondhi, S.L.},
	doi = {10.1146/annurev-conmatphys-031218-013721},
	journal = {Annual Review of Condensed Matter Physics},
	number = {1},
	pages = {345-368},
	title = {Topology and Broken Symmetry in Floquet Systems},
	url = {https://doi.org/10.1146/annurev-conmatphys-031218-013721},
	volume = {11},
	year = {2020}}

@article{Polkovnikov2011,
	author = {Polkovnikov, Anatoli and Sengupta, Krishnendu and Silva, Alessandro and Vengalattore, Mukund},
	doi = {10.1103/RevModPhys.83.863},
	issue = {3},
	journal = {Rev. Mod. Phys.},
	month = {Aug},
	numpages = {0},
	pages = {863--883},
	publisher = {American Physical Society},
	title = {Colloquium: Nonequilibrium dynamics of closed interacting quantum systems},
	url = {https://link.aps.org/doi/10.1103/RevModPhys.83.863},
	volume = {83},
	year = {2011}}

@article{Ueda2020,
	author = {Ueda, Masahito},
	date = {2020/12/01},
	doi = {10.1038/s42254-020-0237-x},
	id = {Ueda2020},
	isbn = {2522-5820},
	journal = {Nature Reviews Physics},
	number = {12},
	pages = {669--681},
	title = {Quantum equilibration, thermalization and prethermalization in ultracold atoms},
	url = {https://doi.org/10.1038/s42254-020-0237-x},
	volume = {2},
	year = {2020}}

@article{Langen2015,
	author = {Langen, Tim and Geiger, Remi and Schmiedmayer, J{\"o}rg},
	doi = {https://doi.org/10.1146/annurev-conmatphys-031214-014548},
	journal = {Annual Review of Condensed Matter Physics},
	number = {Volume 6, 2015},
	pages = {201-217},
	publisher = {Annual Reviews},
	title = {Ultracold Atoms Out of Equilibrium},
	type = {Journal Article},
	url = {https://www.annualreviews.org/content/journals/10.1146/annurev-conmatphys-031214-014548},
	volume = {6},
	year = {2015}}

@article{Ren2022,
	author = {Ren, Zejian and Liu, Dong and Zhao, Entong and He, Chengdong and Pak, Ka Kwan and Li, Jensen and Jo, Gyu-Boong},
	date = {2022/04/01},
	doi = {10.1038/s41567-021-01491-x},
	id = {Ren2022},
	isbn = {1745-2481},
	journal = {Nature Physics},
	number = {4},
	pages = {385--389},
	title = {Chiral control of quantum states in non-Hermitian spin--orbit-coupled fermions},
	url = {https://doi.org/10.1038/s41567-021-01491-x},
	volume = {18},
	year = {2022}}

@article{ZhangH2024,
	author = {Han Zhang and Wen-Wei Wang and Chang Qiao and Long Zhang and Ming-Cheng Liang and Rui Wu and Xu-Jie Wang and Xiong-Jun Liu and Xibo Zhang},
	doi = {https://doi.org/10.1016/j.scib.2024.01.018},
	journal = {Science Bulletin},
	number = {6},
	pages = {747-755},
	title = {Topological spin-orbit-coupled fermions beyond rotating wave approximation},
	url = {https://www.sciencedirect.com/science/article/pii/S2095927324000367},
	volume = {69},
	year = {2024}}

@article{Zhao2025,
	author = {Zhao, Entong and Wang, Zhiyuan and He, Chengdong and Poon, Ting Fung Jeffrey and Pak, Ka Kwan and Liu, Yu-Jun and Ren, Peng and Liu, Xiong-Jun and Jo, Gyu-Boong},
	date = {2025/01/01},
	doi = {10.1038/s41586-024-08347-3},
	id = {Zhao2025},
	isbn = {1476-4687},
	journal = {Nature},
	number = {8046},
	pages = {565--573},
	title = {Two-dimensional non-Hermitian skin effect in an ultracold Fermi gas},
	url = {https://doi.org/10.1038/s41586-024-08347-3},
	volume = {637},
	year = {2025}}

@article{Okuma2023_Review,
	author = {Okuma, Nobuyuki and Sato, Masatoshi},
	doi = {10.1146/annurev-conmatphys-040521-033133},
	journal = {Annual Review of Condensed Matter Physics},
	number = {1},
	pages = {83-107},
	title = {Non-Hermitian Topological Phenomena: A Review},
	url = {https://doi.org/10.1146/annurev-conmatphys-040521-033133},
	volume = {14},
	year = {2023}}

@article{ZhangLin2022,
	author = {Lin Zhang and Wei Jia and Xiong-Jun Liu},
	doi = {https://doi.org/10.1016/j.scib.2022.04.019},
	journal = {Science Bulletin},
	number = {12},
	pages = {1236-1242},
	title = {Universal topological quench dynamics for $\mathbb{Z}_2$ topological phases},
	url = {https://www.sciencedirect.com/science/article/pii/S2095927322001554},
	volume = {67},
	year = {2022}}

@article{Gyger2024,
	author = {Gyger, Flavien and Ammenwerth, Maximilian and Tao, Renhao and Timme, Hendrik and Snigirev, Stepan and Bloch, Immanuel and Zeiher, Johannes},
	doi = {10.1103/PhysRevResearch.6.033104},
	issue = {3},
	journal = {Phys. Rev. Res.},
	month = {Jul},
	numpages = {9},
	pages = {033104},
	publisher = {American Physical Society},
	title = {Continuous operation of large-scale atom arrays in optical lattices},
	url = {https://link.aps.org/doi/10.1103/PhysRevResearch.6.033104},
	volume = {6},
	year = {2024}}

@article{Tao2024,
	author = {Tao, Renhao and Ammenwerth, Maximilian and Gyger, Flavien and Bloch, Immanuel and Zeiher, Johannes},
	doi = {10.1103/PhysRevLett.133.013401},
	issue = {1},
	journal = {Phys. Rev. Lett.},
	month = {Jul},
	numpages = {7},
	pages = {013401},
	publisher = {American Physical Society},
	title = {High-Fidelity Detection of Large-Scale Atom Arrays in an Optical Lattice},
	url = {https://link.aps.org/doi/10.1103/PhysRevLett.133.013401},
	volume = {133},
	year = {2024}}

@article{Spar2022,
	author = {Spar, Benjamin M. and Guardado-Sanchez, Elmer and Chi, Sungjae and Yan, Zoe Z. and Bakr, Waseem S.},
	doi = {10.1103/PhysRevLett.128.223202},
	issue = {22},
	journal = {Phys. Rev. Lett.},
	month = {Jun},
	numpages = {6},
	pages = {223202},
	publisher = {American Physical Society},
	title = {Realization of a Fermi-Hubbard Optical Tweezer Array},
	url = {https://link.aps.org/doi/10.1103/PhysRevLett.128.223202},
	volume = {128},
	year = {2022}}

@article{Young2024,
	author = {Young, Aaron W. and Geller, Shawn and Eckner, William J. and Schine, Nathan and Glancy, Scott and Knill, Emanuel and Kaufman, Adam M.},
	date = {2024/05/01},
	doi = {10.1038/s41586-024-07304-4},
	id = {Young2024},
	isbn = {1476-4687},
	journal = {Nature},
	number = {8011},
	pages = {311--316},
	title = {An atomic boson sampler},
	url = {https://doi.org/10.1038/s41586-024-07304-4},
	volume = {629},
	year = {2024}}

@article{Young2022,
	author = {Aaron W. Young and William J. Eckner and Nathan Schine and Andrew M. Childs and Adam M. Kaufman},
	doi = {10.1126/science.abo0608},
	journal = {Science},
	number = {6608},
	pages = {885-889},
	title = {Tweezer-programmable 2D quantum walks in a Hubbard-regime lattice},
	url = {https://www.science.org/doi/abs/10.1126/science.abo0608},
	volume = {377},
	year = {2022}}

@article{Trisnadi2022,
	author = {Trisnadi, Jonathan and Zhang, Mingjiamei and Weiss, Lauren and Chin, Cheng},
	doi = {10.1063/5.0100088},
	journal = {Review of Scientific Instruments},
	month = {08},
	number = {8},
	pages = {083203},
	title = {Design and construction of a quantum matter synthesizer},
	url = {https://doi.org/10.1063/5.0100088},
	volume = {93},
	year = {2022}}

@article{Morgado2021_review,
	author = {Morgado, M. and Whitlock, S.},
	doi = {10.1116/5.0036562},
	journal = {AVS Quantum Science},
	month = {05},
	number = {2},
	pages = {023501},
	title = {Quantum simulation and computing with Rydberg-interacting qubits},
	url = {https://doi.org/10.1116/5.0036562},
	volume = {3},
	year = {2021}}

@article{Browaeys2020_review,
	author = {Browaeys, Antoine and Lahaye, Thierry},
	date = {2020/02/01},
	doi = {10.1038/s41567-019-0733-z},
	id = {Browaeys2020},
	isbn = {1745-2481},
	journal = {Nature Physics},
	number = {2},
	pages = {132--142},
	title = {Many-body physics with individually controlled {R}ydberg atoms},
	url = {https://doi.org/10.1038/s41567-019-0733-z},
	volume = {16},
	year = {2020}}

@article{Cheng2024,
	author = {Cheng, Yanting and Zhai, Hui},
	date = {2024/09/01},
	doi = {10.1038/s42254-024-00749-6},
	id = {Cheng2024},
	isbn = {2522-5820},
	journal = {Nature Reviews Physics},
	number = {9},
	pages = {566--576},
	title = {Emergent {U(1)} lattice gauge theory in {R}ydberg atom arrays},
	url = {https://doi.org/10.1038/s42254-024-00749-6},
	volume = {6},
	year = {2024}}

@article{Wu2021,
	author = {Wu, Xiaoling and Liang, Xinhui and Tian, Yaoqi and Yang, Fan and Chen, Cheng and Liu, Yong-Chun and Tey, Meng Khoon and You, Li},
	doi = {10.1088/1674-1056/abd76f},
	journal = {Chinese Physics B},
	month = {feb},
	number = {2},
	pages = {020305},
	publisher = {Chinese Physical Society and IOP Publishing Ltd},
	title = {A concise review of Rydberg atom based quantum computation and quantum simulation*},
	url = {https://doi.org/10.1088/1674-1056/abd76f},
	volume = {30},
	year = {2021}}

@article{Ding2022_Review,
	author = {Ding, Kun and Fang, Chen and Ma, Guancong},
	date = {2022/12/01},
	doi = {10.1038/s42254-022-00516-5},
	id = {Ding2022},
	isbn = {2522-5820},
	journal = {Nature Reviews Physics},
	number = {12},
	pages = {745--760},
	title = {Non-Hermitian topology and exceptional-point geometries},
	url = {https://doi.org/10.1038/s42254-022-00516-5},
	volume = {4},
	year = {2022}}

@article{Bergholtz2021_Review,
	author = {Bergholtz, Emil J. and Budich, Jan Carl and Kunst, Flore K.},
	doi = {10.1103/RevModPhys.93.015005},
	issue = {1},
	journal = {Rev. Mod. Phys.},
	month = {Feb},
	numpages = {31},
	pages = {015005},
	publisher = {American Physical Society},
	title = {Exceptional topology of non-{H}ermitian systems},
	url = {https://link.aps.org/doi/10.1103/RevModPhys.93.015005},
	volume = {93},
	year = {2021}}

@article{Wang2017,
	author = {Wang, Ce and Zhang, Pengfei and Chen, Xin and Yu, Jinlong and Zhai, Hui},
	doi = {10.1103/PhysRevLett.118.185701},
	issue = {18},
	journal = {Phys. Rev. Lett.},
	month = {May},
	numpages = {5},
	pages = {185701},
	publisher = {American Physical Society},
	title = {Scheme to Measure the Topological Number of a Chern Insulator from Quench Dynamics},
	url = {https://link.aps.org/doi/10.1103/PhysRevLett.118.185701},
	volume = {118},
	year = {2017}}

@article{ZhangLin2018,
	author = {Lin Zhang and Long Zhang and Sen Niu and Xiong-Jun Liu},
	doi = {https://doi.org/10.1016/j.scib.2018.09.018},
	journal = {Science Bulletin},
	number = {21},
	pages = {1385-1391},
	title = {Dynamical classification of topological quantum phases},
	url = {https://www.sciencedirect.com/science/article/pii/S209592731830478X},
	volume = {63},
	year = {2018}}

@article{McGinley2019a,
	author = {McGinley, Max and Cooper, Nigel R.},
	doi = {10.1103/PhysRevB.99.075148},
	issue = {7},
	journal = {Phys. Rev. B},
	month = {Feb},
	numpages = {17},
	pages = {075148},
	publisher = {American Physical Society},
	title = {Classification of topological insulators and superconductors out of equilibrium},
	url = {https://link.aps.org/doi/10.1103/PhysRevB.99.075148},
	volume = {99},
	year = {2019}}

@article{Gong2018a,
	author = {Gong, Zongping and Ueda, Masahito},
	doi = {10.1103/PhysRevLett.121.250601},
	issue = {25},
	journal = {Phys. Rev. Lett.},
	month = {Dec},
	numpages = {6},
	pages = {250601},
	publisher = {American Physical Society},
	title = {Topological Entanglement-Spectrum Crossing in Quench Dynamics},
	url = {https://link.aps.org/doi/10.1103/PhysRevLett.121.250601},
	volume = {121},
	year = {2018}}

@article{McGinley2018,
	author = {McGinley, Max and Cooper, Nigel R.},
	doi = {10.1103/PhysRevLett.121.090401},
	issue = {9},
	journal = {Phys. Rev. Lett.},
	month = {Aug},
	numpages = {6},
	pages = {090401},
	publisher = {American Physical Society},
	title = {Topology of One-Dimensional Quantum Systems Out of Equilibrium},
	url = {https://link.aps.org/doi/10.1103/PhysRevLett.121.090401},
	volume = {121},
	year = {2018}}

@article{Vale2021_review,
	author = {Vale, Chris J. and Zwierlein, Martin},
	date = {2021/12/01},
	doi = {10.1038/s41567-021-01434-6},
	id = {Vale2021},
	isbn = {1745-2481},
	journal = {Nature Physics},
	number = {12},
	pages = {1305--1315},
	title = {Spectroscopic probes of quantum gases},
	url = {https://doi.org/10.1038/s41567-021-01434-6},
	volume = {17},
	year = {2021}}

@article{Kuhr2016_review,
	author = {Kuhr, Stefan},
	doi = {10.1093/nsr/nww023},
	journal = {National Science Review},
	month = {04},
	number = {2},
	pages = {170-172},
	title = {Quantum-gas microscopes: a new tool for cold-atom quantum simulators},
	url = {https://doi.org/10.1093/nsr/nww023},
	volume = {3},
	year = {2016}}

@article{Gross2021_review,
	author = {Gross, Christian and Bakr, Waseem S.},
	date = {2021/12/01},
	doi = {10.1038/s41567-021-01370-5},
	id = {Gross2021},
	isbn = {1745-2481},
	journal = {Nature Physics},
	number = {12},
	pages = {1316--1323},
	title = {Quantum gas microscopy for single atom and spin detection},
	url = {https://doi.org/10.1038/s41567-021-01370-5},
	volume = {17},
	year = {2021}}

@article{Goldman2014_Review,
	author = {N Goldman and G Juzeli{\=u}nas and P {\"O}hberg and I B Spielman},
	doi = {10.1088/0034-4885/77/12/126401},
	journal = {Reports on Progress in Physics},
	month = {nov},
	number = {12},
	pages = {126401},
	publisher = {IOP Publishing},
	title = {Light-induced gauge fields for ultracold atoms},
	url = {https://dx.doi.org/10.1088/0034-4885/77/12/126401},
	volume = {77},
	year = {2014}}

@article{Dalibard2011_Review,
	author = {Dalibard, Jean and Gerbier, Fabrice and Juzeli\ifmmode \bar{u}\else \={u}\fi{}nas, Gediminas and \"Ohberg, Patrik},
	doi = {10.1103/RevModPhys.83.1523},
	issue = {4},
	journal = {Rev. Mod. Phys.},
	month = {Nov},
	numpages = {0},
	pages = {1523--1543},
	publisher = {American Physical Society},
	title = {Colloquium: Artificial gauge potentials for neutral atoms},
	url = {https://link.aps.org/doi/10.1103/RevModPhys.83.1523},
	volume = {83},
	year = {2011}}

@article{Gross2017_Review,
	author = {Christian Gross and Immanuel Bloch},
	doi = {10.1126/science.aal3837},
	journal = {Science},
	number = {6355},
	pages = {995-1001},
	title = {Quantum simulations with ultracold atoms in optical lattices},
	url = {https://www.science.org/doi/abs/10.1126/science.aal3837},
	volume = {357},
	year = {2017}}

@article{Pesin2012_review,
	author = {Pesin, Dmytro and MacDonald, Allan H.},
	date = {2012/05/01},
	doi = {10.1038/nmat3305},
	id = {Pesin2012},
	isbn = {1476-4660},
	journal = {Nature Materials},
	number = {5},
	pages = {409--416},
	title = {Spintronics and pseudospintronics in graphene and topological insulators},
	url = {https://doi.org/10.1038/nmat3305},
	volume = {11},
	year = {2012}}

@article{Igor2004_review,
	author = {\ifmmode \check{Z}\else \v{Z}\fi{}uti\ifmmode \acute{c}\else \'{c}\fi{}, Igor and Fabian, Jaroslav and Das Sarma, S.},
	doi = {10.1103/RevModPhys.76.323},
	issue = {2},
	journal = {Rev. Mod. Phys.},
	month = {Apr},
	numpages = {0},
	pages = {323--410},
	publisher = {American Physical Society},
	title = {Spintronics: Fundamentals and applications},
	url = {https://link.aps.org/doi/10.1103/RevModPhys.76.323},
	volume = {76},
	year = {2004}}

@article{Lin2019_review,
	author = {Lin, Xiaoyang and Yang, Wei and Wang, Kang L. and Zhao, Weisheng},
	date = {2019/07/01},
	doi = {10.1038/s41928-019-0273-7},
	id = {Lin2019},
	isbn = {2520-1131},
	journal = {Nature Electronics},
	number = {7},
	pages = {274--283},
	title = {Two-dimensional spintronics for low-power electronics},
	url = {https://doi.org/10.1038/s41928-019-0273-7},
	volume = {2},
	year = {2019}}

@article{Sarma2015,
	author = {Sarma, Sankar Das and Freedman, Michael and Nayak, Chetan},
	date = {2015/10/27},
	doi = {10.1038/npjqi.2015.1},
	id = {Sarma2015},
	isbn = {2056-6387},
	journal = {npj Quantum Information},
	number = {1},
	pages = {15001},
	title = {Majorana zero modes and topological quantum computation},
	url = {https://doi.org/10.1038/npjqi.2015.1},
	volume = {1},
	year = {2015}}

@article{Armitage2018_Review,
	author = {Armitage, N. P. and Mele, E. J. and Vishwanath, Ashvin},
	doi = {10.1103/RevModPhys.90.015001},
	issue = {1},
	journal = {Rev. Mod. Phys.},
	month = {Jan},
	numpages = {57},
	pages = {015001},
	publisher = {American Physical Society},
	title = {Weyl and {D}irac semimetals in three-dimensional solids},
	url = {https://link.aps.org/doi/10.1103/RevModPhys.90.015001},
	volume = {90},
	year = {2018}}

@article{Yan2017_Review,
	author = {Yan, Binghai and Felser, Claudia},
	doi = {10.1146/annurev-conmatphys-031016-025458},
	journal = {Annual Review of Condensed Matter Physics},
	number = {1},
	pages = {337-354},
	title = {Topological Materials: Weyl Semimetals},
	url = {https://doi.org/10.1146/annurev-conmatphys-031016-025458},
	volume = {8},
	year = {2017}}

@article{Lv2021_Review,
	author = {Lv, B. Q. and Qian, T. and Ding, H.},
	doi = {10.1103/RevModPhys.93.025002},
	issue = {2},
	journal = {Rev. Mod. Phys.},
	month = {Apr},
	numpages = {68},
	pages = {025002},
	publisher = {American Physical Society},
	title = {Experimental perspective on three-dimensional topological semimetals},
	url = {https://link.aps.org/doi/10.1103/RevModPhys.93.025002},
	volume = {93},
	year = {2021}}

@article{Wen2017,
	author = {Wen, Xiao-Gang},
	doi = {10.1103/RevModPhys.89.041004},
	issue = {4},
	journal = {Rev. Mod. Phys.},
	month = {Dec},
	numpages = {17},
	pages = {041004},
	publisher = {American Physical Society},
	title = {Colloquium: Zoo of quantum-topological phases of matter},
	url = {https://link.aps.org/doi/10.1103/RevModPhys.89.041004},
	volume = {89},
	year = {2017}}

@article{Wen1990,
	author = {Wen, X. G.},
	doi = {10.1142/S0217979290000139},
	journal = {International Journal of Modern Physics B},
	number = {02},
	pages = {239-271},
	title = {Topological Orders in Rigid States},
	url = {https://doi.org/10.1142/S0217979290000139},
	volume = {04},
	year = {1990}}

@article{Senthil2015_review,
	author = {Senthil, T.},
	doi = {https://doi.org/10.1146/annurev-conmatphys-031214-014740},
	journal = {Annual Review of Condensed Matter Physics},
	pages = {299-324},
	publisher = {Annual Reviews},
	title = {Symmetry-Protected Topological Phases of Quantum Matter},
	type = {Journal Article},
	url = {https://www.annualreviews.org/content/journals/10.1146/annurev-conmatphys-031214-014740},
	volume = {6},
	year = {2015}}

@article{Chen2013,
	author = {Chen, Xie and Gu, Zheng-Cheng and Liu, Zheng-Xin and Wen, Xiao-Gang},
	doi = {10.1103/PhysRevB.87.155114},
	issue = {15},
	journal = {Phys. Rev. B},
	month = {Apr},
	numpages = {48},
	pages = {155114},
	publisher = {American Physical Society},
	title = {Symmetry protected topological orders and the group cohomology of their symmetry group},
	url = {https://link.aps.org/doi/10.1103/PhysRevB.87.155114},
	volume = {87},
	year = {2013}}

@article{Chen2012,
	author = {Xie Chen and Zheng-Cheng Gu and Zheng-Xin Liu and Xiao-Gang Wen},
	doi = {10.1126/science.1227224},
	journal = {Science},
	number = {6114},
	pages = {1604-1606},
	title = {Symmetry-Protected Topological Orders in Interacting Bosonic Systems},
	url = {https://www.science.org/doi/abs/10.1126/science.1227224},
	volume = {338},
	year = {2012}}

@article{Chiu2016_Review,
	author = {Chiu, Ching-Kai and Teo, Jeffrey C. Y. and Schnyder, Andreas P. and Ryu, Shinsei},
	doi = {10.1103/RevModPhys.88.035005},
	issue = {3},
	journal = {Rev. Mod. Phys.},
	month = {Aug},
	numpages = {63},
	pages = {035005},
	publisher = {American Physical Society},
	title = {Classification of topological quantum matter with symmetries},
	url = {https://link.aps.org/doi/10.1103/RevModPhys.88.035005},
	volume = {88},
	year = {2016}}

@article{Ludwig2015_Review,
	author = {Andreas W W Ludwig},
	doi = {10.1088/0031-8949/2015/T168/014001},
	journal = {Physica Scripta},
	month = {dec},
	number = {T168},
	pages = {014001},
	publisher = {IOP Publishing},
	title = {Topological phases: classification of topological insulators and superconductors of non-interacting fermions, and beyond},
	url = {https://dx.doi.org/10.1088/0031-8949/2015/T168/014001},
	volume = {2016},
	year = {2015}}

@article{Kitaev2009,
	author = {Kitaev, Alexei},
	doi = {10.1063/1.3149495},
	journal = {AIP Conference Proceedings},
	month = {05},
	number = {1},
	pages = {22-30},
	title = {{Periodic table for topological insulators and superconductors}},
	url = {https://doi.org/10.1063/1.3149495},
	volume = {1134},
	year = {2009}}

@article{Schnyder2008,
	author = {Schnyder, Andreas P. and Ryu, Shinsei and Furusaki, Akira and Ludwig, Andreas W. W.},
	doi = {10.1103/PhysRevB.78.195125},
	issue = {19},
	journal = {Phys. Rev. B},
	month = {Nov},
	numpages = {22},
	pages = {195125},
	publisher = {American Physical Society},
	title = {Classification of topological insulators and superconductors in three spatial dimensions},
	url = {https://link.aps.org/doi/10.1103/PhysRevB.78.195125},
	volume = {78},
	year = {2008}}

@article{Nayak2008,
	author = {Nayak, Chetan and Simon, Steven H. and Stern, Ady and Freedman, Michael and Das Sarma, Sankar},
	doi = {10.1103/RevModPhys.80.1083},
	issue = {3},
	journal = {Rev. Mod. Phys.},
	month = {Sep},
	numpages = {0},
	pages = {1083--1159},
	publisher = {American Physical Society},
	title = {Non-Abelian anyons and topological quantum computation},
	url = {https://link.aps.org/doi/10.1103/RevModPhys.80.1083},
	volume = {80},
	year = {2008}}

@article{Windpassinger2013_Review,
	author = {Patrick Windpassinger and Klaus Sengstock},
	doi = {10.1088/0034-4885/76/8/086401},
	journal = {Reports on Progress in Physics},
	month = {jul},
	number = {8},
	pages = {086401},
	publisher = {IOP Publishing},
	title = {Engineering novel optical lattices},
	url = {https://dx.doi.org/10.1088/0034-4885/76/8/086401},
	volume = {76},
	year = {2013}}

@article{De2019,
	author = {Sylvain de L{\'e}s{\'e}leuc and Vincent Lienhard and Pascal Scholl and Daniel Barredo and Sebastian Weber and Nicolai Lang and Hans Peter B{\"u}chler and Thierry Lahaye and Antoine Browaeys},
	doi = {10.1126/science.aav9105},
	journal = {Science},
	number = {6455},
	pages = {775-780},
	title = {Observation of a symmetry-protected topological phase of interacting bosons with Rydberg atoms},
	url = {https://www.science.org/doi/abs/10.1126/science.aav9105},
	volume = {365},
	year = {2019}}

@article{Meier2018,
	author = {Eric J. Meier and Fangzhao Alex An and Alexandre Dauphin and Maria Maffei and Pietro Massignan and Taylor L. Hughes and Bryce Gadway},
	doi = {10.1126/science.aat3406},
	journal = {Science},
	number = {6417},
	pages = {929-933},
	title = {Observation of the topological Anderson insulator in disordered atomic wires},
	url = {https://www.science.org/doi/abs/10.1126/science.aat3406},
	volume = {362},
	year = {2018}}

@article{Xiao2021,
	author = {Teng Xiao and Dizhou Xie and Zhaoli Dong and Tao Chen and Wei Yi and Bo Yan},
	doi = {https://doi.org/10.1016/j.scib.2021.07.025},
	journal = {Science Bulletin},
	number = {21},
	pages = {2175-2180},
	title = {Observation of topological phase with critical localization in a quasi-periodic lattice},
	url = {https://www.sciencedirect.com/science/article/pii/S2095927321005065},
	volume = {66},
	year = {2021}}

@article{Bloch2005_review,
	author = {Bloch, Immanuel},
	date = {2005/10/01},
	doi = {10.1038/nphys138},
	id = {Bloch2005},
	isbn = {1745-2481},
	journal = {Nature Physics},
	number = {1},
	pages = {23--30},
	title = {Ultracold quantum gases in optical lattices},
	url = {https://doi.org/10.1038/nphys138},
	volume = {1},
	year = {2005}}

@article{Schafer2020_Review,
	author = {Sch{\"a}fer, Florian and Fukuhara, Takeshi and Sugawa, Seiji and Takasu, Yosuke and Takahashi, Yoshiro},
	date = {2020/08/01},
	doi = {10.1038/s42254-020-0195-3},
	id = {Sch{\"a}fer2020},
	isbn = {2522-5820},
	journal = {Nature Reviews Physics},
	number = {8},
	pages = {411--425},
	title = {Tools for quantum simulation with ultracold atoms in optical lattices},
	url = {https://doi.org/10.1038/s42254-020-0195-3},
	volume = {2},
	year = {2020}}

@article{Bloch2008_Review,
	author = {Bloch, Immanuel and Dalibard, Jean and Zwerger, Wilhelm},
	doi = {10.1103/RevModPhys.80.885},
	issue = {3},
	journal = {Rev. Mod. Phys.},
	month = {Jul},
	numpages = {0},
	pages = {885--964},
	publisher = {American Physical Society},
	title = {Many-body physics with ultracold gases},
	url = {https://link.aps.org/doi/10.1103/RevModPhys.80.885},
	volume = {80},
	year = {2008}}

@article{Qi2011_Review,
	author = {Qi, Xiao-Liang and Zhang, Shou-Cheng},
	doi = {10.1103/RevModPhys.83.1057},
	issue = {4},
	journal = {Rev. Mod. Phys.},
	month = {Oct},
	numpages = {0},
	pages = {1057--1110},
	publisher = {American Physical Society},
	title = {Topological insulators and superconductors},
	url = {https://link.aps.org/doi/10.1103/RevModPhys.83.1057},
	volume = {83},
	year = {2011}}

@article{Hasan2010_Review,
	author = {Hasan, M. Z. and Kane, C. L.},
	doi = {10.1103/RevModPhys.82.3045},
	issue = {4},
	journal = {Rev. Mod. Phys.},
	month = {Nov},
	numpages = {0},
	pages = {3045--3067},
	publisher = {American Physical Society},
	title = {Colloquium: Topological insulators},
	url = {https://link.aps.org/doi/10.1103/RevModPhys.82.3045},
	volume = {82},
	year = {2010}}

@article{Halimeh2025,
	author = {Halimeh, Jad C. and Aidelsburger, Monika and Grusdt, Fabian and Hauke, Philipp and Yang, Bing},
	date = {2025/01/01},
	doi = {10.1038/s41567-024-02721-8},
	id = {Halimeh2025},
	isbn = {1745-2481},
	journal = {Nature Physics},
	number = {1},
	pages = {25--36},
	title = {Cold-atom quantum simulators of gauge theories},
	url = {https://doi.org/10.1038/s41567-024-02721-8},
	volume = {21},
	year = {2025}}

@article{Zhang2022,
	author = {Zhang, Long and Liu, Xiong-Jun},
	doi = {10.1103/PRXQuantum.3.040312},
	issue = {4},
	journal = {PRX Quantum},
	month = {Oct},
	numpages = {18},
	pages = {040312},
	publisher = {American Physical Society},
	title = {Unconventional Floquet Topological Phases from Quantum Engineering of Band-Inversion Surfaces},
	url = {https://link.aps.org/doi/10.1103/PRXQuantum.3.040312},
	volume = {3},
	year = {2022}}

@article{Zhang2020,
	author = {Zhang, Long and Zhang, Lin and Liu, Xiong-Jun},
	doi = {10.1103/PhysRevLett.125.183001},
	issue = {18},
	journal = {Phys. Rev. Lett.},
	month = {Oct},
	numpages = {6},
	pages = {183001},
	publisher = {American Physical Society},
	title = {Unified Theory to Characterize Floquet Topological Phases by Quench Dynamics},
	url = {https://link.aps.org/doi/10.1103/PhysRevLett.125.183001},
	volume = {125},
	year = {2020}}

@article{Wang2021,
	author = {Zong-Yao Wang and Xiang-Can Cheng and Bao-Zong Wang and Jin-Yi Zhang and Yue-Hui Lu and Chang-Rui Yi and Sen Niu and Youjin Deng and Xiong-Jun Liu and Shuai Chen and Jian-Wei Pan},
	doi = {10.1126/science.abc0105},
	journal = {Science},
	number = {6539},
	pages = {271-276},
	title = {Realization of an ideal Weyl semimetal band in a quantum gas with 3D spin-orbit coupling},
	url = {https://www.science.org/doi/abs/10.1126/science.abc0105},
	volume = {372},
	year = {2021}}

@article{Song2018,
	author = {Bo Song and Long Zhang and Chengdong He and Ting Fung Jeffrey Poon and Elnur Hajiyev and Shanchao Zhang and Xiong-Jun Liu and Gyu-Boong Jo},
	doi = {10.1126/sciadv.aao4748},
	journal = {Science Advances},
	number = {2},
	pages = {eaao4748},
	title = {Observation of symmetry-protected topological band with ultracold fermions},
	url = {https://www.science.org/doi/abs/10.1126/sciadv.aao4748},
	volume = {4},
	year = {2018}}

@article{Cooper2019,
	author = {Cooper, N. R. and Dalibard, J. and Spielman, I. B.},
	doi = {10.1103/RevModPhys.91.015005},
	issue = {1},
	journal = {Rev. Mod. Phys.},
	month = {Mar},
	numpages = {55},
	pages = {015005},
	publisher = {American Physical Society},
	title = {Topological bands for ultracold atoms},
	url = {https://link.aps.org/doi/10.1103/RevModPhys.91.015005},
	volume = {91},
	year = {2019}}

@article{Bouhiron2024,
	author = {Jean-Baptiste Bouhiron and Aur{\'e}lien Fabre and Qi Liu and Quentin Redon and Nehal Mittal and Tanish Satoor and Raphael Lopes and Sylvain Nascimbene},
	doi = {10.1126/science.adf8459},
	journal = {Science},
	number = {6692},
	pages = {223-227},
	title = {Realization of an atomic quantum {H}all system in four dimensions},
	url = {https://www.science.org/doi/abs/10.1126/science.adf8459},
	volume = {384},
	year = {2024}}

@article{Weitenberg2021_review,
	author = {Weitenberg, Christof and Simonet, Juliette},
	date = {2021/12/01},
	doi = {10.1038/s41567-021-01316-x},
	id = {Weitenberg2021},
	isbn = {1745-2481},
	journal = {Nature Physics},
	number = {12},
	pages = {1342--1348},
	title = {Tailoring quantum gases by Floquet engineering},
	url = {https://doi.org/10.1038/s41567-021-01316-x},
	volume = {17},
	year = {2021}}

@article{Rudner2020_review,
	author = {Rudner, Mark S. and Lindner, Netanel H.},
	date = {2020/05/01},
	doi = {10.1038/s42254-020-0170-z},
	id = {Rudner2020},
	isbn = {2522-5820},
	journal = {Nature Reviews Physics},
	number = {5},
	pages = {229--244},
	title = {Band structure engineering and non-equilibrium dynamics in Floquet topological insulators},
	url = {https://doi.org/10.1038/s42254-020-0170-z},
	volume = {2},
	year = {2020}}

@article{Eckardt2017_review,
	author = {Eckardt, Andr\'e},
	doi = {10.1103/RevModPhys.89.011004},
	issue = {1},
	journal = {Rev. Mod. Phys.},
	month = {Mar},
	numpages = {30},
	pages = {011004},
	publisher = {American Physical Society},
	title = {Colloquium: Atomic quantum gases in periodically driven optical lattices},
	url = {https://link.aps.org/doi/10.1103/RevModPhys.89.011004},
	volume = {89},
	year = {2017}}

@article{Stuhl2015,
	author = {B. K. Stuhl and H.-I. Lu and L. M. Aycock and D. Genkina and I. B. Spielman},
	doi = {10.1126/science.aaa8515},
	journal = {Science},
	number = {6255},
	pages = {1514-1518},
	title = {Visualizing edge states with an atomic Bose gas in the quantum Hall regime},
	url = {https://www.science.org/doi/abs/10.1126/science.aaa8515},
	volume = {349},
	year = {2015}}

@article{Liu2014,
	author = {Liu, Xiong-Jun and Law, K. T. and Ng, T. K.},
	doi = {10.1103/PhysRevLett.112.086401},
	issue = {8},
	journal = {Phys. Rev. Lett.},
	month = {Feb},
	numpages = {5},
	pages = {086401},
	publisher = {American Physical Society},
	title = {Realization of 2D Spin-Orbit Interaction and Exotic Topological Orders in Cold Atoms},
	url = {https://link.aps.org/doi/10.1103/PhysRevLett.112.086401},
	volume = {112},
	year = {2014}}

@article{Goldman2016,
	author = {Goldman, N. and Budich, J. C. and Zoller, P.},
	doi = {10.1038/nphys3803},
	journal = {Nature Physics},
	publisher = {Macmillan Publishers Limited},
	title = {Topological quantum matter with ultracold gases in optical lattices},
	url = {https://doi.org/10.1038/nphys3803},
	volume = {12},
	year = {2016}}

@article{ZhangDW2018,
	author = {Dan-Wei Zhang and Yan-Qing Zhu and Y. X. Zhao and Hui Yan and Shi-Liang Zhu},
	doi = {10.1080/00018732.2019.1594094},
	journal = {Advances in Physics},
	number = {4},
	pages = {253--402},
	publisher = {Taylor \& Francis},
	title = {Topological quantum matter with cold atoms},
	url = {https://doi.org/10.1080/00018732.2019.1594094},
	volume = {67},
	year = {2018}}

@article{Kaufman2021,
	author = {Kaufman, Adam M. and Ni, Kang-Kuen},
	doi = {10.1038/s41567-021-01357-2},
	journal = {Nature Physics},
	pages = {1324--1333},
	publisher = {Springer Nature},
	title = {Quantum science with optical tweezer arrays of ultracold atoms and molecules},
	url = {https://doi.org/10.1038/s41567-021-01357-2},
	volume = {17},
	year = {2021}}

@article{Jaksch2003,
	author = {Jaksch, D and Zoller, P},
	doi = {10.1088/1367-2630/5/1/356},
	journal = {New Journal of Physics},
	month = {may},
	number = {1},
	pages = {56},
	title = {Creation of effective magnetic fields in optical lattices: the Hofstadter butterfly for cold neutral atoms},
	url = {https://doi.org/10.1088/1367-2630/5/1/356},
	volume = {5},
	year = {2003}}

@article{Aidelsburger2013,
	author = {Aidelsburger, M. and Atala, M. and Lohse, M. and Barreiro, J. T. and Paredes, B. and Bloch, I.},
	doi = {10.1103/PhysRevLett.111.185301},
	issue = {18},
	journal = {Phys. Rev. Lett.},
	month = {Oct},
	numpages = {5},
	pages = {185301},
	publisher = {American Physical Society},
	title = {Realization of the {H}ofstadter Hamiltonian with Ultracold Atoms in Optical Lattices},
	url = {https://link.aps.org/doi/10.1103/PhysRevLett.111.185301},
	volume = {111},
	year = {2013}}

@article{Miyake2013,
	author = {Miyake, Hirokazu and Siviloglou, Georgios A. and Kennedy, Colin J. and Burton, William Cody and Ketterle, Wolfgang},
	doi = {10.1103/PhysRevLett.111.185302},
	issue = {18},
	journal = {Phys. Rev. Lett.},
	month = {Oct},
	numpages = {5},
	pages = {185302},
	publisher = {American Physical Society},
	title = {Realizing the Harper Hamiltonian with Laser-Assisted Tunneling in Optical Lattices},
	url = {https://link.aps.org/doi/10.1103/PhysRevLett.111.185302},
	volume = {111},
	year = {2013}}

@article{Aidelsburger2015,
	author = {Aidelsburger, M. and Lohse, M. and Schweizer, C. and Atala, M. and Barreiro, J. T. and Nascimb{\`e}ne, S. and Cooper, N. R. and Bloch, I. and Goldman, N.},
	doi = {10.1038/nphys3171},
	journal = {Nature Physics},
	pages = {162--166},
	publisher = {Macmillan Publishers Limited},
	title = {Measuring the {C}hern number of {H}ofstadter bands with ultracold bosonic atoms},
	url = {https://doi.org/10.1038/nphys3171},
	volume = {11},
	year = {2015}}

@article{Liu2013,
	author = {Liu, Xiong-Jun and Liu, Zheng-Xin and Cheng, Meng},
	doi = {10.1103/PhysRevLett.110.076401},
	issue = {7},
	journal = {Phys. Rev. Lett.},
	month = {Feb},
	numpages = {5},
	pages = {076401},
	publisher = {American Physical Society},
	title = {Manipulating Topological Edge Spins in a One-Dimensional Optical Lattice},
	url = {https://link.aps.org/doi/10.1103/PhysRevLett.110.076401},
	volume = {110},
	year = {2013}}

@article{Wu2016,
	author = {Wu, Zhan and Zhang, Long and Sun, Wei and Xu, Xiao-Tian and Wang, Bao-Zong and Ji, Si-Cong and Deng, Youjin and Chen, Shuai and Liu, Xiong-Jun and Pan, Jian-Wei},
	doi = {10.1126/science.aaf6689},
	journal = {Science},
	number = {6308},
	pages = {83--88},
	publisher = {American Association for the Advancement of Science (AAAS)},
	title = {Realization of two-dimensional spin-orbit coupling for Bose-Einstein condensates},
	url = {https://doi.org/10.1126/science.aaf6689},
	volume = {354},
	year = {2016}}

@article{Sun2018a,
	author = {Sun, Wei and Wang, Bao-Zong and Xu, Xiao-Tian and Yi, Chang-Rui and Zhang, Long and Wu, Zhan and Deng, Youjin and Liu, Xiong-Jun and Chen, Shuai and Pan, Jian-Wei},
	doi = {10.1103/PhysRevLett.121.150401},
	issue = {15},
	journal = {Phys. Rev. Lett.},
	month = {Oct},
	numpages = {6},
	pages = {150401},
	publisher = {American Physical Society},
	title = {Highly Controllable and Robust 2D Spin-Orbit Coupling for Quantum Gases},
	url = {https://link.aps.org/doi/10.1103/PhysRevLett.121.150401},
	volume = {121},
	year = {2018}}

@article{Yuan2025,
	author = {Yuan, Huan and Yi, Chang-Rui and Guo, Jia-Yu and Cheng, Xiang-Can and Jiao, Rui-Heng and Zhang, Jinyi and Chen, Shuai and Pan, Jian-Wei},
	doi = {10.1103/5j5s-946k},
	issue = {6},
	journal = {Phys. Rev. Lett.},
	month = {Aug},
	numpages = {6},
	pages = {063403},
	publisher = {American Physical Society},
	title = {Observation of Kibble-Zurek Behavior across Topological Transitions of a Chern Band in Ultracold Atoms},
	url = {https://link.aps.org/doi/10.1103/5j5s-946k},
	volume = {135},
	year = {2025}}

@article{Song2019,
	author = {Song, Bo and He, Chengdong and Niu, Sen and Zhang, Long and Ren, Zejian and Liu, Xiong-Jun and Jo, Gyu-Boong},
	doi = {10.1038/s41567-019-0564-y},
	journal = {Nature Physics},
	number = {9},
	pages = {911--916},
	publisher = {Nature Publishing Group UK London},
	title = {Observation of nodal-line semimetal with ultracold fermions in an optical lattice},
	url = {https://doi.org/10.1038/s41567-019-0564-y},
	volume = {15},
	year = {2019}}

@article{Yi2019,
	author = {Yi, Chang-Rui and Zhang, Long and Zhang, Lin and Jiao, Rui-Heng and Cheng, Xiang-Can and Wang, Zong-Yao and Xu, Xiao-Tian and Sun, Wei and Liu, Xiong-Jun and Chen, Shuai and Pan, Jian-Wei},
	doi = {10.1103/PhysRevLett.123.190603},
	issue = {19},
	journal = {Phys. Rev. Lett.},
	month = {Nov},
	numpages = {6},
	pages = {190603},
	publisher = {American Physical Society},
	title = {Observing Topological Charges and Dynamical Bulk-Surface Correspondence with Ultracold Atoms},
	url = {https://link.aps.org/doi/10.1103/PhysRevLett.123.190603},
	volume = {123},
	year = {2019}}

@article{Liang2023,
	author = {Liang, Ming-Cheng and Wei, Yu-Dong and Zhang, Long and Wang, Xu-Jie and Zhang, Han and Wang, Wen-Wei and Qi, Wei and Liu, Xiong-Jun and Zhang, Xibo},
	doi = {10.1103/PhysRevResearch.5.L012006},
	issue = {1},
	journal = {Phys. Rev. Res.},
	month = {Jan},
	numpages = {7},
	pages = {L012006},
	publisher = {American Physical Society},
	title = {Realization of Qi-Wu-Zhang model in spin-orbit-coupled ultracold fermions},
	url = {https://link.aps.org/doi/10.1103/PhysRevResearch.5.L012006},
	volume = {5},
	year = {2023}}

@article{Sun2018b,
	author = {Sun, Wei and Yi, Chang-Rui and Wang, Bao-Zong and Zhang, Wei-Wei and Sanders, Barry C. and Xu, Xiao-Tian and Wang, Zong-Yao and Schmiedmayer, Joerg and Deng, Youjin and Liu, Xiong-Jun and Chen, Shuai and Pan, Jian-Wei},
	doi = {10.1103/PhysRevLett.121.250403},
	issue = {25},
	journal = {Phys. Rev. Lett.},
	month = {Dec},
	numpages = {5},
	pages = {250403},
	publisher = {American Physical Society},
	title = {Uncover Topology by Quantum Quench Dynamics},
	url = {https://link.aps.org/doi/10.1103/PhysRevLett.121.250403},
	volume = {121},
	year = {2018}}

@article{Boada2012,
	author = {Boada, O. and Celi, A. and Latorre, J. I. and Lewenstein, M.},
	doi = {10.1103/PhysRevLett.108.133001},
	issue = {13},
	journal = {Phys. Rev. Lett.},
	month = {Mar},
	numpages = {6},
	pages = {133001},
	publisher = {American Physical Society},
	title = {Quantum Simulation of an Extra Dimension},
	url = {https://link.aps.org/doi/10.1103/PhysRevLett.108.133001},
	volume = {108},
	year = {2012}}

@article{Gadway2015,
	author = {Gadway, Bryce},
	doi = {10.1103/PhysRevA.92.043606},
	issue = {4},
	journal = {Phys. Rev. A},
	month = {Oct},
	numpages = {9},
	pages = {043606},
	publisher = {American Physical Society},
	title = {Atom-optics approach to studying transport phenomena},
	url = {https://link.aps.org/doi/10.1103/PhysRevA.92.043606},
	volume = {92},
	year = {2015}}

@article{Meier2016,
	author = {Meier, Eric J. and An, Fangzhao Alex and Gadway, Bryce},
	doi = {10.1038/ncomms13986},
	journal = {Nature Communications},
	number = {1},
	pages = {13986},
	publisher = {Springer Nature},
	title = {Observation of the topological soliton state in the Su--Schrieffer--Heeger model},
	url = {https://doi.org/10.1038/ncomms13986},
	volume = {7},
	year = {2016}}

@article{Paladugu2024,
	author = {Paladugu, Sai Naga Manoj and Chen, Tao and An, Fangzhao Alex and Yan, Bo and Gadway, Bryce},
	doi = {10.1038/s42005-024-01526-8},
	journal = {Communications Physics},
	number = {1},
	pages = {39},
	publisher = {Nature Publishing Group UK London},
	title = {Injection spectroscopy of momentum state lattices},
	url = {https://doi.org/10.1038/s42005-024-01526-8},
	volume = {7},
	year = {2024}}

@article{Dong2025,
	author = {Dong, Zhaoli and Li, Hang and Wang, Hongru and Pan, Yichen and Yi, Wei and Yan, Bo},
	journal = {arXiv preprint arXiv:2501.13499},
	title = {Observation of higher-order topological bound states in the continuum using ultracold atoms},
	url = {https://arxiv.org/abs/2501.13499},
	year = {2025}}

@article{Xie2019,
	author = {Xie, Dizhou and Gou, Wei and Xiao, Teng and Gadway, Bryce and Yan, Bo},
	doi = {10.1038/s41534-019-0159-6},
	journal = {npj Quantum Information},
	number = {1},
	pages = {55},
	publisher = {Springer Nature},
	title = {Topological characterizations of an extended Su--Schrieffer--Heeger model},
	url = {https://doi.org/10.1038/s41534-019-0159-6},
	volume = {5},
	year = {2019}}

@article{Yuan2023,
	author = {Yuan, Tao and Zeng, Chao and Mao, Yi-Yi and Wu, Fei-Fei and Xie, Yan-Jun and Zhang, Wen-Zhuo and Dai, Han-Ning and Chen, Yu-Ao and Pan, Jian-Wei},
	doi = {10.1103/PhysRevResearch.5.L032005},
	issue = {3},
	journal = {Phys. Rev. Res.},
	month = {Jul},
	numpages = {7},
	pages = {L032005},
	publisher = {American Physical Society},
	title = {Realizing robust edge-to-edge transport of atomic momentum states in a dynamically modulated synthetic lattice},
	url = {https://link.aps.org/doi/10.1103/PhysRevResearch.5.L032005},
	volume = {5},
	year = {2023}}

@article{Zeng2024,
	author = {Zeng, Chao and Shi, Yue-Ran and Mao, Yi-Yi and Wu, Fei-Fei and Xie, Yan-Jun and Yuan, Tao and Zhang, Wei and Dai, Han-Ning and Chen, Yu-Ao and Pan, Jian-Wei},
	doi = {10.1103/PhysRevLett.132.063401},
	issue = {6},
	journal = {Phys. Rev. Lett.},
	month = {Feb},
	numpages = {6},
	pages = {063401},
	publisher = {American Physical Society},
	title = {Transition from Flat-Band Localization to Anderson Localization in a One-Dimensional Tasaki Lattice},
	url = {https://link.aps.org/doi/10.1103/PhysRevLett.132.063401},
	volume = {132},
	year = {2024}}

@article{Celi2014,
	author = {Celi, A. and Massignan, P. and Ruseckas, J. and Goldman, N. and Spielman, I. B. and Juzeli\ifmmode \bar{u}\else \={u}\fi{}nas, G. and Lewenstein, M.},
	doi = {10.1103/PhysRevLett.112.043001},
	issue = {4},
	journal = {Phys. Rev. Lett.},
	month = {Jan},
	numpages = {6},
	pages = {043001},
	publisher = {American Physical Society},
	title = {Synthetic Gauge Fields in Synthetic Dimensions},
	url = {https://link.aps.org/doi/10.1103/PhysRevLett.112.043001},
	volume = {112},
	year = {2014}}

@article{Mancini2015,
	author = {Mancini, Marco and Pagano, Guido and Cappellini, Giacomo and Livi, Lorenzo and Rider, Marie and Catani, Jacopo and Sias, Carlo and Zoller, Peter and Inguscio, Massimo and Dalmonte, Marcello and others},
	doi = {10.1126/science.aab1748},
	journal = {Science},
	number = {6255},
	pages = {1510--1513},
	publisher = {American Association for the Advancement of Science},
	title = {Observation of chiral edge states with neutral fermions in synthetic Hall ribbons},
	url = {https://science.sciencemag.org/content/349/6255/1510},
	volume = {349},
	year = {2015}}

@article{Roell2023,
	author = {Roell, Roberto Vittorio and Laskar, Arif Warsi and Huybrechts, Franz Richard and Weitz, Martin},
	doi = {10.1103/PhysRevA.107.043302},
	journal = {Physical Review A},
	number = {4},
	pages = {043302},
	publisher = {APS},
	title = {Chiral edge dynamics and quantum Hall physics in synthetic dimensions with an atomic erbium Bose-Einstein condensate},
	url = {https://link.aps.org/doi/10.1103/PhysRevA.107.043302},
	volume = {107},
	year = {2023}}

@article{Chalopin2020,
	author = {Chalopin, Thomas and Satoor, Tanish and Evrard, Alexandre and Makhalov, Vasiliy and Dalibard, Jean and Lopes, Raphael and Nascimbene, Sylvain},
	doi = {10.1038/s41567-020-0942-5},
	journal = {Nature Physics},
	number = {10},
	pages = {1017--1021},
	publisher = {Nature Publishing Group UK London},
	title = {Probing chiral edge dynamics and bulk topology of a synthetic Hall system},
	url = {https://doi.org/10.1038/s41567-020-0942-5},
	volume = {16},
	year = {2020}}

@article{Kanungo2022,
	author = {Kanungo, SK and Whalen, JD and Lu, Y and Yuan, M and Dasgupta, S and Dunning, FB and Hazzard, KRA and Killian, TC},
	doi = {10.1038/s41467-022-28550-y},
	journal = {Nature communications},
	number = {1},
	pages = {972},
	publisher = {Nature Publishing Group UK London},
	title = {Realizing topological edge states with Rydberg-atom synthetic dimensions},
	url = {https://doi.org/10.1038/s41467-022-28550-y},
	volume = {13},
	year = {2022}}

@article{Han2019,
	author = {Han, Jeong Ho and Kang, Jin Hyoun and Shin, Yong-il},
	doi = {10.1103/PhysRevLett.122.065303},
	journal = {Physical Review Letters},
	number = {6},
	pages = {065303},
	publisher = {APS},
	title = {Band gap closing in a synthetic Hall tube of neutral fermions},
	url = {https://link.aps.org/doi/10.1103/PhysRevLett.122.065303},
	volume = {122},
	year = {2019}}

@article{Li2022bose,
	author = {Li, Chuan-Hsun and Yan, Yangqian and Feng, Shih-Wen and Choudhury, Sayan and Blasing, David B and Zhou, Qi and Chen, Yong P},
	doi = {10.1103/PRXQuantum.3.010316},
	journal = {PRX Quantum},
	number = {1},
	pages = {010316},
	publisher = {APS},
	title = {Bose-Einstein condensate on a synthetic topological Hall cylinder},
	url = {https://link.aps.org/doi/10.1103/PRXQuantum.3.010316},
	volume = {3},
	year = {2022}}

@article{Fabre2022laughlin,
	author = {Fabre, Aur{\'e}lien and Bouhiron, Jean-Baptiste and Satoor, Tanish and Lopes, Raphael and Nascimbene, Sylvain},
	doi = {10.1103/PhysRevLett.128.173202},
	journal = {Physical Review Letters},
	number = {17},
	pages = {173202},
	publisher = {APS},
	title = {Laughlin's topological charge pump in an atomic Hall cylinder},
	url = {https://link.aps.org/doi/10.1103/PhysRevLett.128.173202},
	volume = {128},
	year = {2022}}

@article{Lohse2018,
	author = {Lohse, Michael and Schweizer, Christian and Price, Hannah M and Zilberberg, Oded and Bloch, Immanuel},
	doi = {10.1038/nature25000},
	journal = {Nature},
	number = {7686},
	pages = {55--58},
	publisher = {Nature Publishing Group UK London},
	title = {Exploring 4D quantum Hall physics with a 2D topological charge pump},
	url = {https://doi.org/10.1038/nature25000},
	volume = {553},
	year = {2018}}

@article{Zhou2023,
	author = {Zhou, T.-W. and Cappellini, G. and Tusi, D. and Franchi, L. and Parravicini, J. and Repellin, C. and Greschner, S. and Inguscio, M. and Giamarchi, T. and Filippone, M. and Catani, J. and Fallani, L.},
	doi = {10.1126/science.add1969},
	journal = {Science},
	number = {6656},
	pages = {427--430},
	title = {Observation of universal Hall response in strongly interacting Fermions},
	url = {https://www.science.org/doi/10.1126/science.add1969},
	volume = {381},
	year = {2023}}

@article{Zhou2025,
	author = {Zhou, T.-W. and Beller, T. and Masini, G. and Parravicini, J. and Cappellini, G. and Repellin, C. and Giamarchi, T. and Catani, J. and Filippone, M. and Fallani, L.},
	doi = {10.1038/s41467-025-65083-6},
	journal = {Nature Communications},
	number = {1},
	pages = {10247},
	title = {Measuring Hall voltage and Hall resistance in an atom-based quantum simulator},
	url = {https://doi.org/10.1038/s41467-025-65083-6},
	volume = {16},
	year = {2025}}

@article{Zheng2014,
	author = {Zheng, Wei and Zhai, Hui},
	doi = {10.1103/PhysRevA.89.061603},
	issue = {6},
	journal = {Phys. Rev. A},
	month = {Jun},
	numpages = {5},
	pages = {061603},
	publisher = {American Physical Society},
	title = {Floquet topological states in shaking optical lattices},
	url = {https://link.aps.org/doi/10.1103/PhysRevA.89.061603},
	volume = {89},
	year = {2014}}

@article{Jotzu2014,
	author = {Jotzu, Gregor and Messer, Michael and Desbuquois, R{\'e}mi and Lebrat, Martin and Uehlinger, Thomas and Greif, Daniel and Esslinger, Tilman},
	doi = {10.1038/nature13915},
	journal = {Nature},
	number = {7526},
	pages = {237--240},
	publisher = {Nature Publishing Group UK London},
	title = {Experimental realization of the topological Haldane model with ultracold fermions},
	url = {https://doi.org/10.1038/nature13915},
	volume = {515},
	year = {2014}}

@article{Flaschner2016,
	author = {Fl{\"a}schner, N. and Rem, B. S. and Tarnowski, M. and Vogel, D. and L{\"u}hmann, D.-S. and Sengstock, K. and Weitenberg, C.},
	doi = {10.1126/science.aad4568},
	journal = {Science},
	number = {6289},
	pages = {1091--1094},
	title = {Experimental reconstruction of the Berry curvature in a Floquet Bloch band},
	url = {https://www.science.org/doi/10.1126/science.aad4568},
	volume = {352},
	year = {2016}}

@article{Tarnowski2019,
	author = {Tarnowski, M. and {\"U}nal, F. N. and Fl{\"a}schner, N. and Rem, B. S. and Eckardt, A. and Sengstock, K. and Weitenberg, C.},
	doi = {10.1038/s41467-019-09668-y},
	journal = {Nature Communications},
	number = {1},
	pages = {1728},
	title = {Measuring topology from dynamics by obtaining the Chern number from a linking number},
	url = {https://www.nature.com/articles/s41467-019-09668-y},
	volume = {10},
	year = {2019}}

@article{Kang2020a,
	author = {Kang, Jin Hyoun and Han, Jeong Ho and Shin, Y},
	doi = {10.1088/1367-2630/ab61d7},
	journal = {New Journal of Physics},
	month = {jan},
	number = {1},
	pages = {013023},
	publisher = {IOP Publishing},
	title = {Creutz ladder in a resonantly shaken 1D optical lattice},
	url = {https://doi.org/10.1088/1367-2630/ab61d7},
	volume = {22},
	year = {2020}}

@article{Leonard2023FQH,
	author = {L{\'e}onard, Julian and Kim, Sooshin and Kwan, Joyce and Segura, Perrin and Grusdt, Fabian and Repellin, C{\'e}cile and Goldman, Nathan and Greiner, Markus},
	doi = {10.1038/s41586-023-06122-4},
	journal = {Nature},
	number = {7970},
	pages = {495--499},
	publisher = {Nature Publishing Group UK London},
	title = {Realization of a fractional quantum Hall state with ultracold atoms},
	url = {https://doi.org/10.1038/s41586-023-06122-4},
	volume = {619},
	year = {2023}}

@article{Haldane1988,
	author = {Haldane, F. D. M.},
	doi = {10.1103/PhysRevLett.61.2015},
	issue = {18},
	journal = {Phys. Rev. Lett.},
	month = {Oct},
	numpages = {0},
	pages = {2015--2018},
	publisher = {American Physical Society},
	title = {Model for a Quantum Hall Effect without Landau Levels: Condensed-Matter Realization of the "Parity Anomaly"},
	url = {https://link.aps.org/doi/10.1103/PhysRevLett.61.2015},
	volume = {61},
	year = {1988}}

@article{Goldman2014,
	author = {Goldman, N. and Dalibard, J.},
	doi = {10.1103/PhysRevX.4.031027},
	issue = {3},
	journal = {Phys. Rev. X},
	month = {Aug},
	numpages = {29},
	pages = {031027},
	publisher = {American Physical Society},
	title = {Periodically Driven Quantum Systems: Effective Hamiltonians and Engineered Gauge Fields},
	url = {https://link.aps.org/doi/10.1103/PhysRevX.4.031027},
	volume = {4},
	year = {2014}}

@article{Bukov2015,
	author = {Marin Bukov and Luca D'Alessio and Anatoli Polkovnikov},
	doi = {10.1080/00018732.2015.1055918},
	journal = {Advances in Physics},
	number = {2},
	pages = {139--226},
	publisher = {Taylor \& Francis},
	title = {Universal high-frequency behavior of periodically driven systems: from dynamical stabilization to {F}loquet engineering},
	url = {https://doi.org/10.1080/00018732.2015.1055918},
	volume = {64},
	year = {2015}}

@article{Rudner2013,
	author = {Rudner, Mark S. and Lindner, Netanel H. and Berg, Erez and Levin, Michael},
	doi = {10.1103/PhysRevX.3.031005},
	issue = {3},
	journal = {Phys. Rev. X},
	month = {Jul},
	numpages = {15},
	pages = {031005},
	publisher = {American Physical Society},
	title = {Anomalous Edge States and the Bulk-Edge Correspondence for Periodically Driven Two-Dimensional Systems},
	url = {https://link.aps.org/doi/10.1103/PhysRevX.3.031005},
	volume = {3},
	year = {2013}}

@article{Quelle2017,
	author = {Quelle, A and Weitenberg, C and Sengstock, K and Smith, C Morais},
	doi = {10.1088/1367-2630/aa8646},
	journal = {New Journal of Physics},
	month = {nov},
	number = {11},
	pages = {113010},
	publisher = {IOP Publishing},
	title = {Driving protocol for a Floquet topological phase without static counterpart},
	url = {https://doi.org/10.1088/1367-2630/aa8646},
	volume = {19},
	year = {2017}}

@article{Wintersperger2020,
	author = {Wintersperger, Karen and Braun, Christoph and {\"U}nal, F Nur and Eckardt, Andr{\'e} and Liberto, Marco Di and Goldman, Nathan and Bloch, Immanuel and Aidelsburger, Monika},
	doi = {10.1038/s41567-020-0949-y},
	journal = {Nature Physics},
	number = {10},
	pages = {1058--1063},
	publisher = {Nature Publishing Group UK London},
	title = {Realization of an anomalous Floquet topological system with ultracold atoms},
	url = {https://doi.org/10.1038/s41567-020-0949-y},
	volume = {16},
	year = {2020}}

@article{Lu2022,
	author = {Lu, Mingwu and Reid, G. H. and Fritsch, A. R. and Pi\~neiro, A. M. and Spielman, I. B.},
	doi = {10.1103/PhysRevLett.129.040402},
	issue = {4},
	journal = {Phys. Rev. Lett.},
	month = {Jul},
	numpages = {6},
	pages = {040402},
	publisher = {American Physical Society},
	title = {Floquet Engineering Topological Dirac Bands},
	url = {https://link.aps.org/doi/10.1103/PhysRevLett.129.040402},
	volume = {129},
	year = {2022}}

@article{Zhang2023tuning,
	author = {Zhang, Jin-Yi and Yi, Chang-Rui and Zhang, Long and Jiao, Rui-Heng and Shi, Kai-Ye and Yuan, Huan and Zhang, Wei and Liu, Xiong-Jun and Chen, Shuai and Pan, Jian-Wei},
	doi = {10.1103/PhysRevLett.130.043201},
	journal = {Physical Review Letters},
	number = {4},
	pages = {043201},
	publisher = {APS},
	title = {Tuning anomalous Floquet topological bands with ultracold atoms},
	url = {https://link.aps.org/doi/10.1103/PhysRevLett.130.043201},
	volume = {130},
	year = {2023}}

@article{Braun2024,
	author = {Braun, Christoph and Saint-Jalm, Rapha{\"e}l and Hesse, Alexander and Arceri, Johannes and Bloch, Immanuel and Aidelsburger, Monika},
	doi = {10.1038/s41567-024-02506-z},
	journal = {Nature Physics},
	number = {8},
	pages = {1306--1312},
	publisher = {Nature Publishing Group UK London},
	title = {Real-space detection and manipulation of topological edge modes with ultracold atoms},
	url = {https://doi.org/10.1038/s41567-024-02506-z},
	volume = {20},
	year = {2024}}

@article{Hesse2025,
	author = {Hesse, Alexander and Arceri, Johannes and Hornung, Moritz and Braun, Christoph and Aidelsburger, Monika},
	journal = {arXiv preprint arXiv:2508.20154},
	title = {Probing disorder-driven topological phase transitions via topological edge modes with ultracold atoms in Floquet-engineered honeycomb lattices},
	url = {https://arxiv.org/abs/2508.20154},
	year = {2025}}

@article{Xie2020,
	author = {Xie, Dizhou and Deng, Tian-Shu and Xiao, Teng and Gou, Wei and Chen, Tao and Yi, Wei and Yan, Bo},
	doi = {10.1103/PhysRevLett.124.050502},
	issue = {5},
	journal = {Phys. Rev. Lett.},
	month = {Feb},
	numpages = {6},
	pages = {050502},
	publisher = {American Physical Society},
	title = {Topological Quantum Walks in Momentum Space with a Bose-Einstein Condensate},
	url = {https://link.aps.org/doi/10.1103/PhysRevLett.124.050502},
	volume = {124},
	year = {2020}}

@article{Kitagawa2010,
	author = {Kitagawa, Takuya and Berg, Erez and Rudner, Mark and Demler, Eugene},
	doi = {10.1103/PhysRevB.82.235114},
	issue = {23},
	journal = {Phys. Rev. B},
	month = {Dec},
	numpages = {12},
	pages = {235114},
	publisher = {American Physical Society},
	title = {Topological characterization of periodically driven quantum systems},
	url = {https://link.aps.org/doi/10.1103/PhysRevB.82.235114},
	volume = {82},
	year = {2010}}

@article{Weber2022,
	author = {Weber, S. and Bai, R. and Makki, N. and M\"ogerle, J. and Lahaye, T. and Browaeys, A. and Daghofer, M. and Lang, N. and B\"uchler, H. P.},
	doi = {10.1103/PRXQuantum.3.030302},
	issue = {3},
	journal = {PRX Quantum},
	month = {Jul},
	numpages = {11},
	pages = {030302},
	publisher = {American Physical Society},
	title = {Experimentally Accessible Scheme for a Fractional Chern Insulator in Rydberg Atoms},
	url = {https://link.aps.org/doi/10.1103/PRXQuantum.3.030302},
	volume = {3},
	year = {2022}}

@article{Verresen2021,
	author = {Verresen, Ruben and Lukin, Mikhail D. and Vishwanath, Ashvin},
	doi = {10.1103/PhysRevX.11.031005},
	issue = {3},
	journal = {Phys. Rev. X},
	month = {Jul},
	numpages = {23},
	pages = {031005},
	publisher = {American Physical Society},
	title = {Prediction of Toric Code Topological Order from Rydberg Blockade},
	url = {https://link.aps.org/doi/10.1103/PhysRevX.11.031005},
	volume = {11},
	year = {2021}}

@article{Kalinowski2023,
	author = {Kalinowski, Marcin and Maskara, Nishad and Lukin, Mikhail D.},
	doi = {10.1103/PhysRevX.13.031008},
	issue = {3},
	journal = {Phys. Rev. X},
	month = {Jul},
	numpages = {22},
	pages = {031008},
	publisher = {American Physical Society},
	title = {Non-Abelian Floquet Spin Liquids in a Digital Rydberg Simulator},
	url = {https://link.aps.org/doi/10.1103/PhysRevX.13.031008},
	volume = {13},
	year = {2023}}

@article{Chen2024,
	author = {Chen, Yi-Hong and Wang, Bao-Zong and Poon, Ting-Fung Jeffrey and Zhou, Xin-Chi and Liu, Zheng-Xin and Liu, Xiong-Jun},
	doi = {10.1103/PhysRevResearch.6.L042054},
	issue = {4},
	journal = {Phys. Rev. Res.},
	month = {Dec},
	numpages = {7},
	pages = {L042054},
	publisher = {American Physical Society},
	title = {Proposal for realization and detection of {K}itaev quantum spin liquid with {R}ydberg atoms},
	url = {https://link.aps.org/doi/10.1103/PhysRevResearch.6.L042054},
	volume = {6},
	year = {2024}}

@article{Semeghini2021,
	author = {Semeghini, G. and Levine, H. and Keesling, A. and Ebadi, S. and Wang, T. T. and Bluvstein, D. and Verresen, R. and Pichler, H. and Kalinowski, M. and Samajdar, R. and Omran, A. and Sachdev, S. and Vishwanath, A. and Greiner, M. and Vuleti{\'c}, V. and Lukin, M. D.},
	doi = {10.1126/science.abi8794},
	journal = {Science},
	pages = {1242--1247},
	publisher = {American Association for the Advancement of Science},
	title = {Probing topological spin liquids on a programmable quantum simulator},
	url = {https://doi.org/10.1126/science.abi8794},
	volume = {374},
	year = {2021}}

@article{Evered2025,
	author = {Evered, Simon J. and Kalinowski, Marcin and Geim, Alexandra A. and Manovitz, Tom and Bluvstein, Dolev and Li, Sophie H. and Maskara, Nishad and Zhou, Hengyun and Ebadi, Sepehr and Xu, Muqing and Campo, Joseph and Cain, Madelyn and Ostermann, Stefan and Yelin, Susanne F. and Sachdev, Subir and Greiner, Markus and Vuleti{\'c}, Vladan and Lukin, Mikhail D.},
	day = {11},
	doi = {10.1038/s41586-025-09475-0},
	journal = {Nature},
	month = {9},
	pages = {341},
	title = {Probing the Kitaev honeycomb model on a neutral-atom quantum computer},
	url = {https://doi.org/10.1038/s41586-025-09475-0},
	volume = {645},
	year = {2025}}

@article{huang2025,
	author = {Chenxi Huang and Tao Chen and Qian Liang and Matthew A. Krebs and Ethan Springhorn and Ruiyu Li and Mingsheng Tian and Kaden R. A. Hazzard and Jacob P. Covey and Bryce Gadway},
	journal = {arXiv preprint arXiv:2512.12364},
	primaryclass = {cond-mat.quant-gas},
	title = {Interaction-assisted topological pumping in few- and many-atom Rydberg arrays},
	url = {https://arxiv.org/abs/2512.12364},
	year = {2025}}

\end{document}